\documentclass[reprint,prb,longbibliography,superscriptaddress]{revtex4-2}

\usepackage{hyperref}
\hypersetup{colorlinks=true, citecolor=blue, urlcolor=teal, linkcolor=blue}

\usepackage{amssymb, amsmath}
\usepackage[english]{babel}
\usepackage{bm}
\usepackage{graphicx}
\usepackage[utf8]{inputenc}
\usepackage{bbold}

\usepackage[dvipsnames]{xcolor}

\newcommand{\red}[1]{{ #1}}

\newcommand{\blue}[1]{{ #1}}

\mathchardef\myminus="2D

\usepackage{letltxmacro}
\LetLtxMacro{\ORIGselectlanguage}{\selectlanguage}
\makeatletter
\DeclareRobustCommand{\selectlanguage}[1]{%
  \@ifundefined{alias@\string#1}
    {\ORIGselectlanguage{#1}}
    {\begingroup\edef\x{\endgroup
       \noexpand\ORIGselectlanguage{\@nameuse{alias@#1}}}\x}%
}
\newcommand{\definelanguagealias}[2]{%
  \@namedef{alias@#1}{#2}%
}
\newcommand{\opvec}[1]{\hat{\mathbf{#1}}}
\newcommand{\qnm}[2]{\tilde{\mathbf{#1}}_{#2}}

\makeatother

\definelanguagealias{en}{english}
\definelanguagealias{EN}{english}
\definelanguagealias{eng}{english}
\definelanguagealias{de}{ngerman}

\begin{document}
\author{Robert Meiners Fuchs}
\email[]{r.meiners.fuchs@tu-berlin.de}
\affiliation{Institut für Physik und Astronomie, Nichtlineare Optik und
Quantenelektronik, Technische Universität Berlin, Hardenbergstr. 36, EW 7-1, 10623 Berlin, Germany}

\author{Juanjuan Ren}
\affiliation{Department of Physics, Engineering Physics, and Astronomy,
Queen’s University, Kingston, Ontario K7L 3N6, Canada}

\author{Sebastian Franke}
\affiliation{Institut für Physik und Astronomie, Nichtlineare Optik und
Quantenelektronik, Technische Universität Berlin, Hardenbergstr. 36, EW 7-1, 10623
Berlin, Germany}
\affiliation{Department of Physics, Engineering Physics, and Astronomy,
Queen’s University, Kingston, Ontario K7L 3N6, Canada}

\author{Stephen Hughes}
\affiliation{Department of Physics, Engineering Physics, and Astronomy,
Queen’s University, Kingston, Ontario K7L 3N6, Canada}

\author{Marten Richter}
\email[]{marten.richter@tu-berlin.de}
\affiliation{Institut für Physik und Astronomie, Nichtlineare Optik und
Quantenelektronik, Technische Universität Berlin, Hardenbergstr. 36, EW 7-1, 10623
Berlin, Germany}

\title{\red{Correlation functions for q}uantum dynamics of coupled quasinormal modes and quantum emitters interacting via 
finite-delay propagating photons}



\begin{abstract}
A time-dependent theory for the interactions between spatially separated lossy cavities in a homogeneous background medium using quantized quasinormal modes (QNMs) is presented. The cavities interact via a bath of traveling photons, described by non-bosonic operators that are orthogonal to the open-cavity QNMs. The retarded (i.e., time-delayed) inter-cavity dynamics are fully described by system-bath correlation functions, in which the emission from one cavity appears as the input field for another. Coupling between quantum emitters (described as two-level systems), placed inside a cavity or embedded in an external medium, and the electromagnetic field (cavity modes and bath photons) is included in the theory, which gives rise to both bath-mediated and QNM-mediated interactions between the emitters. 
\end{abstract}


\date{\today}
\maketitle

\section{Introduction}
\label{sec:intro}
A photonic cavity interacting with a quantum emitter such as an atom or quantum dot is a fundamental system in cavity quantum electrodynamics \cite{walther2006cavity, carmichael2009statistical}. Tuning the cavity frequency near resonance with the emitter allows for an increase of the spontaneous emission via a Purcell enhancement \cite{purcell1946resonance, purcell1946spontaneous, walther2006cavity}, making coupled cavity-emitter systems key to many quantum technologies, such as \red{lasers} \cite{meschede1985one, scully1988two, an1994microlaser, pscherer2021single, wang2017solution, rivero2021non}, \red{spasers} \red{(surface plasmon amplification by stimulated emission of radiation)} \cite{bergman2003surface, noginov2009demonstration, suh2012plasmonic, song2018three, hsieh2020perovskite, epstein2020far}, and 
quantum information processing
\cite{pellizzari1995decoherence, cirac1997quantum, pellizzari1997quantum, imamog1999quantum, duan2001long, blais2004cavity, benito2019optimized, borjans2020resonant, li2021quantum}. For many applications, including quantum networks, the transmission of quantum states between spatially separated systems is important, and quantum dynamical calculations must include {\it time delays} to account for photon propagation between the subsystems \cite{pichler2016photonic, regidor2021modeling, richter2022enhanced,Bundgaard-Nielsen2024WaveguideQED.jl:Electrodynamics,arranzRegidor2026Qwavemps}.

Theoretically, such cavity systems are usually treated within a rotating-wave approximation using the Jaynes-Cummings model \cite{jaynes1963comparison}, which describes the interaction of a quantum emitter with the modes of a {\it closed} cavity (with quantized normal modes). Losses to the cavity and coupling to a surrounding bath or other systems are typically added through phenomenological coupling, e.g., with Lindblad superoperators \cite{cirac1997quantum, blais2004cavity, cirac1991two}.
Such approaches assume that the losses act only as a small perturbation to the cavity-emitter system so that the ``normal modes'' of a closed cavity are a good approximation of the field inside the open cavity. Open cavity effects, such as the non-Hermitian coupling between modes \cite{franke2019quantization, ren2022connecting} are often neglected, and coupling elements and decay rates must be obtained from additional calculations or experimental data. 

The natural modes (i.e., source-free) of an open optical or plasmonic resonator are termed quasinormal modes (QNMs), which are subject to open boundary conditions, and they account for radiative and possible non-radiative losses that are inherent to the system \cite{leung1994completeness, muljarov2011brillouin, kristensen2012generalized, sauvan2013theory}. 
Conveniently, QNMs of practical cavity systems allow for an expansion of the physical fields in terms of only a few (and often just one) dominant modes, which makes them an intuitive and rigorous tool for modeling lossy resonators \cite{muljarov2011brillouin, lalanne2018light, kristensen2020modeling, el2020quasinormal, ren2021quasinormal,PhysRevA.107.013722, gustin2025what}.

Quantized QNMs yield rigorously defined decay rates and coupling elements for quantum dynamics calculations \cite{ho1998second, franke2019quantization, fuchs2023hierarchical, fuchs2024quantization}. However, the current quantization schemes from Refs.~\cite{franke2019quantization, franke2020quantized, fuchs2024quantization} yield creation/annihilation operators for \textit{quasibound} QNMs at the resonator and a photonic bath that the QNMs interact with \cite{fuchs2024quantization, franke2020quantized}. The quantized QNMs are quasibound in the sense that they are primarily localized at their respective cavity, and the overlap with modes from distant cavities decreases exponentially with increasing separation \cite{fuchs2024quantization}. Information about causality and the propagation of the QNMs is removed from the QNM operators and is instead contained in the bath and enters the quantum dynamics via the system-bath coupling \cite{fuchs2024quantization, franke2020quantized}.

So far, the focus of the quantized QNM theory has mostly been on short-range interactions \cite{franke2019quantization, ren2022connecting} and coupling without time delay \cite{fuchs2024quantization} (i.e., instantaneous coupling), where the influence of the bath can be largely neglected while the QNMs couple directly to each other and quantum emitters near the resonator. In separated systems, the bath is expected to play a critical role as a mediator of time-delayed interactions between the subsystems.
Preliminary work on quantum dynamics, including the system-bath interaction, has yielded accurate system decay rates \cite{franke2020quantized} as well as time-delayed dynamics for structures with well-separated systems \cite{fuchs2023hierarchical}. However, a thorough treatment of the system-bath interactions and QNM time dynamics is,
to the best of our knowledge, still lacking for these quantum field theories.

\red{In this paper, we extend the multi-cavity quantized QNM theory from Ref.~\onlinecite{fuchs2024quantization} into a general theory for the time dynamics of multiple QNM cavities and quantum emitters coupled to propagating photons in a homogeneous background medium. Towards this, three steps must be taken: (i) quantum emitters must be included in the Hamiltonian and the direct coupling between the bound QNMs and emitters must be formulated; (ii) the bath operators which arise from the QNM quantization are by construction non-bosonic. The resulting complex dynamics of the bath operators must be solved rigorously; and (iii),  we require system-bath correlation functions which are needed as input to most standard quantum dynamics schemes, such as the time-convolutionless (TCL) \cite{breuer2002theory} method or the hierarchical equations of motion (HEOM) \cite{tanimura2020numerically}. Similar QNM correlation functions were used in quantum dynamics calculations, e.g., in Refs.~\onlinecite{franke2020quantized, fuchs2023hierarchical, gustin2025dissipation}. To make these correlation functions feasible to calculate numerically, we must formulate approximate expression in terms of coupling parameters that can be obtained from standard Maxwell solvers.}

\red{We note that, while} coupling in a homogeneous medium has limited practical applications (e.g., the transmission of quantum states via satellites \cite{yin2017satellite, yin2017satellite1200}), it is an established model that allows one to focus on the details of the quantization scheme and derivation of coupling elements. Extensions to structured environments (e.g., coupling via open waveguides) are \red{non-trivial, since the different dimensionality must be taken into account in the boundary conditions and in the construction and quantization of the QNMs. A discussion of the theory for different kinds of background structures can be found in Ref.~\onlinecite{fuchs2026quasinormal}.}

Our theoretical analysis is split into two parts: In Sec.~\ref{sec:quanti}, we discuss the non-retarded coupling, i.e., the direct interaction between quasibound QNMs and quantum emitters without mediation by the bath photons\red{, i.e., the first step from above.} We model the emitters as two-level systems (TLSs), which couple to the electromagnetic field via dipole coupling. The extension to multi-level emitters is straightforward. We define an \textit{area of direct influence} \(P_i(\mathbf{R}_a)\) around the cavity that can be used to quantify the strength of the interaction between the QNMs and TLS. For a TLS outside this area of direct influence, the coupling to the QNMs of that cavity is negligible, unless there is a time delay to account for photon propagation through the background medium. 

Subsequently, in Sec.~\ref{sec:corrfunc}, we \red{give a dynamical description of} the interactions mediated by the non-bosonic photonic bath. We \red{evaluate the} dynamics of the bath\red{, thus addressing the second step from above.} \red{We then define} system-bath correlation functions that can be straightforwardly applied to standard open quantum systems approaches \red{and derive} retarded 
(i.e., finite-time) coupling parameters based on the QNM theory that can be obtained from numerical calculations for QNM-QNM, QNM-TLS, and TLS-TLS interactions\red{, thus taking the third step. See} Tab.~\ref{tab:summary} for an overview of the different kinds of couplings and correlation functions. 
We provide examples of the numerical calculation of the correlation functions using two spatially separated 3D metal dimers (including material dispersion and losses through a Drude model) as QNM cavities, and TLSs in the gap centers of the dimers, serving as an example of quantum emitters.

\begin{table}
    \centering
    \caption{Overview of the couplings and correlation functions appearing between multiple cavities and QNMs and reference to the manuscript parts.}
    \begin{tabular}{c c c c}
       \hline\hline Coupling & time delay & Section & Equation\\ \hline
         QNM-QNM & \(\bm \times\) & \ref{sec:multicavityquanti} & \eqref{eq:chiplusdef} \\
         QNM-QNM & \checkmark & \ref{sec:qnmcorrfunc} & \eqref{eq:QNMcorrfunc} \\
         TLS-TLS & \checkmark & \ref{sec:tlscorrfunc} & \eqref{eq:tlscorrfunc}\\ 
         QNM-TLS & \(\bm \times\) & \ref{sec:qnmtlscoup} & \eqref{eq:qnmtlscoup}\\
         QNM-TLS & \checkmark & \ref{sec:qnmtlscorrfunc} & \eqref{eq:QTcorr}\\\hline\hline
    \end{tabular}
    \label{tab:summary}
\end{table}

\section{Non-retarded coupling between QNMs and quantum emitters\label{sec:quanti}}
\subsection{Quantum electrodynamics in dissipative media}
\begin{figure}
    \centering
    \includegraphics[width=0.95\columnwidth]{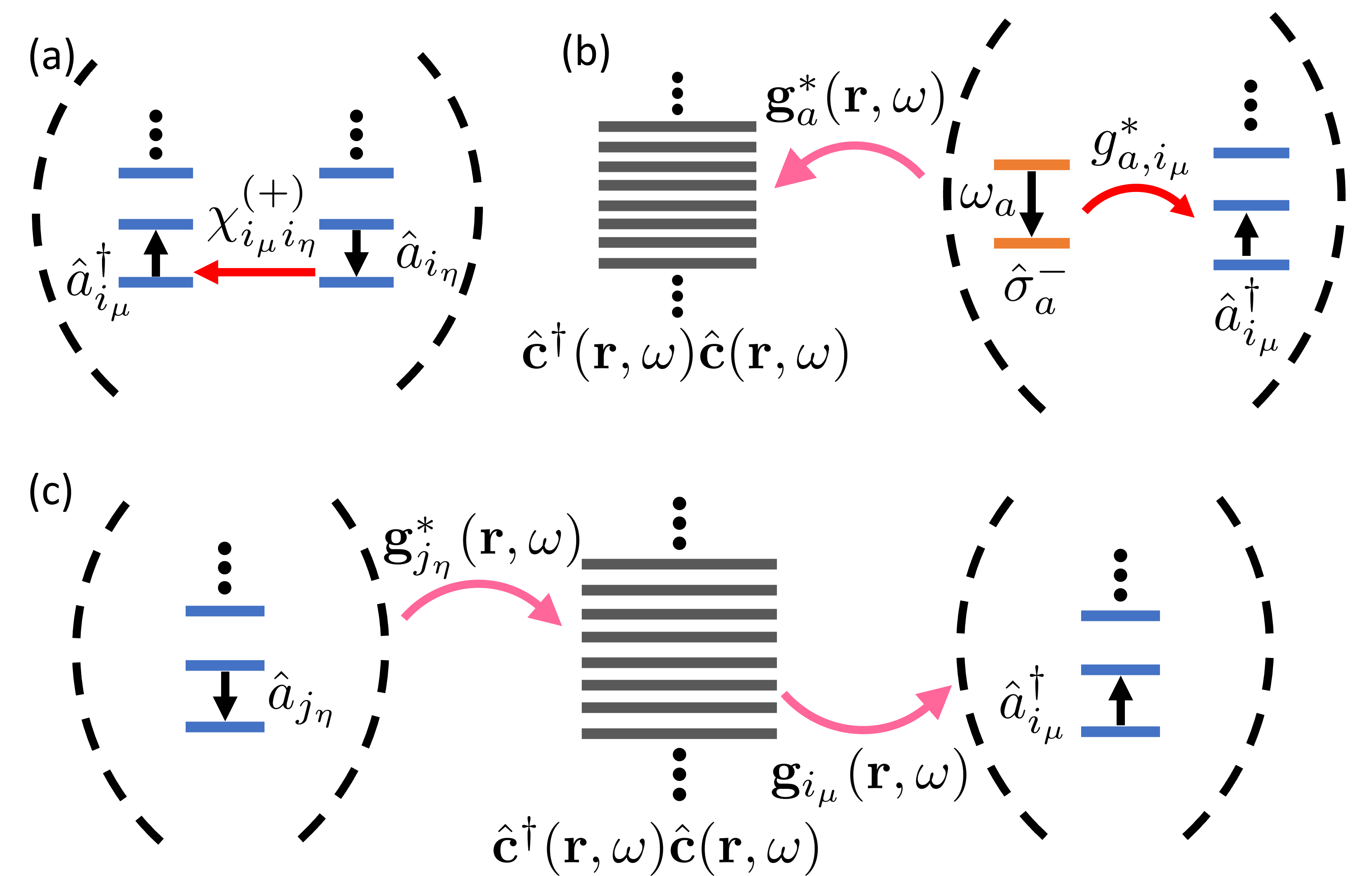}
    \caption{Schemes of the different \red{kinds of interactions considered in this work [cf.~Eq.~\eqref{eq:qnmham}].} (a) QNM part of the system Hamiltonian, with the annihilation of a photon in mode \(i_{\eta}\) and creation of a photon in mode \(i_{\mu}\). (b) TLS Hamiltonian, where a TLS with frequency \(\omega_a\) interacts directly with the QNMs and the continuum of bath modes. (c) QNM-bath interaction, where two different cavities couple to the same photonic bath.}
    \label{fig:Hamiltonian}
\end{figure}

We consider a system of TLSs \red{interacting with the full electric field} in a spatially inhomogeneous, dispersive, and absorptive medium. \red{The medium consists of optical cavities or plasmonic resonators as well as a homogeneous background medium. The different kinds of interactions for this type of setup are sketched in Fig.~\ref{fig:Hamiltonian}.} 
In a rotating wave approximation, the \red{general} Hamiltonian for this interaction reads
\begin{align}\label{eq:welschham} 
    H =& \hbar\sum_a \omega_a\hat{\sigma}^+_a\hat{\sigma}^-_a + \hbar\int_0^{\infty}\mathrm{d}\omega\int\mathrm{d}^3r\, \omega \opvec{b}^{\dagger}(\mathbf{r},\omega)\cdot \opvec{b}(\mathbf{r},\omega)\nonumber\\
    &-\left(\sum_a \hat{\sigma}^+_a\int_0^{\infty}\mathrm{d}\omega\, \mathbf{d}_a\cdot \opvec{E}(\mathbf{r}_a,\omega) + \mathrm{H.a.}\right),
\end{align}
where \(\omega_a\) is the transition frequency between the lower and upper state of TLS \(a\), while \(\hat{\sigma}^+_a\)(\(\hat{\sigma}^-_a\)) is the raising (lowering) operator of the TLS. The electric field operator in the dipole coupling reads \cite{dung1998three, suttorp2004field, philbin2010canonical, scheel1998qed, matloob1999electromagnetic, PhysRevA.53.1818},
\begin{align} \label{eq:welschquanti}
    \opvec{E}(\mathbf{r},\omega) = \frac{i}{\omega\epsilon_0}\int\mathrm{d}^3r'\mathbf{G}(\mathbf{r},\mathbf{r}',\omega)\cdot\opvec{j}_N(\mathbf{r}',\omega),
\end{align}
where \(\opvec{j}_N(\mathbf{r},\omega)=\omega\sqrt{(\hbar\epsilon_0/\pi)\epsilon_I(\mathbf{r},\omega)}\opvec{b}(\mathbf{r},\omega)\) is the noise-current density that results from the absorption of radiation by the dissipative media. The operators \(\opvec{b}(\mathbf{r},\omega)\) are bosonic, with continuous spatial and frequency indices,
\begin{align*}
    \left[\hat{b}_p(\mathbf{r},\omega),\hat{b}_q^{\dagger}(\mathbf{r}',\omega')\right]_- = \delta_{pq}\delta(\mathbf{r}-\mathbf{r}')\delta(\omega-\omega').
\end{align*}

The Green's function \(\mathbf{G}\) fulfills the Helmholtz equation
(satisfying the same open boundary conditions as the QNMs)
\begin{align}\label{eq:helmgreen}
    \left[\nabla\times\nabla\times-\frac{\omega^2}{c^2}\epsilon(\mathbf{r},\omega)\right] \mathbf{G}(\mathbf{r},\mathbf{r}',\omega) = \frac{\omega^2}{c^2}\mathbb{1}\delta(\mathbf{r}-\mathbf{r}'),
\end{align}
where \(\epsilon(\mathbf{r},\omega) = \epsilon_R(\mathbf{r},\omega)+i\epsilon_I(\mathbf{r},\omega)\). In this paper, we assume purely absorptive media for the cavity regions (\(\epsilon_I(\mathbf{r},\omega) > 0\)), but cases with gain have been discussed, e.g., in Refs.~\onlinecite{ren2021quasinormal, franke2022quantized, vanDrummen2025gain}. 

\red{While the Hamiltonian from Eq.~\eqref{eq:welschham} is formally exact, it has several limitations for the kinds of systems we are interested in here. Most notably, it lacks the modal structure that is required for describing rigorous open cavity-QED,
including unique quantum optical effects such as the quantum Rabi model \cite{guimond2016rabi, zueco2019ultrastrongly}, 
as well as ultrastrong and broadband cavity-emitter coupling \cite{gustin2023gauge, gustin2025what, gustin2025dissipation}, or multi-photon propagation effects \cite{cirac1997quantum, pellizzari1997quantum, fuchs2023hierarchical}. Furthermore, it is not directly applicable to systems where the bound cavity field is an observable of interest (e.g., in photonic quantum computing), and instead, separate, commuting operators for the bound cavity modes and propagating photons are desirable. Lastly, a description of the coupling between different cavities or between emitters and cavities in terms of coupling parameters has conceptual and practical advantages over a formulation in terms of the full electromagnetic Green's function as in Eq.~\eqref{eq:welschham}, which can become unfeasible to calculate for large, complex networks of many cavities. In this way, the quantized QNM formalism which we present here can be applied to rigorous quantum optics schemes \cite{franke2019quantization, gustin2025dissipation}, and connects directly to common quantum dynamics methods \cite{franke2020quantized, fuchs2023hierarchical, gustin2025what}.}

\subsection{QNMs in classical calculations}
We consider a photonic structure containing spatially separated electromagnetic resonators in a homogeneous, non-absorptive background medium. The whole system is divided into the resonator volumes \(V_i\), and the {\it outside volume} \(V_{\rm out}\), where \(\epsilon(\mathbf{r},\omega)|_{r\in V_{\rm out}}= \epsilon_{\rm B}\). The lossy resonators (i.e., cavity structures) have two main loss channels: non-radiative absorption by the resonator material and radiative losses into the surrounding medium. The
QNMs are solutions to the source-free Helmholtz equation \cite{lee1999dyadic, muljarov2011brillouin, lalanne2018light, kristensen2020modeling},
\begin{align} \label{eq:helmquasi}
    \nabla\times\nabla\times\Tilde{\mathbf{f}}_{i_{\mu}}(\mathbf{r}) -\frac{\Tilde{\omega}^2_{i_{\mu}}}{c^2}\epsilon(\mathbf{r},\Tilde{\omega}_{i_{\mu}})\Tilde{\mathbf{f}}_{i_{\mu}}(\mathbf{r}) = 0,
\end{align}
under open boundary conditions, such as the Silver-Müller radiation condition \cite{muller1948grundzuge, silver1984microwave, gumerov2005fast}
\begin{align} \label{eq:qnmsilvmüll}
    \frac{\mathbf{r}}{r}\times\nabla\times\Tilde{\mathbf{f}}_{i_{\mu}}(\mathbf{r}) \to -in_{\rm B}\frac{\Tilde{\omega}_{i_{\mu}}}{c}\Tilde{\mathbf{f}}_{i_{\mu}}(\mathbf{r}),\quad r\to \infty,
\end{align}
which accounts for the emission of radiation into the surrounding medium. Note that \(\Tilde{\mathbf{f}}_{i_{\mu}}(\mathbf{r})\) is the \(\mu\)-th QNM of the \(i\)-th cavity. The presence of the other cavities is neglected in the Helmholtz equation (setting \(\epsilon(\mathbf{r},\omega) = \epsilon_{\rm B}\) at the other cavity volumes) or treated as a perturbation \cite{franke2023impact}. When this approximation does not hold, the cavities are treated as a single system using hybridized QNMs \cite{franke2019quantization, ren2022connecting}. 

The QNMs have complex eigenfrequencies \(\Tilde{\omega}_{i_{\mu}} = \omega_{i_{\mu}}-i\gamma_{i_{\mu}}\) with \(\gamma_{i_{\mu}}>0\), since we assume a purely absorptive (lossy) resonator medium. Due to the separation of the cavities (usually more than one QNM wavelength), an expansion of the Green's function in terms of the QNMs of the \(i\)-th cavity is given by \cite{leung1996completeness, ge2014quasinormal}
\begin{align}\label{eq:qnmgreen}
    \mathbf{G}^i_{ff}(\mathbf{r},\mathbf{r}',\omega) = \sum_{\mu}A_{i_{\mu}}(\omega)\qnm{f}{i_{\mu}}(\mathbf{r})\qnm{f}{i_{\mu}}(\mathbf{r}'),
\end{align}
for \(\mathbf{r},\mathbf{r}'\in V_i\), is assumed to be a good approximation of the photonic Green's function. Here, \(A_{i_{\mu}}(\omega) = \omega/(2(\Tilde{\omega}_{i_{\mu}}-\omega))\). 
As remarked earlier, for most practical applications, the sum in Eq.~\eqref{eq:qnmgreen} can be restricted to a few dominant modes or even just a single QNM. 

Due to the complex eigenfrequencies, the QNMs diverge spatially in the far field (corresponding to temporal decay in the time domain). While properly normalized QNMs can possibly be used to expand physical properties in terms of QNMs in the outside region, the fields outside the cavities are generally continuous, and expansions require many QNMs \cite{abdelrahman2018completeness}. 
However, for quantum dynamical calculations, where the Hilbert space scales exponentially with the number of modes, expansions in terms of a few dominant modes are necessary. Hence, we replace the QNMs outside the resonator with frequency-dependent, non-divergent {\it regularized fields}, which are obtained via a Dyson equation \cite{ge2014quasinormal}, or alternatively, via the field equivalence principle \cite{franke2020quantized, franke2023impact, fuchs2026greens} 
\begin{align}\label{eq:regQNM}
    &\qnm{F}{i_{\mu}}(\mathbf{R},\omega) \nonumber\\
    &= \frac{c^2}{\omega^2}\oint_{\mathcal{S}_i}\mathrm{d}A_s \Big\{\Big[\nabla_s\times\mathbf{G}_B(\mathbf{s},\mathbf{R},\omega)\Big]^T\cdot\Big[\opvec{n}_s\times\qnm{f}{i_{\mu}}(\mathbf{s})\Big]\nonumber\\
    &\qquad\qquad-\Big[\opvec{n}_s\times\mathbf{G}_B(\mathbf{s},\mathbf{R},\omega)\Big]^T\cdot\Big[\nabla_s\times\qnm{f}{i_{\mu}}(\mathbf{s})\Big]\Big\},
\end{align}
where \(\mathcal{S}_i\) is a closed near-field surface around the cavity volume \(V_i\), \(\mathbf{G}_B\) is the Green's function of the homogeneous background medium, which solves Eq.~\eqref{eq:helmgreen} for \(\epsilon(\mathbf{r},\omega)=\epsilon_{\rm B}\), and \(\opvec{n}_s\) is the surface vector on \(\mathcal{S}_i\) pointing outwards with respect to \(V_i\).

Note that, \red{for explicit numerical calculations}, the regularized fields, \(\qnm{F}{i_{\mu}}(\mathbf{R},\omega)\), give an accurate expansion of the fields only for some minimum distance away from the resonator, usually on the order of half the QNM wavelength \cite{ren2020near}. Close to the surface, the QNMs, \(\qnm{f}{i_{\mu}}(\mathbf{r})\), are far better suited for expansions of the fields. To distinguish these cases, we use the lowercase \(\mathbf{r}\) as a generic position vector, while using the uppercase \(\mathbf{R}\) in places where the difference is essential to denote positions sufficiently far away from the resonator for the expansion using the regularized fields to hold.

\subsection{Multi-cavity quantization of QNMs}\label{sec:multicavityquanti}
\red{We now summarize the QNM quantization from Ref.~\onlinecite{fuchs2024quantization} assuming multiple cavities with a weak equal-time intercavity modal overlap. If the overlap between two cavities is not weak, they are treated as a single system with hybridized modes and joint operators \cite{franke2019quantization, ren2022connecting, fuchs2024quantization}.}

We construct QNM operators by projecting the noise operators, \(\opvec{b}(\mathbf{r},\omega)\), onto the subspace of symmetrized QNMs, \(\qnm{f}{i_{\mu}}^s(\mathbf{r}) = \sum_{\eta}\left(S^{1/2}\right)_{i_{\eta}i_{\mu}}\sqrt{\omega_{i_{\eta}}/\omega_{i_{\mu}}}\qnm{f}{i_{\eta}}(\mathbf{r})\) \cite{franke2019quantization, franke2020quantized, fuchs2024quantization}, via
\begin{align}\label{eq:adef}
    \hat{a}_{i_{\mu}} = \int_0^{\infty}\text{d}\omega\int\text{d}^3r\, \mathbf{L}_{i_{\mu}}(\mathbf{r},\omega)\cdot\opvec{b}(\mathbf{r},\omega),
\end{align}
where the QNM projectors are defined as
\begin{align} \label{eq:Ldef}
    &\mathbf{L}_{i_{\mu}}(\mathbf{r},\omega) = \sum_{\eta}\left(S^{-1/2}\right)_{i_{\mu}i_{\eta}}\sqrt{\frac{2}{\pi\omega_{i_{\eta}}}}A_{i_{\eta}}(\omega)\nonumber\\
    &\qquad\times\left[\chi_{V_i}(\mathbf{r})\sqrt{\epsilon_I(\mathbf{r},\omega)}\qnm{f}{i_{\eta}}(\mathbf{r})+\chi_{\overline{V_i}}(\mathbf{r})\sqrt{\epsilon_{B,I}}\qnm{F}{i_{\eta}}(\mathbf{r},\omega)\right],
\end{align}
and the function \(\chi_{V_i}(\mathbf{r})\) is unity if \({\bf r}\in V_i\), and zero elsewhere.

By setting \(\epsilon(\mathbf{r},\omega) = \epsilon_{\rm B}\) everywhere except inside cavity \(i\) in the projector \(\mathbf{L}_{i_{\mu}}(\mathbf{r},\omega)\), we ensure that the operators \(\hat{a}_{i_{\mu}}\) only act on the QNMs of one specific cavity, in contrast to approaches using hybridized QNMs (cf.~Refs.~\onlinecite{franke2019quantization, franke2020quantized, fuchs2024quantization}).
The overlap matrix \(S_{i_{\mu}j_{\eta}}= \delta_{ij}S^{\rm intra}_{i_{\mu}j_{\eta}}+(1-\delta_{ij})S^{\rm inter}_{i_{\mu}j_{\eta}}\), contains intracavity contributions of modes of the same cavity (\(i=j\)) and intercavity contributions between modes of different cavities (\(i\neq j\)). The intracavity contributions read \cite{franke2019quantization, fuchs2024quantization},
\begin{align}  \label{eq:Sintradef}
    &S^{\rm intra}_{i_{\mu}i_{\eta}} = \frac{2}{\pi\sqrt{\omega_{i_{\mu}}\omega_{i_{\eta}}}}\int_0^{\infty}\mathrm{d}\omega\, A_{i_{\mu}}(\omega)A^*_{i_{\eta}}(\omega)\nonumber\\
    &\quad\qquad\times\Big\{\int_{V_i}\mathrm{d}^3r\epsilon_I(\mathbf{r},\omega)\qnm{f}{i_{\mu}}(\mathbf{r})\cdot\qnm{f}{i_{\eta}}^*(\mathbf{r})\nonumber\\
    &\qquad\quad+\frac{1}{2\omega\epsilon_0}\oint_{\mathcal{S}_i}\mathrm{d}A_s\Big[\left(\qnm{H}{i_{\mu}}(\mathbf{s},\omega)\times\opvec{n}_s\right)\cdot\qnm{F}{i_{\eta}}^*(\mathbf{s},\omega)\nonumber\\
    &\quad\qquad\qquad\qquad\qquad\qquad\qquad+\mathrm{c.c.}(\mu\leftrightarrow \eta)\Big]\Big\},
\end{align}
where the first term in the curly brackets are related to non-radiative absorption losses, while the second term is related to radiative losses through the cavity surface (photon decay to the far field). 

The intercavity overlap contributions between modes of separate cavities reads,
\begin{align} \label{eq:Sinterdef}
    &S^{\rm inter}_{i_{\mu}j_{\eta}}\Big|_{i\neq j} = \frac{2}{\pi\sqrt{\omega_{i_{\mu}}\omega_{j_{\eta}}}}\int_0^{\infty}\mathrm{d}\omega\,\frac{A_{i_{\mu}}(\omega)A^*_{j_{\eta}}(\omega)}{2\omega \epsilon_0}\nonumber\\
    &\quad\times\Big\{\oint_{\mathcal{S}_i}\mathrm{d}A_s\Big[\left(\qnm{H}{i_{\mu}}(\mathbf{s},\omega)\times\opvec{n}_s\right)\cdot\qnm{F}{j_{\eta}}^*(\mathbf{s},\omega)\nonumber\\
    &\qquad\qquad\qquad\qquad\qquad\qquad\qquad{+}\mathrm{c.c.}(i_{\mu}\leftrightarrow j_{\eta})\Big]\nonumber\\
    &\qquad+\oint_{\mathcal{S}_j}\mathrm{d}A_s\Big[\left(\qnm{H}{i_{\mu}}(\mathbf{s},\omega)\times\opvec{n}_s\right)\cdot\qnm{F}{j_{\eta}}^*(\mathbf{s},\omega)\nonumber\\
    &\qquad\qquad\qquad\qquad\qquad\qquad\qquad{+}\mathrm{c.c.}(i_{\mu}\leftrightarrow j_{\eta})\Big]\Big\},
\end{align}
where \(\opvec{n}_s\) is the normal vector on the cavity surface \(\mathcal{S}_i\) pointing outwards with respect to the cavity volume \(V_i\), and \(\qnm{H}{i_{\mu}}(\mathbf{r},\omega) = \nabla\times\qnm{F}{i_{\mu}}(\mathbf{r},\omega)/(i\omega\mu_0)\) are the regularized magnetic field QNMs. Practical calculations of the surface integrals are usually carried out by integrating over an equivalent far-field surface, where the regularized fields are obtained via an efficient near-field to far-field transformation \cite{ren2020near, fuchs2024quantization}.

In Ref.~\onlinecite{fuchs2024quantization}, it was shown that the intercavity overlap decreases exponentially with the separation of the cavities, characterized by the \textit{cavity separation parameter},
\begin{align} \label{eq:cavsepparam}
    P_{i_{\mu}j_{\eta}} = \gamma_{i_{\mu}j_{\eta}}^{\min}n_{\rm B}R_{ij}/c -\mathrm{ln}(D_{i_{\mu}j_{\eta}}^{\max}),
\end{align}
where \(\gamma_{i_{\mu}j_{\eta}}^{\min} = \min[\gamma_{i_{\mu}},\gamma_{j_{\eta}}]\), \(R_{ij}\) is the distance between the cavities, and \(D_{i_{\mu}j_{\eta}}^{\max} = \max[D_{i_{\mu}}(\Omega_{ij}), D_{j_{\eta}}(\Omega_{ji})]\), where \(D_{i_{\mu}}(\Omega_{ij})\) is the directionality \cite{balanis2016antenna} of the emission from mode \(i_{\mu}\) into the solid angle \(\Omega_{ij}\) between the cavities \(i\) and \(j\). For high-\(Q\) cavities, the separation parameter is amended by an additional term \(+\mathrm{ln}(Q_{i_{\mu}}Q_{j_{\eta}})/2\) \cite{fuchs2024quantization}, where \(Q_{i_{\mu}} = \omega_{i_{\mu}}/(2\gamma_{i_{\mu}})\) is the quality factor of the mode \(i_{\mu}\), but we omit this correction here for generality.

For well-separated cavities (large \(P_{i_{\mu}j_{\eta}}\)), the inter-cavity overlap is negligible, and the cavities can be treated as separate systems. \red{Under this assumption,} the QNM projector kernels \(\mathbf{L}_{i_{\mu}}(\mathbf{r},\omega)\) are orthogonal under the quantized QNM subspace inner product:
\begin{align} \label{eq:Linnerproduct}
    \int\mathrm{d}^3r\int_0^{\infty}\mathrm{d}\omega \mathbf{L}_{i_{\mu}}(\mathbf{r},\omega)\cdot\mathbf{L}^*_{j_{\eta}}(\mathbf{r},\omega) = \delta_{ij}\delta_{\mu\eta},
\end{align}
and the QNM operators from Eq.~\eqref{eq:adef} are bosonic \([\hat{a}_{i_{\mu}},\hat{a}^{\dagger}_{j_{\eta}}] = \delta_{ij}\delta_{\mu\eta}\). \red{Note that the inverse square root \(\left(S^{-1/2}\right)_{i_{\mu}i_{\eta}}\) of the overlap matrix in Eq.~\eqref{eq:Ldef} refers to the intracavity overlap from Eq.~\eqref{eq:Sintradef} only, meaning the QNM operators are diagonalized \textit{with respect to the single cavity}. Therefore, Eq.~\eqref{eq:Linnerproduct} holds exactly for \(i=j\) and for \(i\neq j\) if the cavity separation parameter \(P_{i_{\mu}j_{\nu}}\gg 0\) is large.}

The \(\omega\)-integration reduces the information about causality and propagation contained in the quantized QNMs. Instead, the QNM operators create and annihilate {\it quasibound} QNMs that are mostly concentrated at the cavity \cite{fuchs2024quantization}. Information about the propagation of states between systems is contained in a photonic bath that is orthogonal to the QNM projectors,
\begin{align} \label{eq:cortho}
    \int_0^{\infty}\text{d}\omega\int\text{d}^3r\, \mathbf{L}_{i_{\mu}}(\mathbf{r},\omega)\cdot \opvec{c}(\mathbf{r},\omega) = 0,
\end{align}
so that the QNM operators \(\hat{a}_{i_{\mu}}\) and the bath operators \(\opvec{c}(\mathbf{r},\omega)\) commute. We formally decompose the noise operators \(\opvec{b}(\mathbf{r},\omega)\),
\begin{align}\label{eq:bdecomp}
    \opvec{b}(\mathbf{r},\omega) = \sum_{i,\mu}\mathbf{L}_{i_{\mu}}^*(\mathbf{r}, \omega)\hat{a}_{i_{\mu}} + \opvec{c}(\mathbf{r},\omega).
\end{align}

The presence of the bath ensures causality \cite{fuchs2024quantization}, and the system-bath interaction yields the temporal decay of the quantized modes \cite{franke2020quantized} as well as the propagation of states between separated systems \cite{fuchs2023hierarchical}. 
Inserting the decomposition from Eq.~\eqref{eq:bdecomp} into Eq.~\eqref{eq:welschquanti}, we obtain the electric field operator in system-bath form:
\begin{align}
    \opvec{E}(\mathbf{r},\omega) =& \sum_{i,\mu}\mathbf{E}_{i_{\mu}}(\mathbf{r},\omega)\hat{a}_{i_{\mu}} \nonumber\\
    &+ i\sqrt{\frac{\hbar}{\epsilon_0\pi}}\int\mathrm{d}^3r'\sqrt{\epsilon_I(\mathbf{r}',\omega)}\mathbf{G}(\mathbf{r},\mathbf{r}',\omega)\cdot\opvec{c}(\mathbf{r}',\omega).
\end{align}

We have inserted the \textit{QNM-generated electric field} \(\mathbf{E}_{i_{\mu}}(\mathbf{r},\omega)\) here, which is defined below in Eq.~\eqref{eq:qnmgenefeld} and will be discussed in the following section. 
With this decomposition of the electric field operator, we also bring the Hamiltonian from Eq.~\eqref{eq:welschham} into system-bath form, which reads \red{(a sketch of the different terms is given in Fig.~\ref{fig:Hamiltonian})}
\begin{align}\label{eq:qnmham}
    H &= H_S+H_B+H_{SB}\nonumber\\
    &= \hbar\sum_a \omega_a\hat{\sigma}^+_a\hat{\sigma}^-_a + \hbar\sum_{i,\mu\eta}\chi^{(+)}_{i_{\mu}i_{\eta}}\hat{a}^{\dagger}_{i_{\mu}}\hat{a}_{i_{\eta}}\nonumber\\
    &-i\hbar\sum_a\sum_{i,\mu}g_{a,i_{\mu}}\hat{\sigma}_a^+\hat{a}_{i_{\mu}} +\mathrm{H.a.}\nonumber\\
    &+\hbar\int_0^{\infty}\mathrm{d}\omega\int\mathrm{d}^3r\, \omega \opvec{c}^{\dagger}(\mathbf{r},\omega)\cdot \opvec{c}(\mathbf{r},\omega)\nonumber\\
    &+\hbar\sum_{i,\mu}\int_0^{\infty}\mathrm{d}\omega\int\mathrm{d}^3r \omega\mathbf{L}_{i_{\mu}}(\mathbf{r},\omega)\cdot \opvec{c}(\mathbf{r},\omega)\hat{a}^{\dagger}_{i_{\mu}}+ \mathrm{H.a.}\nonumber\\
    &-i\hbar\sum_a \int_0^{\infty}\mathrm{d}\omega\int\mathrm{d}^3r \mathbf{g}_a(\mathbf{r},\omega)\cdot \opvec{c}(\mathbf{r},\omega)\hat{\sigma}^+_a + \mathrm{H.a.},
\end{align}
where the system Hamiltonian \(H_S\) contains the free TLS and QNM contributions and TLS-QNM coupling. Here, the matrix
\begin{align} \label{eq:chiplusdef}
    &\chi^{(+)}_{i_{\mu}i_{\eta}} = \left(\int_0^{\infty}\text{d}\omega\int\text{d}^3r\, \omega\, \mathbf{L}_{i_{\mu}}(\mathbf{r},\omega)\cdot\mathbf{L}^*_{i_{\eta}}(\mathbf{r},\omega)\right)\nonumber\\ 
    &\approx \frac{1}{2}\sum_{\mu'\eta'} \left(S^{-1/2}\right)_{i_{\mu}i_{\mu'}}(\Tilde{\omega}_{i_{\mu'}}+\Tilde{\omega}^*_{i_{\eta'}})  S_{i_{\mu'}i_{\eta'}}\left(S^{-1/2}\right)_{i_{\eta'}i_{\eta}}
\end{align}
is the real part of the symmetrized eigenfrequency \(\chi_{i_{\mu}i_{\eta}} = \sum_{\mu'} \left(S^{-1/2}\right)_{i_{\mu}i_{\mu'}}\Tilde{\omega}_{i_{\mu'}}\left(S^{1/2}\right)_{i_{\mu'}i_{\eta}}\) (a derivation is given in Ref.~\onlinecite{franke2020quantized}), and 
\begin{align}\label{eq:fullqnmtlscoup}
    g_{a,i_{\mu}} = -\frac{i}{\hbar}\int_0^{\infty}\mathrm{d}\omega\, \mathbf{d}_a\cdot \mathbf{E}_{i_{\mu}}(\mathbf{r}_a,\omega),
\end{align}
is the full instantaneous dipole coupling between the TLS \(a\) and local electric field generated by the QNM \(i_{\mu}\). 

Meanwhile, the bath Hamiltonian \(H_B\) contains the bath contributions. The system-bath coupling \(H_{SB}\) describes the coupling of the TLS and QNM to the bath, where 
\begin{align}\label{eq:tlsbathcoup}
    \mathbf{g}_a(\mathbf{r},\omega) = \sqrt{\epsilon_I(\mathbf{r},\omega)/(\pi\hbar\epsilon_0)}\, \mathbf{d}_a\cdot\mathbf{G}(\mathbf{r}_a,\mathbf{r},\omega),
\end{align}
is the coupling between the TLS and the photonic bath. Note that there is no instantaneous coupling between different cavities in the system part since the cavities are assumed to be well-separated, and the direct (non-bath-mediated) coupling is, therefore, negligible. When this assumption does not hold, the cavities must be quantized together and treated as a single system. The interaction with a shared photonic bath gives rise to an indirect, retarded interaction between the distant systems, which will be discussed in Sec.~\ref{sec:corrfunc}.

\subsection{Direct QNM-TLS coupling}\label{sec:qnmtlscoup}
First, we discuss the direct (non-bath-mediated) interaction between the QNMs and TLSs. \red{We again assume well-separated cavities with negligible instantaneous intercavity overlap (\(P_{i_{\mu}j_{\nu}}\gg 0\)).
The TLSs couple to the} QNM-generated electric field,
\begin{align}\label{eq:qnmgenefeld}
    \mathbf{E}_{i_{\mu}}(\mathbf{r},\omega) = \frac{i}{\omega\epsilon_0}\int\mathrm{d}^3r'\mathbf{G}(\mathbf{r},\mathbf{r}',\omega)\cdot\mathbf{j}_{i_{\mu}}(\mathbf{r}',\omega).
\end{align}
In analogy to the noise-current density operator \(\mathbf{\hat j}_N\) from Eq.~\eqref{eq:welschquanti}, we defined here a (classical) current density generated by the QNM, \(\mathbf{j}_{i_{\mu}}(\mathbf{r}, \omega) = \omega\sqrt{\frac{\hbar \epsilon_0}{\pi}\epsilon_I(\mathbf{r}, \omega)}\mathbf{L}^*_{i_{\mu}}(\mathbf{r}, \omega)\) which is the source of the QNM-generated electric field that the TLS interacts with. 
Equation \eqref{eq:qnmgenefeld} is the formal solution to the Helmholtz equation
\begin{align} \label{eq:helmqnmgenfield}
    \nabla\times\nabla\times\mathbf{E}_{i_{\mu}}(\mathbf{r},\omega)-\frac{\omega^2}{c^2}\epsilon(\mathbf{r},\omega)\mathbf{E}_{i_{\mu}}(\mathbf{r},\omega) = i\omega\mu_0\mathbf{j}_{i_{\mu}}(\mathbf{r},\omega).
\end{align}

The QNM projectors, \(\mathbf{L}_{i_{\mu}}\), contain the QNMs and regularized QNM fields, which can be calculated using established numerical software 
for solving (classical) Maxwell's equations with finite 
cavity structures \cite{lalanne2018light, franke2019quantization, kristensen2020modeling, ren2020near}. Thus, the QNM-current density \(\mathbf{j}_{i_{\mu}}\) as a source in the Helmholtz equation can be used to obtain the QNM-generated electric field numerically. It is clear, however, that solving this equation is not numerically feasible for large and complex structures. Instead, the QNM-generated electric field can be approximated by expanding the Green's function in Eq.~\eqref{eq:qnmgenefeld} in terms of few, dominant QNMs \cite{fuchs2026greens}. 

For a TLS inside one of the cavities (\( \mathbf{r}_a\in V_i\)), the modes of that cavity are dominant, with only small corrections due to incoming fields from other cavities \cite{fuchs2024quantization}. These incoming fields are negligible for well-separated cavities (cf.~Sec.~\ref{sec:multicavityquanti}), and the direct (non-time delayed) coupling for a TLS inside a cavity reads (see Appendix~\ref{appsec:qnmtlscoup} for details),
\begin{align} \label{eq:qnmtlscoup}
    g^{\rm in}_{a,i_{\mu}} = \sqrt{\frac{\omega_{i_{\mu}}}{2\hbar\epsilon_0}}\chi_{V_i}(\mathbf{r}_a)\mathbf{d}_a\cdot\qnm{f}{i_{\mu}}^s(\mathbf{r}_a),
\end{align}
where \(\chi_{V_i}(\mathbf{r}_a)\) ensures that only coupling to the QNMs of the cavity in which the TLS is located contributes. Equation~\eqref{eq:qnmtlscoup} is identical to the single-cavity case in Ref.~\onlinecite{franke2019quantization}, since we neglected the influence of other cavities. 

\begin{figure}
    \centering
    \includegraphics[width=0.9\linewidth]{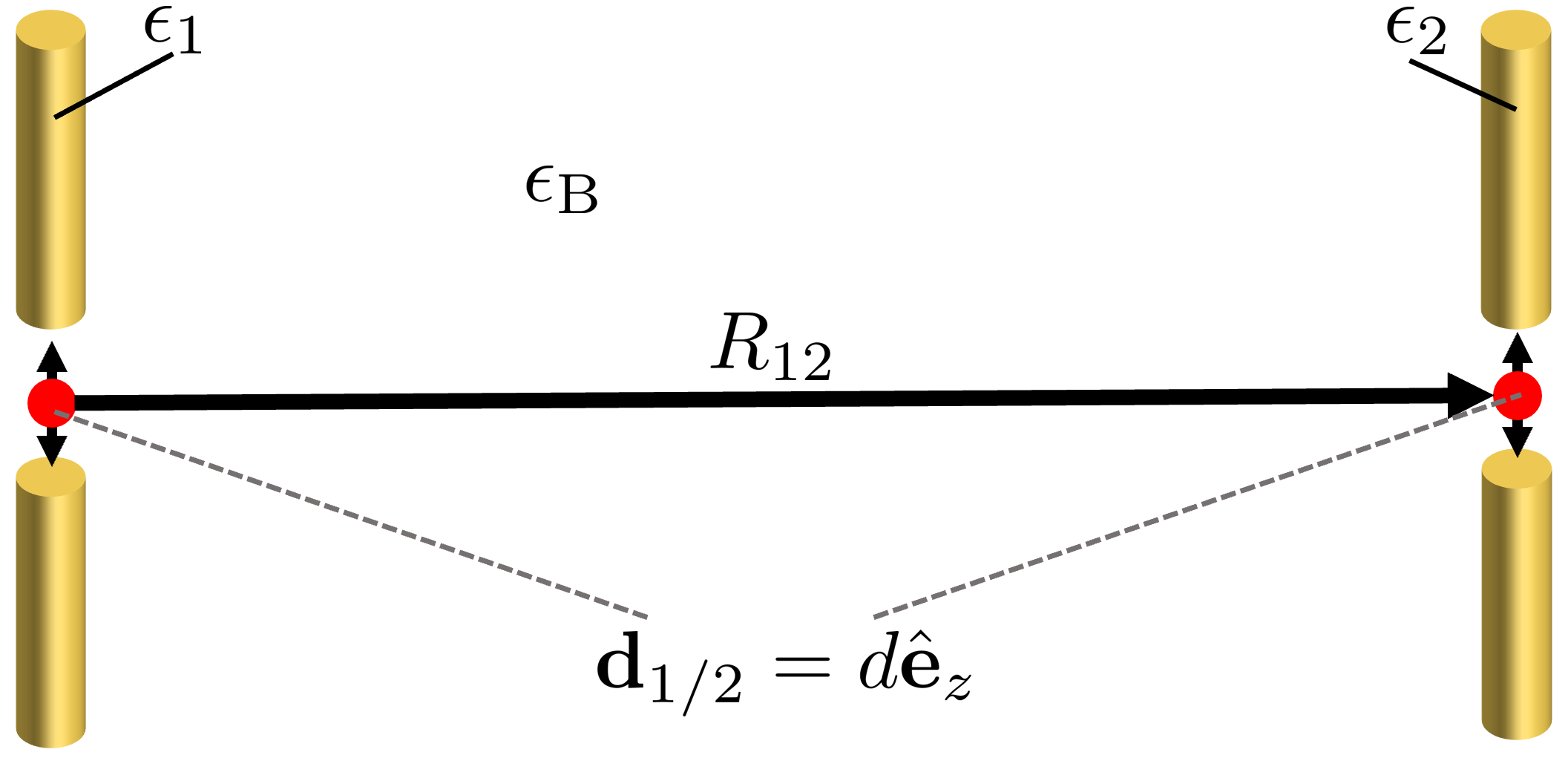}
    \caption{Sketch of two coupled metal dimers in vacuum (\(\epsilon_{\rm B} = 1\)), with \(z\)-polarized dipoles in the dimer gaps. The dimers are described by a Drude permittivity (cf.~Appendix~\ref{appsec:numerics}), and are separated by a distance of \(R_{12} = 2020\,{\rm nm}\). In the frequency regime of interst, the dimers have two dominant QNMs each, with \(\hbar\tilde{\omega}_1 = (1.6904-0.0652i)\,{\rm eV}\) and \(\hbar\tilde{\omega}_2 = (1.6482-0.0388i)\,{\rm eV}\).}
    \label{fig:dimer_sketch}
\end{figure}

To test the approximation from Eq.~\eqref{eq:qnmtlscoup} in the following discussion, we consider a system consisting of two metal dimers in vacuum (\(n_{\rm B}=1\)) serving as QNM cavities with one dominant QNM each and a TLS in each dimer gap (cf.~Fig.~\ref{fig:dimer_sketch}). Dimer 1 consists of two identical cylindrical gold-like nanorods with base radius \(r_1 = 10\,{\rm nm}\) and length \(L_1 = 80\,{\rm nm}\) each. The gap distance between the rods is \(d_1 = 10\,{\rm nm}\). In the frequency regime of interest, the dominant QNM of dimer 1 has the eigenfrequency \(\hbar\tilde{\omega}_1 = (1.6904-0.0652i)\,{\rm eV}\), which corresponds to a wavelength of \(\lambda_1 \approx 734\,{\rm nm}\). Accordingly, dimer 2 consists of two identical nanorods with \(r_2 = 10\,{\rm nm}\), \(L_2=90\,{\rm nm}\) and \(d_2 = 20\,{\rm nm}\), yielding a dominant QNM with \(\hbar\tilde{\omega}_2 = (1.6482-0.0388i)\,{\rm eV}\) and \(\lambda_2 \approx 752\,{\rm nm}\). The dimers are separated by a center-to-center distance of \(R_{12}=2020\,{\rm nm}\). We stress that 
all calculations are fully three-dimensional and include dispersive losses at the level of a Drude model (see Appendix~\ref{appsec:numerics} for details on the numerical calculations).

In Tab.~\ref{tab:g_in}, we show the normalized difference between the full TLS-QNM coupling from Eq.~\eqref{eq:fullqnmtlscoup} (using the Green's function expansion from Appendix~\ref{appsec:greensexp}) and the approximated coupling from Eq.~\eqref{eq:qnmtlscoup}. The difference is negligible in this case (fractions of a percent).

\begin{table}
    \centering
    \caption{Normalized difference between the full TLS-QNM coupling [cf.~Eq.~\eqref{eq:fullqnmtlscoup}] and approximated coupling [cf.~Eq.~\eqref{eq:qnmtlscoup}] for a dipole inside a cavity (\(\mathbf{r}_{a_i}\in V_i\)) for the model system from Fig.~\ref{fig:dimer_sketch} with two TLSs in the gap centers of metal dimers serving as QNM cavities with a center-to-center distance of \(R_{12} = 2020\,{\rm nm}\). Each dimer is assumed to be dominated by a single QNM in the frequency regime of interest.}
    \begin{tabular}{c c c}
      \hline\hline  & \(i=1\) & \(i=2\)  \\ \hline
      \(|(g_{a_i,i}-g^{\rm in}_{a_i,i})/g_{a_i,i}|\) & \(4.08\cdot 10^{-8}\) & \(3.94\cdot 10^{-8}\)\\ \hline\hline
    \end{tabular}
    \label{tab:g_in}
\end{table}

For a TLS outside the cavities (\(\mathbf{R}_a\in V_{\rm out}\)), the field is generally a complicated overlap of emissions from the different cavities as well as scattered fields \cite{fuchs2026greens}. On the other hand, propagation delays are relevant for the interaction between a TLS and a distant cavity. Since the coupling \(g_{a,i_{\mu}}\) from Eq.~\eqref{eq:fullqnmtlscoup} contains no time delay, the coupling between distant systems is suppressed and instead, these systems only couple indirectly via the propagating bath photons (cf.~Sec.~\ref{sec:corrfunc}).

\begin{figure}
    \centering
    \includegraphics[width=0.9\columnwidth]{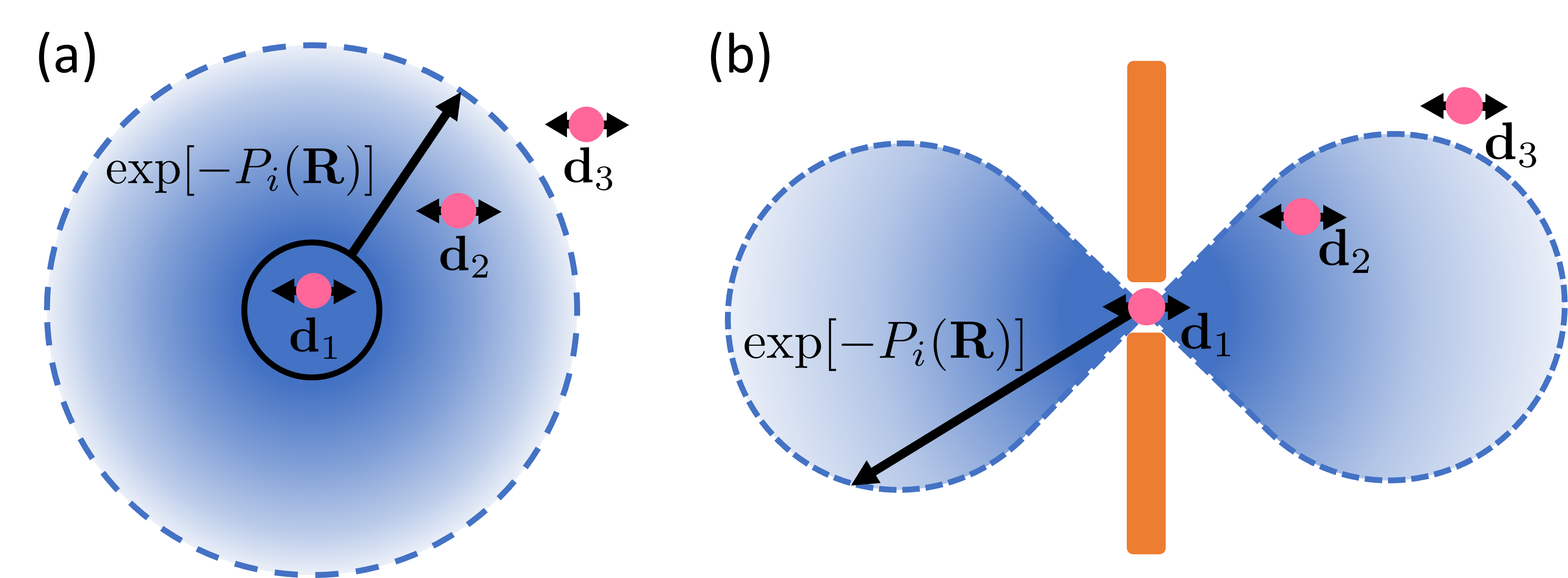}
    \caption{Sketch of the area of direct influence for the instantaneous interaction of a TLS with different cavity types from Eq.~\eqref{eq:tlsqnmseparation}. (a) Spherical cavity with isotropic emission. (b) Dimer nanoantenna with dipole-like emission. The TLS \(\mathbf{d}_1\) inside the cavity couples to the quasibound cavity modes, while TLS \(\mathbf{d}_2\) in the outside medium couples to the emission from the cavity. The coupling strength decreases exponentially with the separation of the TLS from the cavity [cf. Eq.~\eqref{eq:tlsqnmseparation}]. The TLS  \(\mathbf{d}_3\) lies outside the area of direct influence and does not significantly feel the presence of the cavity, unless there is a time delay to match the photon retardation.}
    \label{fig:cavity_tls}
\end{figure}

As we show in Appendix~\ref{appsec:qnmtlscoup}, the direct (non-retarded) coupling between a QNM and TLS outside decreases exponentially with the distance between the TLS and the cavity, as a result of the quasibound nature of the quantized QNMs. Furthermore, the coupling is impacted by the directionality, \(D_{i_{\lambda}}(\mathbf{R}_a)\), of the emission from the QNM cavity in the direction of the TLS. 
Taken together, these considerations allow us to define an {\it area of direct influence} around the cavity, via
\begin{align}\label{eq:tlsqnmseparation}
    P_{i}(\mathbf{R}_a)= \gamma^{\min}_i n_{\rm B}R_a/c - \mathrm{ln}[D^{\max}_i(\hat{\mathbf{R}}_a)],
\end{align}
with \(\gamma^{\min}_i = \min_{\mu}(\gamma_{i_{\mu}})\), and \(D_i^{\max}(\hat{\mathbf{R}}_a) =\max_{\mu}[D_{i_{\mu}}(\hat{\mathbf{R}}_a)]\), for all \(\mu\) within the frequency regime of interest. Here, \(\hat{\mathbf{R}}_a\) is the unit vector in direction of \(\mathbf{R}_a\). The area of direct influence is sketched in Fig.~\ref{fig:cavity_tls} for different cavity types. Similar to the cavity separation parameter from Eq.~\eqref{eq:cavsepparam}, the area of direct influence characterizes the separation of a QNM cavity and TLS in the outside medium. For an emitter outside this area of influence, the direct coupling is negligible (\(g^{\rm out}_{a,i_{\mu}}\approx 0\)). Therefore, an expansion of the Green's function in Eq.~\eqref{eq:qnmgenefeld} in terms of the QNMs of a few, nearby cavities is sufficient to obtain the direct coupling between the QNMs and a TLS in the outside medium.

\section{Photon propagation effects between spatially separated QNM cavities and quantum emitters\label{sec:corrfunc}}
Bound cavity modes and quantum emitters coupled to a bath of propagating photons are typical examples of an open quantum system. In methods used for calculating the dynamics of open quantum systems such as Nakajima-Zwanzig, TCL equations \cite{breuer2002theory}, HEOM or quantum path integrals (including modern implementations with tensor networks) \cite{tanimura2020numerically, fuchs2023hierarchical, vagov2011real, strathearn2017efficient, richter2022enhanced} the coupling between system and bath is connected to system-bath correlation functions  \cite{caldeira1983path, grabert1988quantum, chernyak1996collective, garraway1997nonperturbative}.

For the setup considered here, the correlation functions are derived from the system-bath coupling and are of the form
\begin{align} \label{eq:gencorrfunc}
    &C_{\mathbb{x}\mathbb{y}}(t,t')= \sum_{pq}\int\mathrm{d}^3r\int_0^{\infty}\mathrm{d}\omega\int\mathrm{d}^3r'\int_0^{\infty}\mathrm{d}\omega'\nonumber\\
    &\times g_{\mathbb{x},p}(\mathbf{r},\omega)g^*_{\mathbb{y},q}(\mathbf{r}',\omega') \mathrm{tr}_B\left[\hat{c}_p(\mathbf{r},\omega,t)\hat{c}^{\dagger}_q(\mathbf{r}',\omega',t')\rho_B\right],
\end{align}
where \(\mathrm{tr}_B[...]\) is the trace over the bath degrees of freedom, and \(\rho_B = |{\rm vac}\rangle\langle {\rm vac}|\) is assumed to be the initial vacuum bath state. 

The bath operators are time-dependent because the open quantum system methods mentioned above are routinely formulated in the interaction picture, i.e., the operators carry the free evolution of the system \(H_S\) and the bath \(H_B\). The correlation function describes the emission of a photon into the bath at time \(t'\) and reabsorption at a later time \(t\) (or the other way around, depending on the time order), yielding an effective interaction between the systems. The indices \(\mathbb{x},\mathbb{y}\) are placeholders for the QNM or TLS indices, and \(p,q\) are the polarization degrees of freedom of the photon. The interaction strength is determined by the integrals over the system-bath coupling. In contrast to the direct (non-mediated) coupling discussed thus far, this bath-mediated coupling includes a time delay and, therefore, allows for time-delayed interactions ($t \approx t' + t_{\rm delay}$)  as well as time-local processes ($ t \approx t'$).

\subsection{Time evolution of a photonic bath disturbed by the presence of open cavities}\label{sec:bathdyn}
The noise operators from Eq.~\eqref{eq:welschquanti} describe the electromagnetic response of a dissipative or dispersive medium that creates the electromagnetic field. By separating the QNM contributions, we implicitly defined bath operators in Eq.~\eqref{eq:bdecomp} that mainly account for the homogeneous background medium. Without cavities, the bath photons would freely propagate through the medium. The presence of the cavities perturbs the propagation and leads to a non-bosonic commutation relation for the bath operators, similar to the non-bosonic commutator of the electromagnetic vector potential due to the restriction of transversality \cite{franke2020quantized},
\begin{align} \label{eq:ccomm}
    \left[\hat{c}_p(\mathbf{r},\omega),\hat{c}^{\dagger}_q(\mathbf{r}',\omega')\right]_- &= \delta_{pq}\delta(\mathbf{r}-\mathbf{r}')\delta(\omega-\omega')\nonumber\\
    &-\sum_{k,\mu}L^*_{k_{\mu},p}(\mathbf{r},\omega)L_{k_{\mu},q}(\mathbf{r}',\omega').
\end{align}

\red{The non-bosonic contribution was neglected in previous works on QNM quantum dynamics \cite{franke2020quantized, fuchs2023hierarchical}, implicitly assuming highly localized modes. For a general theory, a rigorous treatment of the non-bosonic dynamics is required. We derive a general solution to the bath dynamics in the following, under the assumption that instantaneous coupling betwee well-separated cavities (\(P_{i_{\mu}j_{\nu}}\gg 0\)) is negligible.}

The sum over the QNM projectors on the right-hand side is the completeness relation on the quantized QNM subspace, which is here removed from the full space \(\delta\) distribution (analog to a transverse $\delta$ function in quantum optics). Calculating the dynamics with respect to the free bath Hamiltonian \(H_B\) thus yields,
\begin{align} \label{eq:cDGL}
    &\partial_t\opvec{c}(\mathbf{r},\omega,t) = \frac{i}{\hbar}\left[H_B,\opvec{c}(\mathbf{r},\omega,t)\right]\nonumber\\
    &\;= -i\omega \opvec{c}(\mathbf{r},\omega,t) +i\sum_{i,\mu}\mathbf{L}^*_{i_{\mu}}(\mathbf{r},\omega)\nonumber\\
    &\qquad\times\int\mathrm{d}^3r'\int_0^{\infty}\mathrm{d}\omega'\mathbf{g}_{i_{\mu}}(\mathbf{r}',\omega')\cdot\opvec{c}(\mathbf{r}',\omega',t),
\end{align}
where, in the last term, we used the orthogonality relation from Eq.~\eqref{eq:cortho} to replace \(\omega'\mathbf{L}_{i_{\mu}}(\mathbf{r}',\omega')\) from Eq.~\eqref{eq:qnmham} with the noise coupling elements (cf.~Ref.~\onlinecite{franke2020quantized}):
\begin{align}\label{eq:gcoup}
    &\mathbf{g}_{i_{\mu}}(\mathbf{r},\omega) \nonumber\\
    &= \sum_{\mu'\eta}\left(S^{-1/2}\right)_{i_{\mu}i_{\mu'}}(\omega-\tilde{\omega}_{i_{\mu'}})\left(S^{1/2}\right)_{i_{\mu'}i_{\eta}}\mathbf{L}_{i_{\eta}}(\mathbf{r},\omega),
\end{align}
i.e., the pole at the QNM eigenfrequency is removed from the projectors, yielding a coupling for a broad range of frequencies. We introduce the noise-coupling elements at this stage in the derivation for later convenience, but a derivation using the original coupling \(\omega'\mathbf{L}_{i_{\mu}}(\mathbf{r}',\omega')\) yields the same results. 

Equation~\eqref{eq:cDGL} has the form of a scattering process. The (bosonic) first term on the right-hand side describes the undisturbed propagation of a photon through the homogeneous background medium. The (non-bosonic) second term represents the scattering, either at the quasibound QNMs at a cavity or at the outward-propagating regularized QNM fields, with the noise coupling elements acting as a scattering potential.
The non-bosonic form ensures the orthogonality from Eq.~\eqref{eq:cortho} for all times \(t\). In previous works, the non-bosonic nature was neglected in favor of a simple bosonic time evolution \(\opvec{c}(\mathbf{r},\omega,t) \approx \opvec{c}(\mathbf{r},\omega,t_0)\mathrm{e}^{-i\omega (t-t_0)}\) \cite{franke2020quantized, fuchs2023hierarchical} for an initial time \(t_0\), which violates the orthogonality for \(t>t_0\). As we show below, the bosonic approximation of the bath still yields exact results for the single-cavity case \cite{franke2020quantized} when using the noise coupling elements \(\mathbf{g}_{i_{\mu}}(\mathbf{r},\omega)\) to characterize QNM-bath coupling. For more complex multi-cavity cases with retardation, however, the bosonic approximation is generally not justified.

To investigate the influence of the non-bosonic time evolution, we formally integrate Eq.~\eqref{eq:cDGL}:
\begin{align}\label{eq:cFormal}
    \opvec{c}(\mathbf{r},\omega,t) &= \opvec{c}(\mathbf{r},\omega,t_0)\mathrm{e}^{-i\omega(t-t_0)}\nonumber\\
    &+\sum_{i,\mu}\mathbf{L}_{i_{\mu}}^*(\mathbf{r},\omega)\int_{t_0}^t\mathrm{d}t_1\mathrm{e}^{-i\omega(t-t_1)}\nonumber\\
    &\times i\int\mathrm{d}^3r'\int_0^{\infty}\mathrm{d}\omega' \mathbf{g}_{i_{\mu}}(\mathbf{r}',\omega')\cdot\opvec{c}(\mathbf{r}',\omega',t_1),
\end{align}
where \(\opvec{c}(\mathbf{r}',\omega',t_0)\) fulfills the orthogonality relation from Eq.~\eqref{eq:cortho} at the chosen initial time \(t_0\). 
Equation~\eqref{eq:cFormal} is a recursive scattering problem. By inserting the solution into the right-hand side again, we obtain
\begin{align}
     &\opvec{c}(\mathbf{r},\omega,t) = \opvec{c}(\mathbf{r},\omega,t_0)\mathrm{e}^{-i\omega(t-t_0)}\nonumber\\
    &\quad+\sum_{i,\mu}\mathbf{L}_{i_{\mu}}^*(\mathbf{r},\omega)\int_{t_0}^t\mathrm{d}t_1\mathrm{e}^{-i\omega(t-t_1)}\hat{C}_{i_{\mu}}(t_1)\nonumber\\
    &\quad+\sum_{ij,\mu\nu}\mathbf{L}_{i_{\mu}}^*(\mathbf{r},\omega)\int_{t_0}^t\mathrm{d}t_1\int_{t_0}^{t_1}\mathrm{d}t_2 \mathbb{K}^1_{i_{\mu}j_{\nu}}(t_1-t_2)\nonumber\\
    &\qquad \times i\int\mathrm{d}^3r'\int_0^{\infty}\mathrm{d}\omega' \mathbf{g}_{j_{\nu}}(\mathbf{r}',\omega')\cdot\opvec{c}(\mathbf{r}',\omega',t_2),
\end{align}
where
\begin{align}\label{eq:initscat}
    &\hat{C}_{i_{\mu}}(t)=i\int\mathrm{d}^3r\int_0^{\infty}\mathrm{d}\omega \mathbf{g}_{i_{\mu}}(\mathbf{r},\omega)\cdot\opvec{c}(\mathbf{r},\omega,t_0)\mathrm{e}^{-i\omega(t-t_0)}
\end{align}
describes the annihilation of a noise quantum in the bath due to absorption by the QNM \(i_{\mu}\). Conversely, \(\hat{C}^{\dagger}_{i_{\mu}}(t)\), describes the creation of noise due to the emission of a photon from the QNM \(i_{\mu}\) (cf.~Refs.~\onlinecite{franke2019quantization, gustin2025dissipation}). 

Furthermore (leaving \(t>t'\) implicit),
\begin{align}\label{eq:singleK}
    &\mathbb{K}^1_{i_{\mu}j_{\nu}}(t-t') \nonumber\\
    &\qquad= i\int\mathrm{d}^3r\int_0^{\infty}\mathrm{d}\omega \mathbf{g}_{i_{\mu}}(\mathbf{r},\omega)\cdot\mathbf{L}^*_{j_{\nu}}(\mathbf{r},\omega)\mathrm{e}^{-i\omega(t-t')}
\end{align} 
describes single scatter processes between two QNMs \(i_{\mu}\) and \(j_{\nu}\) during the time interval \((t-t')\). Note that \(\mathbf{L}^*_{j_{\nu}}(\mathbf{r},\omega)\) contains poles at the complex conjugated QNM frequencies located in the upper half of the complex plane, while \(\mathbf{g}_{i_{\mu}}(\mathbf{r},\omega)\) contains no poles. Since \(t>t'\), it follows from the residue theorem that \(\int\mathrm{d}\omega \mathrm{e}^{-i\omega(t-t')}/(\omega-\tilde{\omega}^*_{j_{\nu}}) = 0\), and therefore, \(\mathbb{K}^1_{i_{\mu}j_{\nu}}(t-t')\) vanishes for intracavity scattering processes (\(i=j\)) as a consequence of causality \cite{gustin2025dissipation}. For intercavity scattering (\(i\neq j\)), additional phase terms \(\mathrm{e}^{\pm i\omega\tau_{ij}}\) appear caused by retardation, where \(\tau_{ij}\) is the photon travel time between the cavities, and as a consequence, \(\mathbb{K}^1_{i_{\mu}j_{\nu}}(t-t')\) does not vanish.

It is clear that the time-dependent bath operators \(\opvec{c}(\mathbf{r},\omega,t)\) contain these scattering processes to infinite order. Therefore, we define the \(N\)-th order intercavity scattering term \(\mathbb{K}^N_{i_{\mu}j_{\nu}}(t-t')\) via (again assuming \(t>t'\) implicit)
\begin{align}\label{eq:doubleK}
    &\mathbb{K}^N_{i_{\mu}j_{\nu}}(t-t') = \sum_{l,\eta}\int_{t'}^t\mathrm{d}t_1 \mathbb{K}^{N-1}_{i_{\mu}l_{\eta}}(t-t_1)\mathbb{K}^1_{l_{\eta}j_{\nu}}(t_1-t')\nonumber\\
    &\qquad\qquad\quad= \sum_{l,\eta}\int_{t'}^t\mathrm{d}t_1 \mathbb{K}^1_{i_{\mu}l_{\eta}}(t-t_1)\mathbb{K}^{N-1}_{l_{\eta}j_{\nu}}(t_1-t').
\end{align}

For convenience, we also define the zeroth-order intercavity scattering, including no scattering via,
\begin{align}\label{eq:Kzero}
    \mathbb{K}^0_{i_{\mu}j_{\nu}}(t-t') = 2\delta_{ij}\delta_{\mu\nu}\Theta(t-t')\delta(t-t'),
\end{align}
so that the full solution of Eq.~\eqref{eq:cDGL} reads
\begin{align}\label{eq:fullbath}
    \opvec{c}(\mathbf{r},\omega,t) &= \opvec{c}(\mathbf{r},\omega,t_0)\mathrm{e}^{-i\omega(t-t_0)}\nonumber\\
    &\quad+\sum_{ij,\mu\nu}\sum_{N=0}^{\infty}\mathbf{L}^*_{i_{\mu}}(\mathbf{r},\omega)\int_{t_0}^t\mathrm{d}t_1\mathrm{e}^{-i\omega(t-t_1)}\nonumber\\
    &\qquad\times\int_{t_0}^{t_1}\mathrm{d}t_2 \mathbb{K}^N_{i_{\mu}j_{\nu}}(t_1-t_2)\hat{C}_{j_{\nu}}(t_2).
\end{align}

The correlation function from Eq.~\eqref{eq:gencorrfunc} contains three kinds of interactions: the exchange of photons between separate cavities (Sec.~\ref{sec:qnmcorrfunc}), the exchange of excitations between two TLS (Sec.~\ref{sec:tlscorrfunc}), and the interaction between a TLS and QNM cavity (Sec.~\ref{sec:qnmtlscorrfunc}). Using the time evolution of the bath operators from Eq~\eqref{eq:fullbath}, we derive explicit forms of these different couplings. We discuss these three types of interactions separately and then summarize our findings. An overview of the different types of couplings is also given in Tab.~\ref{tab:summary}, shown in Sec.~\ref{sec:intro}. \red{For quantum dynamics calculations using similar correlation functions for quantized QNM cavities, see Refs.~\onlinecite{franke2020quantized, fuchs2023hierarchical, gustin2025dissipation}.}

\subsection{QNM-QNM coupling} \label{sec:qnmcorrfunc}
\subsubsection{General form}
In Sec.~\ref{sec:quanti}, we discussed how the instantaneous (or direct) coupling between different QNM cavities is suppressed for cases with sufficient separation. As a result, the QNM Hamiltonian [Eq.~\eqref{eq:qnmham}] contains no direct coupling between QNMs of separate cavities \(i\neq j\). In a time-dependent theory, propagating photons transmit energy between the cavities via the bath if the time delay is larger than the retardation time.

With the time-dependent bath operators from Eq.~\eqref{eq:fullbath}, the QNM-bath coupling Hamiltonian [cf.~Eq.~\eqref{eq:qnmham} together with Eq.~\eqref{eq:gcoup}] in the interaction picture reads: 
\begin{align}\label{eq:HQNMbath}
    &H_{\rm QNM-bath}(t) \nonumber\\
    &= \hbar\sum_{i,\mu}\int\mathrm{d}^3r\int_0^{\infty}\mathrm{d}\omega\,\mathbf{g}_{i_{\mu}}(\mathbf{r},\omega)\cdot\opvec{c}(\mathbf{r},\omega,t)\hat{a}^{\dagger}_{i_{\mu}}(t)+\mathrm{H.a.}\nonumber\\
    &=-i\hbar\sum_{ij,\mu\nu}\sum_{N=0}^{\infty}\int_{t_0}^t\mathrm{d}t_1\mathbb{K}^N_{i_{\mu}j_{\nu}}(t-t_1)\hat{C}_{j_{\nu}}(t_1)\hat{a}^{\dagger}_{i_{\mu}}(t)+\mathrm{H.a.},
\end{align}
where \(\hat{C}_{i_{\mu}}\) describes the initial scattering of a bath photon at a QNM, and \(\mathbb{K}^N_{i_{\mu}j_{\nu}}\) describes the retarded scattering between QNMs of different cavities. These higher-order terms (\(N>0\)) vanish for a single cavity, so that the exact system-bath coupling in the single-cavity case is precisely the combination of noise-coupling elements \(\mathbf{g}_{i_{\mu}}\) and bosonic bath operators \(\opvec{c}(\mathbf{r},\omega,t_0)\mathrm{e}^{-i\omega(t-t_0)}\) that was obtained heuristically in Ref.~\onlinecite{franke2020quantized}.

Using the Hamiltonian from Eq.~\eqref{eq:HQNMbath}, the correlation function from Eq.~\eqref{eq:gencorrfunc} for the coupling between two QNMs reads (the derivation is shown in Appendix~\ref{appsec:QNMcorr}),
\begin{align}\label{eq:QNMcorrfunc}
    &C^{\rm QNM}_{i_{\mu}j_{\nu}}(t-t') = \delta_{ij}C^{\rm bos}_{i_{\mu}j_{\nu}}(t-t')\nonumber\\
    &\qquad\quad - \sum_{N=1}^{\infty}\sum_{\eta}(\delta_{\eta\nu}\partial_{t'}{-}i\chi_{j_{\eta}j_{\nu}})\mathbb{K}^N_{i_{\mu}j_{\eta}}(t-t') \nonumber\\
    &\qquad\quad - \sum_{N=1}^{\infty}\sum_{\eta}\Big[(\delta_{\eta\mu}\partial_{t}{-}i\chi_{i_{\eta}i_{\mu}})\mathbb{K}^N_{j_{\nu}i_{\eta}}(t'-t)\Big]^*,
\end{align}
where
\begin{align}\label{eq:Cbos}
    &C^{\rm bos}_{i_{\mu}j_{\nu}}(t-t')\nonumber\\
    &\qquad =\int_0^{\infty}\mathrm{d}\omega\int\mathrm{d}^3r \mathbf{g}_{i_{\mu}}(\mathbf{r},\omega)\cdot\mathbf{g}^*_{j_{\nu}}(\mathbf{r},\omega)\mathrm{e}^{-i\omega(t-t')}
\end{align}
is the QNM correlation function for a \textit{bosonic} bath, obtained exactly in Ref.~\onlinecite{franke2020quantized} for the single-cavity case and approximately in Ref.~\onlinecite{fuchs2023hierarchical} for two coupled cavities assumming a bosonic bath [note the relation between \(C^{\rm bos}\) and \(\mathbb{K}^1\) from Eq.~\eqref{appeq:CbosK1}].
 
The first term on the right-hand side in Eq.~\eqref{eq:QNMcorrfunc} describes the scattering between modes of the same cavity only, while the second and third terms describe the transfer of excitation from cavity \(j\) to cavity \(i\) for \(t>t'\) and the opposite process for \(t'>t\), respectively. Since the sums contain the scattering to infinite order, the correlation function includes all possible processes for the transfer of a photon between cavities \(i\) and \(j\). 

\subsubsection{Perturbative intercavity coupling}
The form of the correlation function from Eq.~\eqref{eq:QNMcorrfunc} allows for a perturbative treatment of the intercavity scattering. Since instantaneous intercavity transfer is negligible for well-separated cavities (cf.~Sec.~\ref{sec:multicavityquanti}), only a finite number of intercavity scattering processes contribute significantly within a finite time span \(t-t'\). Furthermore, for the cavities in homogeneous three-dimensional background media considered here, each order of the scattering decreases in magnitude with \(1/R_{ij}\). Hence, for cavities with significant separation, we include only terms \(N\leq 1\) in the correlation, yielding [using the relation between \(C^{\rm bos}\) and \(\mathbb{K}^1\) from Eq.~\eqref{appeq:CbosK1}]:
\begin{align}\label{eq:QNMcorrN1}
    C^{\rm QNM}_{i_{\mu}j_{\nu}}(t-t') &= C^{\rm bos}_{i_{\mu}j_{\nu}}(t-t')\nonumber\\
    &+2\Theta(t-t')\sum_{\eta}\mathbb{K}^1_{i_{\mu}j_{\eta}}(t-t')\chi^{(-)}_{j_{\eta}j_{\nu}}\nonumber\\
    &+2\Theta(t'-t)\sum_{\eta}\Big[\mathbb{K}^1_{j_{\nu}i_{\eta}}(t'-t)\chi^{(-)}_{i_{\eta}i_{\mu}}\Big]^*,
\end{align}
where \(\chi^{(-)}_{i_{\mu}i_{\eta}} = i(\chi_{i_{\mu}i_{\eta}}-\chi^*_{i_{\eta}i_{\mu}})/2\) is the imaginary part of the symmetrized QNM frequency \(\chi_{i_{\mu}i_{\eta}} = \sum_{\mu'} \left(S^{-1/2}\right)_{i_{\mu}i_{\mu'}}\Tilde{\omega}_{i_{\mu'}}\left(S^{1/2}\right)_{i_{\mu'}i_{\eta}}\). Note that in structured environments (e.g., coupled via waveguides), Eq.~\eqref{eq:QNMcorrfunc} can still be truncated at low order if the cavities are well-separated [large cavity separation parameter, cf.~Eq.~\eqref{eq:cavsepparam}]. In such cases, instantaneous intercavity scattering is negligible (cf.~Sec.~\ref{sec:multicavityquanti}), and only a finite number of retarded intercavity scattering processes occur in a finite time \(t-t'\).

\begin{figure}
    \centering
    \includegraphics[width=0.98\columnwidth]{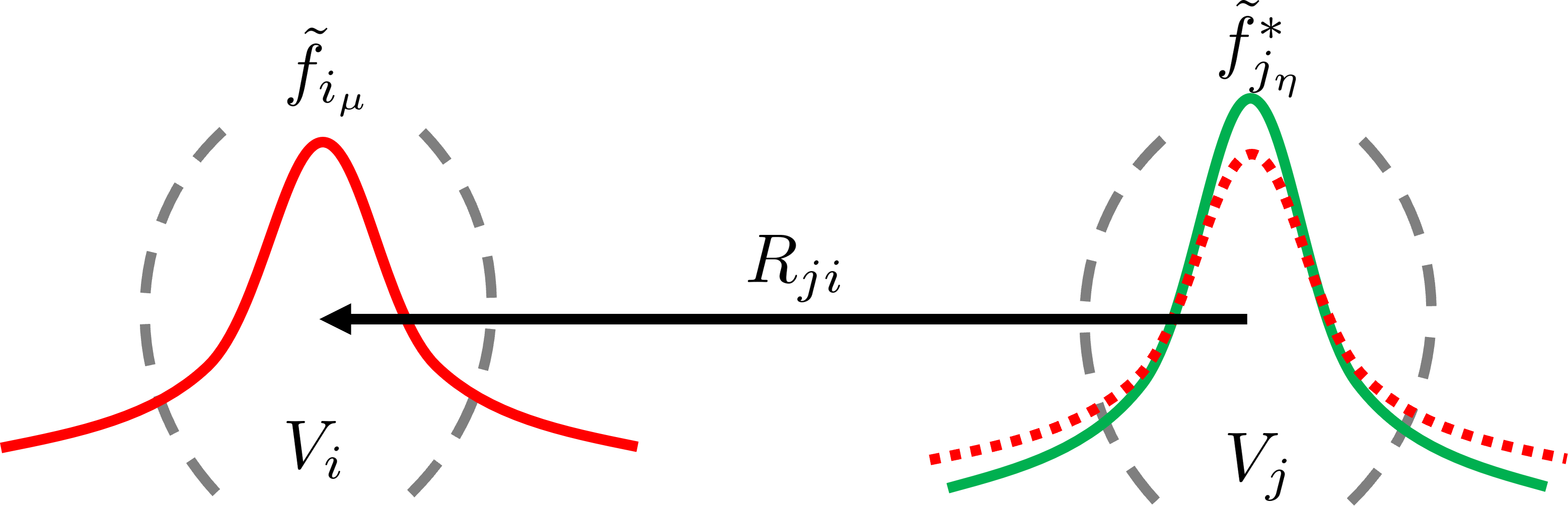}
    \caption{We obtain the retarded overlap matrix \(S_{i_{\mu}\leftarrow j_{\eta}}\) from Eq.~\eqref{eq:retardedSmatrix} by eliminating the propagation from the inter-cavity overlap integral to obtain an effective, frequency-independent overlap.}
    \label{fig:Tilde_S}
\end{figure}

In Eq.~\eqref{eq:QNMcorrN1}, \(C^{\rm bos}_{i_{\mu}j_{\nu}}\) and \(\mathbb{K}^1_{i_{\mu}j_{\eta}}\) contain numerically demanding spatial integrals. For an efficient calculation of QNM-QNM coupling, we formulate the correlation using effective coupling elements (see Appendix~\ref{appsec:QNMscatter} for the derivation),
\begin{align} \label{eq:QNMcoupmatrix}
    &\chi^{(-)}_{i_{\mu}\leftarrow j_{\nu}} \nonumber\\
    &= \sum_{\mu'\nu'}\left(S^{-1/2}\right)_{i_{\mu}i_{\mu'}}i(\Tilde{\omega}_{i_{\mu'}}-\Tilde{\omega}^*_{j_{\nu'}})S_{i_{\mu'}\leftarrow j_{\nu'}}\left(S^{-1/2}\right)_{j_{\nu'}j_{\nu}},
\end{align}
with the retarded overlap matrix
\begin{align} \label{eq:retardedSmatrix}
    &S_{i_{\mu}\leftarrow j_{\eta}} = \frac{\delta_{ij}}{2}\int_{V_j}\mathrm{d}^3r\frac{\sqrt{\omega_{i_{\mu}}\omega_{j_{\eta}}}\sqrt{\epsilon_I(\mathbf{r},\omega_{i_{\mu}})\epsilon_I(\mathbf{r},\omega_{j_{\eta}})}}{i(\Tilde{\omega}_{i_{\mu}}-\Tilde{\omega}^*_{j_{\eta}})}\nonumber\\
    &\qquad\qquad\qquad\qquad\qquad\qquad\times \qnm{f}{i_{\mu}}(\mathbf{r})\cdot\qnm{f}{j_{\eta}}^*(\mathbf{r})\nonumber\\
    &+\left(1-\frac{\delta_{ij}}{2}\right)\oint_{\mathcal{S}_j}\mathrm{d}A_s\Bigg\{\frac{[\qnm{H}{i_{\mu}}'(\mathbf{s},\omega_{i_{\mu}})\times\opvec{n}_s]\cdot\qnm{F}{j_{\eta}}^*(\mathbf{s},\omega_{j_{\eta}})}{2\epsilon_0i(\Tilde{\omega}_{i_{\mu}}-\Tilde{\omega}^*_{j_{\eta}})}\nonumber\\
    &\qquad\qquad\qquad+\frac{[\qnm{H}{j_{\eta}}^*(\mathbf{s},\omega_{j_{\eta}})\times\opvec{n}_s]\cdot\qnm{F}{i_{\mu}}'(\mathbf{s},\omega_{i_{\mu}})}{2\epsilon_0i(\Tilde{\omega}_{i_{\mu}}-\Tilde{\omega}^*_{j_{\eta}})} \Bigg\},
\end{align}
where \(\mathcal{S}_j\) is the surface of the volume \(V_j\) of the \(j\)-th cavity, \(\opvec{n}_s\) is the surface vector on \(\mathcal{S}_j\) that points inwards of \(V_j\), and \(\qnm{F}{i_{\mu}}'(\mathbf{r},\omega)|_{r\in V_j} = \mathrm{e}^{-i\omega\tau_{ij}}\qnm{F}{i_{\mu}}(\mathbf{r},\omega)\) are slow-varying envelope functions of the regularized fields for positions inside the other cavity (see Appendix \ref{appsec:QNMscatter} for a detailed derivation). Thus, \(S_{i_{\mu}\leftarrow j_{\eta}}\) is an effective intercavity overlap matrix with the retardation removed (Fig.~\ref{fig:Tilde_S}), similar to what was done in Ref.~\onlinecite{carmichael1993quantum} in the context of quantum dynamics for two coupled cavities.

\begin{figure*}
    \centering
    \includegraphics[width=0.9\linewidth]{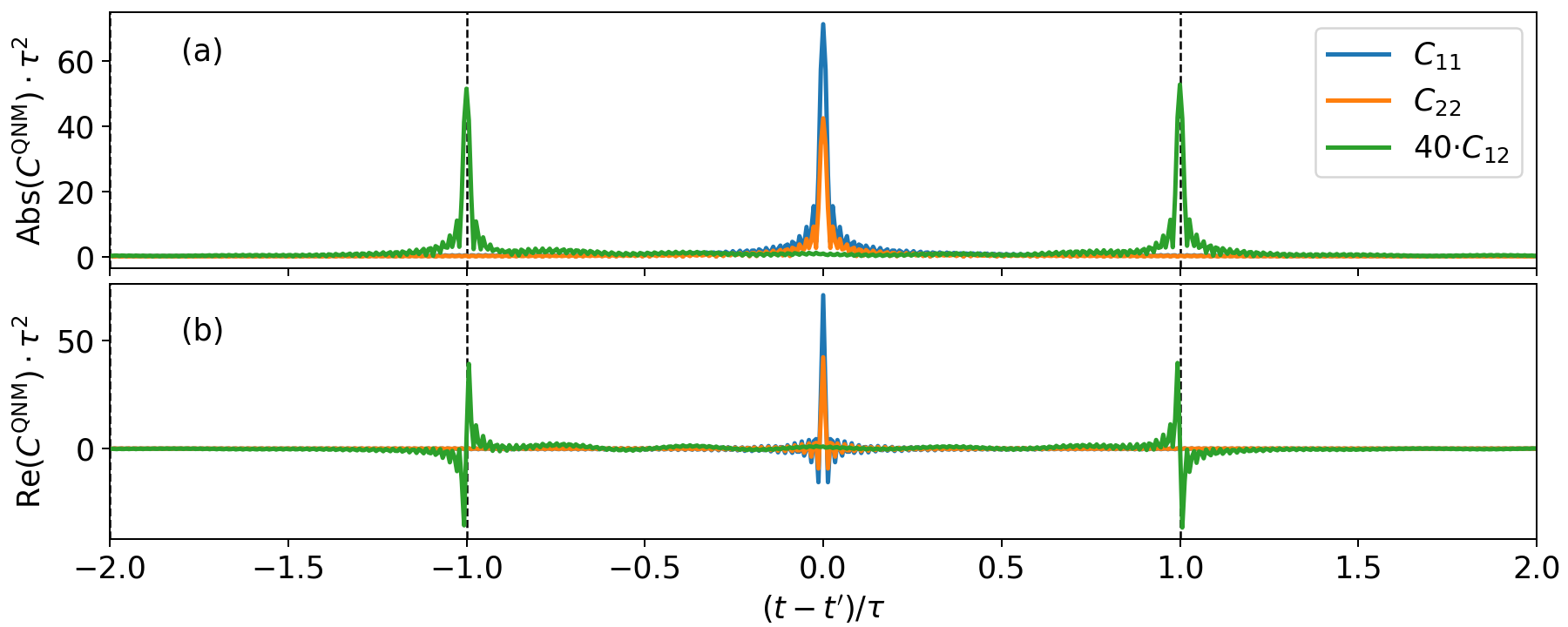}
    \caption{QNM correlation function for two coupled metal dimers \(1\) and \(2\) (cf.~Fig.~\ref{fig:dimer_sketch}) with one dominant QNM each, (a) Absolute value and (b) Real part. Since \(C_{21}(t-t') = [C_{12}(t'-t)]^*\), we only include \(C_{12}\), and enhance it for better visibility. The correlations are dominated by \(\delta\)-like peaks, matching the approximation from Eq.~\eqref{eq:CbosMarkov}. The high-frequency oscillations arise as a consequence of the numerical implementation at a finite bandwidth. }
    \label{fig:QNMcorr}
\end{figure*}

The spatial integrals in \(S_{i_{\mu}\leftarrow j_{\eta}}\) run over the \(j\)-th cavity, while the generalized Poynting vector \(\qnm{F}{j_{\eta}}^*(\mathbf{r},\omega)\times\qnm{H}{i_{\mu}}(\mathbf{r},\omega)\) points outwards so that the matrix is associated with the emission of radiation through the surface \(\mathcal{S}_j\). Similarly, the advanced coupling \(\chi^{(-)}_{i_{\mu}\rightarrow j_{\eta}}\) contains the matrix \(S_{i_{\mu}\rightarrow j_{\eta}}=\left(S_{j_{\eta}\leftarrow i_{\mu} }\right)^*\) with integrals over cavity \(i\). The coupling is generally non-symmetric in the indices \(i_{\mu}\) and \(j_{\eta}\), unless the arrow is switched accordingly: \(\chi^{(-)*}_{i_{\mu}\rightarrow j_{\eta}} \neq \chi^{(-)}_{j_{\eta}\rightarrow i_{\mu}} = \chi^{(-)*}_{i_{\mu}\leftarrow j_{\eta}}\). In the intracavity case \(i=j\), \(S_{i_{\mu}\leftarrow i_{\eta}} = S_{i_{\mu}\rightarrow i_{\eta}} = S_{i_{\mu}i_{\eta}} /2\), where \( S_{i_{\mu}i_{\eta}}\) is the intracavity overlap matrix from Eq.~\eqref{eq:Sintradef}. Thus, in the intracavity case,
\begin{align}
    \chi^{(-)}_{i_{\mu}\rightarrow i_{\eta}} = \chi^{(-)}_{i_{\mu}\leftarrow i_{\eta}} = \chi^{(-)}_{i_{\mu}i_{\eta}}.
\end{align}
The matrix \(\chi^{(-)}_{i_{\mu}i_{\eta}}\) was found previously in Ref.~\onlinecite{franke2019quantization} as the dissipator matrix in the QNM Lindblad master equation, and \(\chi^{(-)}_{i_{\mu}\leftarrow j_{\nu}}\) is the extension of this coupling to cases with retardation.

Using these definitions, we find (Appendix~\ref{appsec:QNMscatter}),
\begin{align}\label{eq:Cbosapprox}
    &C^{\rm bos}_{i_{\mu}j_{\nu}}(t-t') \nonumber\\
    &\quad= \int_0^{\infty}\mathrm{d}\omega \frac{\mathrm{e}^{-i\omega(t-t')}}{2\pi}\Big(\chi^{(-)}_{i_{\mu}\rightarrow j_{\nu}}\mathrm{e}^{-i\omega\tau_{ij}}+\chi^{(-)}_{i_{\mu}\leftarrow j_{\nu}}\mathrm{e}^{i\omega\tau_{ij}}\Big),
\end{align}
and
\begin{align}\label{eq:singleKapprox}
    &\mathbb{K}^1_{i_{\mu}j_{\nu}}(t-t')\nonumber\\
    &\quad= \frac{i}{2\pi} \sum_{\eta\kappa} \int_0^{\infty} \mathrm{d}\omega \Big(\chi^{(-)}_{i_{\mu}\rightarrow j_{\eta}}\mathrm{e}^{-i\omega\tau_{ij}}+\chi^{(-)}_{i_{\mu}\leftarrow j_{\eta}}\mathrm{e}^{i\omega\tau_{ij}}\Big)\nonumber\\
    &\qquad\qquad\qquad\times\left(S^{1/2}\right)_{j_{\eta}j_{\kappa}}\frac{\mathrm{e}^{-i\omega(t-t')}}{\omega-\tilde{\omega}^*_{j_{\kappa}}}\left(S^{-1/2}\right)_{j_{\kappa}j_{\nu}},
\end{align}
where, for the intracavity case \(i=j\), \(\tau_{ii}=0\) holds, so that the results from Ref.~\onlinecite{franke2020quantized} are recovered for a single cavity.

For some applications with sufficiently high-Q cavities, the frequency regime of interest is confined to a small bandwidth around the QNM resonance frequency. In such cases, we can expand the lower limit of the integral in Eq.~\eqref{eq:Cbosapprox} to \(-\infty\) and employ contour integral methods  to obtain
\begin{align}\label{eq:CbosMarkov}
    &C^{\rm bos}_{i_{\mu}j_{\nu}}(t-t') \nonumber\\
    &\quad\approx \Big[\chi^{(-)}_{i_{\mu}\rightarrow j_{\nu}}\delta(t-t'+\tau_{ij})+\chi^{(-)}_{i_{\mu}\leftarrow j_{\nu}}\delta(t-t'-\tau_{ij})\Big],
\end{align}
which is precisely the correlation function that was obtained in Ref.~\onlinecite{fuchs2023hierarchical} under the assumption of a bosonic bath. Hence, in these cases, the QNM-QNM correlation function is dominated by \(\delta\)-like correlations with only small (sometimes negligible) corrections due to non-bosonic bath effects and intercavity scattering.

In the following, we use the more general forms from Eq.~\eqref{eq:Cbosapprox} and Eq.~\eqref{eq:singleKapprox} to calculate the correlation functions.

\subsubsection{Coupled metal dimers}
To illustrate the calculation of the correlation function, we consider an example of two metal dimers in vacuum (\(n_{\rm B} = 1\)) serving as QNM cavities with one dominant QNM each (cf.~Fig.~\ref{fig:dimer_sketch}).

We consider the case \(N\leq 1\) as discussed above. In Fig.~\ref{fig:QNMcorr}, we show the absolute value and real part of the QNM-QNM correlation function from Eq.~\eqref{eq:QNMcorrN1} for the two coupled metal dimers using the effective coupling from Eqs.~\eqref{eq:Cbosapprox} and \eqref{eq:singleKapprox}.
The correlation functions are dominated by \(\delta\)-like peaks, confirming the validity of the approximation from Eq.~\eqref{eq:CbosMarkov} for this setup. The diagonal elements peak at \(t-t' = 0\) and are related to the temporal decay of the quasibound QNMs via photon emission into the bath \cite{franke2020quantized}. The off-diagonal terms peak at the delay time \(t-t' = \pm \tau\) and are related to intercavity transfer between the spatially separated dimers. However, we see a slight deviation from a Lorentzian form, again matching Eq.~\eqref{eq:CbosMarkov}. There are also some oscillations that arise here as a consequence of the numerical implementation with a finite bandwidth. As shown in Ref.~\onlinecite{fuchs2023hierarchical}, the intercavity photon exchange can lead to coherent superpositions of the occupations and state trapping in the cavities.

\red{Note that for this example with low-Q plasmonic resonators, the inclusion of time delay is particularly important since even a short separation leads to delays of a similar magnitude as the QNM lifetime. For higher quality optical cavities, the delay can often be subsumed into an effective Markovian coupling, since it is not relevant on the system dynamics time scale. Still, the general formalism shown here (dynamics of the non-bosonic bath + system-bath correlation functions) applies also to higher-Q cavities.}


\subsection{QNM-TLS coupling}\label{sec:qnmtlscorrfunc}
\subsubsection{General form}
A TLS that transitions from the excited state to the ground state emits a photon into the surrounding medium, described in Eq.~\eqref{eq:qnmham} by the coupling to the bath modes. The resulting bath photon can propagate through space and excite another system (such as a QNM cavity) elsewhere, leading to an effective interaction. In contrast to the QNM-TLS interaction derived in Sec.~\ref{sec:qnmtlscoup}, this interaction is time-dependent and includes retarded coupling between spatially separated systems. In this section, we derive the correlation function for the time-dependent QNM-TLS interaction.

Using the bath operators from Eq.~\eqref{eq:fullbath}, the Hamiltonian for the coupling between a TLS and the bath photons [cf.~Eq.~\eqref{eq:qnmham}] in the interaction picture reads,
\begin{align}\label{eq:HTLSbath}
    &H_{\rm TLS-bath}(t)\nonumber\\
    &= -i\hbar\sum_a\int\mathrm{d}^3r\int_0^{\infty}\mathrm{d}\omega \mathbf{g}_a(\mathbf{r},\omega)\cdot\opvec{c}(\mathbf{r},\omega,t_0)\nonumber\\
    &\qquad\qquad\qquad\qquad\qquad\qquad\qquad\times\mathrm{e}^{-i\omega(t-t_0)}\hat{\sigma}^+_a(t)\nonumber\\
    &\qquad\qquad\qquad\qquad\qquad\qquad\qquad\qquad\qquad+\mathrm{H.a.}\nonumber\\
    &-\sum_a\sum_{ij,\mu\nu}\sum_{N=0}^{\infty} \int_{t_0}^{t}\mathrm{d}t_1\int_{t_0}^{t_1}\mathrm{d}t_2\mathbf{d}_a\cdot\mathbf{E}_{i_{\mu}}(\mathbf{r}_a,t-t_1)\nonumber\\
    &\qquad\qquad\qquad\qquad\qquad\times\mathbb{K}^N_{i_{\mu}j_{\nu}}(t_1-t_2)C_{j_{\nu}}(t_2)\hat{\sigma}^+_a(t)\nonumber\\
    &\qquad\qquad\qquad\qquad\qquad\qquad\qquad\qquad\qquad+\mathrm{H.a.},
\end{align}
where we defined the \textit{time-dependent QNM-generated electric field} via
\begin{align}\label{eq:timedepQNMgenefield}
    \mathbf{E}_{i_{\mu}}(\mathbf{r},t)= \int_0^{\infty}\mathrm{d}\omega\mathbf{E}_{i_{\mu}}(\mathbf{r},\omega)\mathrm{e}^{-i\omega t},
\end{align}
from the QNM-generated electric field \(\mathbf{E}_{i_{\mu}}(\mathbf{r}_a,\omega)\) in Eq.~\eqref{eq:qnmgenefeld}.

In Eq.~\eqref{eq:HTLSbath}, the first term on the right-hand side gives the full dipole coupling of the TLS directly to the (bosonic) bath photon. The coupling elements \(\mathbf{g}_a(\mathbf{r},\omega)\) are defined in Eq.~\eqref{eq:tlsbathcoup}. The second term describes the coupling via QNM scattering, where the TLS couples to the QNM generated electric field \(\mathbf{E}_{i_{\mu}}(\mathbf{r}_a,t)\), followed by retarded QNM-QNM scattering \(\mathbb{K}^N_{i_{\mu}j_{\nu}}(t_1-t_2)\) to infinite order and the QNM-bath scattering \(\hat{C}_{i_{\mu}}(t)\) defined in Eq.~\eqref{eq:initscat}. Using this Hamiltonian together with the QNM-bath coupling from Eq.~\eqref{eq:HQNMbath}, we derive the correlation functions (see Appendix~\ref{appsec:qnmtlscorr} for details)
\begin{align}\label{eq:QTcorr}
    &C^{\rm Q-T}_{i_{\mu}a}(t-t')\nonumber\\
    &=-\Theta(t{-}t')\sum_{j,\eta\nu}\sum_{N=0}^{\infty}(\delta_{\eta\nu}\partial_t{+}i\chi_{j_{\eta}j_{\nu}})\int_{t'}^t\mathrm{d}t_1\mathbb{K}^N_{i_{\mu}j_{\eta}}(t-t_1)\nonumber\\
    &\qquad\qquad\qquad\qquad\qquad\times\Bigg[\frac{\mathbf{d}_a\cdot\mathbf{E}_{j_{\nu}}(\mathbf{r}_a,t'-t_1)}{i\hbar}\Bigg]^*\nonumber\\
    &-\Theta(t'{-}t)\sum_{j,\eta\nu}\sum_{N=0}^{\infty}(\delta_{\eta\mu}\partial_t{+}i\chi^*_{i_{\eta}i_{\mu}})\int_{t}^{t'}\mathrm{d}t_1\Big[\mathbb{K}^N_{j_{\nu}i_{\eta}}(t_1-t)\Big]^*\nonumber\\
    &\qquad\qquad\qquad\qquad\qquad\times\Bigg[\frac{\mathbf{d}_a\cdot\mathbf{E}_{j_{\nu}}(\mathbf{r}_a,t'-t_1)}{i\hbar}\Bigg]^*,
\end{align}
and \(C^{\rm T-Q}_{a i_{\mu}}(t-t')=[C^{\rm Q-T}_{i_{\mu}a}(t'-t)]^*\).

\subsubsection{Perturbative treatment of intercavity transfer} 
For cavities with significant separation (\(P_{ij}\gg 0\), c.f.~Sec.~\ref{sec:multicavityquanti}), we treat the intercavity scattering perturbatively and restrict the number of intercavity scattering processes to \(N\leq 1\). Note, however, that the QNM-TLS coupling \(\mathbf{d}_a\cdot\mathbf{E}_{j_{\nu}}(\mathbf{r}_a,t'-t_1)/(i\hbar)\) may also contain intercavity scattering terms if the TLS is inside one of the cavities. These have to be included in the perturbative treatment. Thus, we obtain:
\begin{align}\label{eq:QTcorrsingletrans}
    &C^{\rm Q-T}_{i_{\mu}a}(t-t')\nonumber\\
    &=-\Theta(t{-}t')\sum_{\nu}(\delta_{\eta\nu}\partial_t{+}i\chi_{i_{\eta}i_{\nu}})\Bigg[\frac{\mathbf{d}_a\cdot\mathbf{E}_{i_{\nu}}(\mathbf{r}_a,t'-t)}{i\hbar}\Bigg]^*\nonumber\\
    &-\Theta(t'{-}t)\sum_{\nu}(\delta_{\eta\nu}\partial_t{+}i\chi^*_{i_{\eta}i_{\mu}})\Bigg[\frac{\mathbf{d}_a\cdot\mathbf{E}_{i_{\nu}}(\mathbf{r}_a,t'-t)}{i\hbar}\Bigg]^*\nonumber\\
    &-\Theta(t{-}t')\sum_{j,\eta\nu}(\delta_{\eta\nu}\partial_t{+}i\chi_{j_{\eta}j_{\nu}})\int_{t'}^t\mathrm{d}t_1\mathbb{K}^1_{i_{\mu}j_{\eta}}(t-t_1)\nonumber\\
    &\qquad\qquad\qquad\qquad\times\chi_{V_j}(\mathbf{r}_a)\Bigg[\frac{\mathbf{d}_a\cdot\mathbf{E}_{j_{\nu}}(\mathbf{r}_a,t'-t_1)}{i\hbar}\Bigg]^*\nonumber\\
    &-\Theta(t'{-}t)\sum_{j,\eta\nu}(\delta_{\eta\mu}\partial_t{+}i\chi^*_{i_{\eta}i_{\mu}})\int_{t}^{t'}\mathrm{d}t_1\Big[\mathbb{K}^1_{j_{\nu}i_{\eta}}(t_1-t)\Big]^*\nonumber\\
    &\qquad\qquad\qquad\qquad\times\chi_{V_j}(\mathbf{r}_a)\Bigg[\frac{\mathbf{d}_a\cdot\mathbf{E}_{j_{\nu}}(\mathbf{r}_a,t'-t_1)}{i\hbar}\Bigg]^*,
\end{align}
where \(\chi_{V_j}(\mathbf{r}_a)\) ensures \(\mathbf{r}_a\in V_j\), so that at most one intercavity transfer is included.

\begin{figure}
    \centering
    \includegraphics[width=0.98\linewidth]{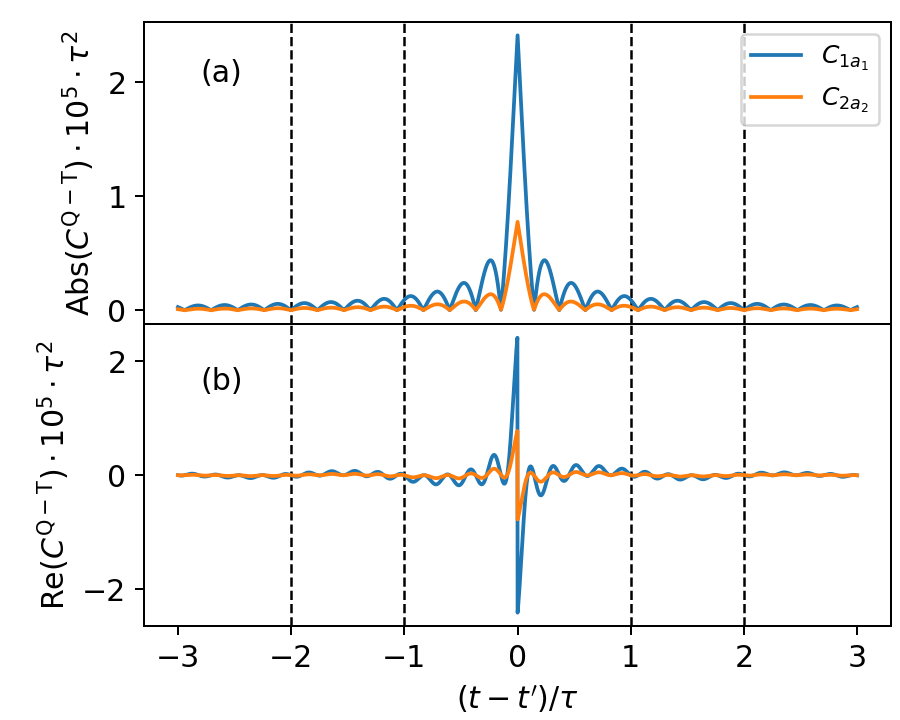}
    \caption{Bath-mediated coupling between the QNM and TLS of the same cavity for the coupled dimers from Fig.~\ref{fig:dimer_sketch}. (a) Absolute value and (b) real part calculated from Eq.~\eqref{eq:QTcorrsingletrans}.}
    \label{fig:QTcorr_diag}
\end{figure}

\subsubsection{TLS coupled to a metal dimer}
We again use metal dimers as QNM cavities. We place a TLS in the gap of each dimer (see Fig.~\ref{fig:dimer_sketch}). The TLS dipoles are polarized along the \(z\)-axis, i.e., along the symmetry axis of the cylindrical dimers. The dipoles are also assumed to be identical with \(\mathbf{d}_1=\mathbf{d}_2 =d\opvec{n}_z\), where \(\opvec{n}_z\) is the unit vector in the \(z\) direction and \(d = 1\,{\rm e}\cdot{\rm nm}\) is the strength of the dipole moment, where \({\rm e}\) is the elementary charge. 
This is at the upper end for a typical quantum dot exciton \cite{PhysRevLett.87.246401}. 

In Fig.~\ref{fig:QTcorr_diag}, we show the diagonal terms of the correlation function from Eq.~\eqref{eq:QTcorrsingletrans}, where a QNM couples to the TLS in the same cavity. As expected, the correlation peaks at \(t=t'\), but the overall coupling strength is very weak (note that the values in Fig.~\ref{fig:QTcorr_diag} are enhanced by a factor of \(10^5\)). This is because, Eq.~\eqref{eq:QTcorrsingletrans} accounts for the \textit{bath-mediated} coupling between the QNM and TLS. As discussed in Sec.~\ref{sec:qnmtlscoup}, for a TLS inside a cavity, the direct coupling to the QNMs of that cavity dominates over coupling to the bath or QNMs of other cavities (cf.~Tab.~\ref{tab:g_in}). 

\begin{figure}
    \centering
    \includegraphics[width=0.98\linewidth]{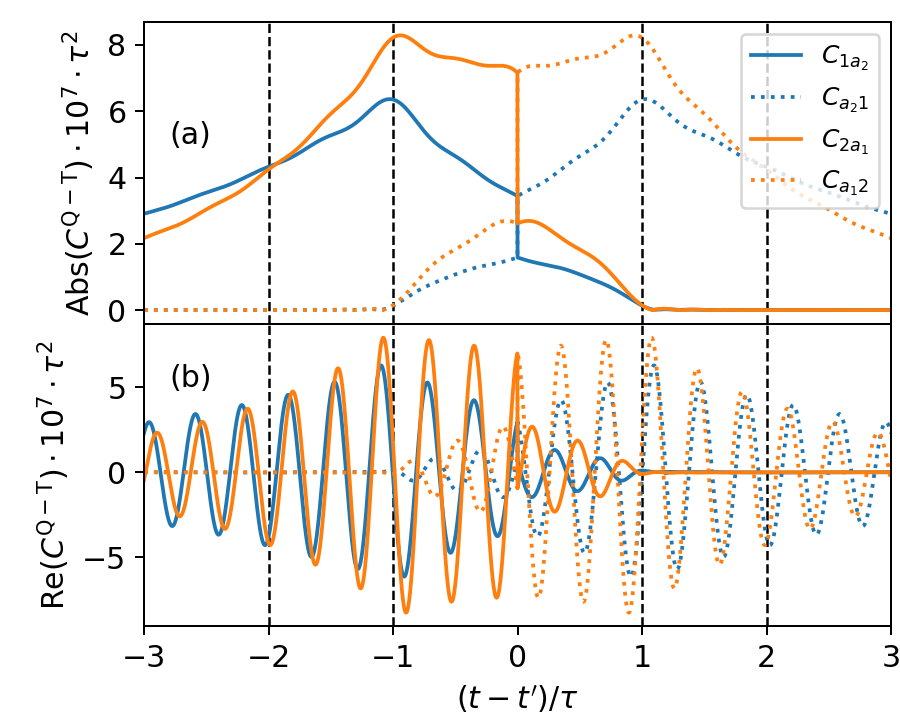}
    \caption{Bath-mediated coupling between the QNM and TLS of the other cavity for the coupled dimers from Fig.~\ref{fig:dimer_sketch}. (a) Absolute value and (b) real part calculated from Eq.~\eqref{eq:QTcorrsingletrans}.}
    \label{fig:QTcorr_ndiag}
\end{figure}

In Fig.~\ref{fig:QTcorr_ndiag}, we show the off-diagonal terms of Eq.~\eqref{eq:QTcorrsingletrans}, where a QNM couples to the TLS inside the other cavity. Again, the coupling is very weak (on the order of the instantaneous intercavity coupling that was neglected in Sec.~\ref{sec:multicavityquanti}), since the direct coupling to the QNMs of the same cavity dominates over coupling to the QNM of the other cavity. However, the coupling is still non-zero, even at \(t=t'\), since the quasibound QNMs extend outside of their respective cavities.
We stress that when calculating the system dynamics, the correlation functions presented here act together to give the overall evolution of the coupled systems, and so, one must refrain from reading physical interpretations directly from the individual correlation functions.

\subsection{TLS-TLS coupling}\label{sec:tlscorrfunc}
\subsubsection{General form}
The photon emitted by a TLS can also excite another TLS (or re-excite the same TLS), leading to an effective TLS-TLS interaction mediated by propagating bath photons. Using the Hamiltonian for the TLS-bath coupling from Eq.~\eqref{eq:HTLSbath} and the results derived in the previous sections, we derive the TLS-TLS correlation function (see Appendix~\ref{appsec:tlscorrfunc} for details):
\begin{widetext}
\begin{align}\label{eq:tlscorrfunc}
    &C_{ab}^{\rm TLS}(t-t')= \frac{1}{\pi\hbar\epsilon_0}\int_0^{\infty}\mathrm{d}\omega \mathbf{d}_{a}\cdot{\rm Im}\Big[\mathbf{G}(\mathbf{r}_{a},\mathbf{r}_{b},\omega)\Big]\cdot\mathbf{d}^*_{b}\mathrm{e}^{-i\omega(t-t')}\nonumber\\
    &+\Theta(t-t')\sum_{i,\mu\nu}(\delta_{\mu\nu}\partial_{t'}{-}i\chi_{i_{\mu}i_{\nu}})\int_{t'}^t\mathrm{d}t_1 \frac{\mathbf{d}_{a}\cdot\mathbf{E}_{i_{\mu}}(\mathbf{r}_{a},t-t_1)}{i\hbar}\left[\frac{\mathbf{d}_{b}\cdot\mathbf{E}_{i_{\nu}}(\mathbf{r}_{b},t'-t_1)}{i\hbar}\right]^*\nonumber\\
    &+\Theta(t'-t)\sum_{i,\mu\nu}(\delta_{\mu\nu}\partial_{t}{+}i\chi^*_{i_{\nu}i_{\mu}})\int_{t}^{t'}\mathrm{d}t_1 \frac{\mathbf{d}_{a}\cdot\mathbf{E}_{i_{\mu}}(\mathbf{r}_{a},t-t_1)}{i\hbar}\left[\frac{\mathbf{d}_{b}\cdot\mathbf{E}_{i_{\nu}}(\mathbf{r}_{b},t'-t_1)}{i\hbar}\right]^*\nonumber\\
    &\;+\Theta(t-t')\sum_{ij,\mu\nu\eta}\sum_{N=1}^{\infty}\int_{t'}^t\mathrm{d}t_1\int_{t'}^{t_1}\mathrm{d}t_2\Bigg[\frac{\mathbf{d}_{a}\cdot\mathbf{E}_{i_{\mu}}(\mathbf{r}_{a},t-t_1)}{i\hbar} (\delta_{\eta\nu}\partial_{t_2}{-}i\chi_{j_{\eta}j_{\nu}})\mathbb{K}^N_{i_{\mu}j_{\nu}}(t_1-t_2)\Bigg]\Bigg[\frac{\mathbf{d}_{b}\cdot\mathbf{E}_{j_{\nu}}(\mathbf{r}_{b},t'-t_2)}{i\hbar}\Bigg]^*\nonumber\\
    &\;+\Theta(t'-t)\sum_{ij,\mu\nu\eta}\sum_{N=1}^{\infty}\int_t^{t'}\mathrm{d}t_1\int_{t}^{t_1}\mathrm{d}t_2 \frac{\mathbf{d}_{a}\cdot\mathbf{E}_{i_{\mu}}(\mathbf{r}_{a},t-t_2)}{i\hbar} \Bigg[\frac{\mathbf{d}_{b}\cdot\mathbf{E}_{j_{\nu}}(\mathbf{r}_{b},t'-t_1)}{i\hbar}(\delta_{\eta\mu}\partial_{t_2}{-}i\chi_{i_{\eta}i_{\mu}})\mathbb{K}^N_{j_{\nu}i_{\mu}}(t_1-t_2)\Bigg]^*.
\end{align}
\end{widetext}

Here, the first line on the right-hand side represents the full dipole coupling of two TLS via the electromagnetic field \cite{PhysRevA.53.1818, richter2022enhanced}.
The other terms are QNM correction terms that account for the QNM-TLS coupling (cf.~Sec.~\ref{sec:qnmtlscoup}) where a TLS couples to a QNM, which then scatters via the bath into the same or another QNM which finally couples to another TLS.

\subsubsection{TLS coupling in the presence of metal dimer QNM cavities}
We again consider the case of two metal dimers serving as QNM cavities with TLSs in the dimer gaps from Fig.~\ref{fig:dimer_sketch}. We also take \(N\leq 1\) in Eq.~\eqref{eq:tlscorrfunc} under the assumption of weak intercavity coupling, as in Eq.~\eqref{eq:QTcorrsingletrans}. For the full Green's function, we use the expansion from Appendix~\ref{appsec:greensexp}.

\begin{figure}
    \centering
    \includegraphics[width=1\linewidth]{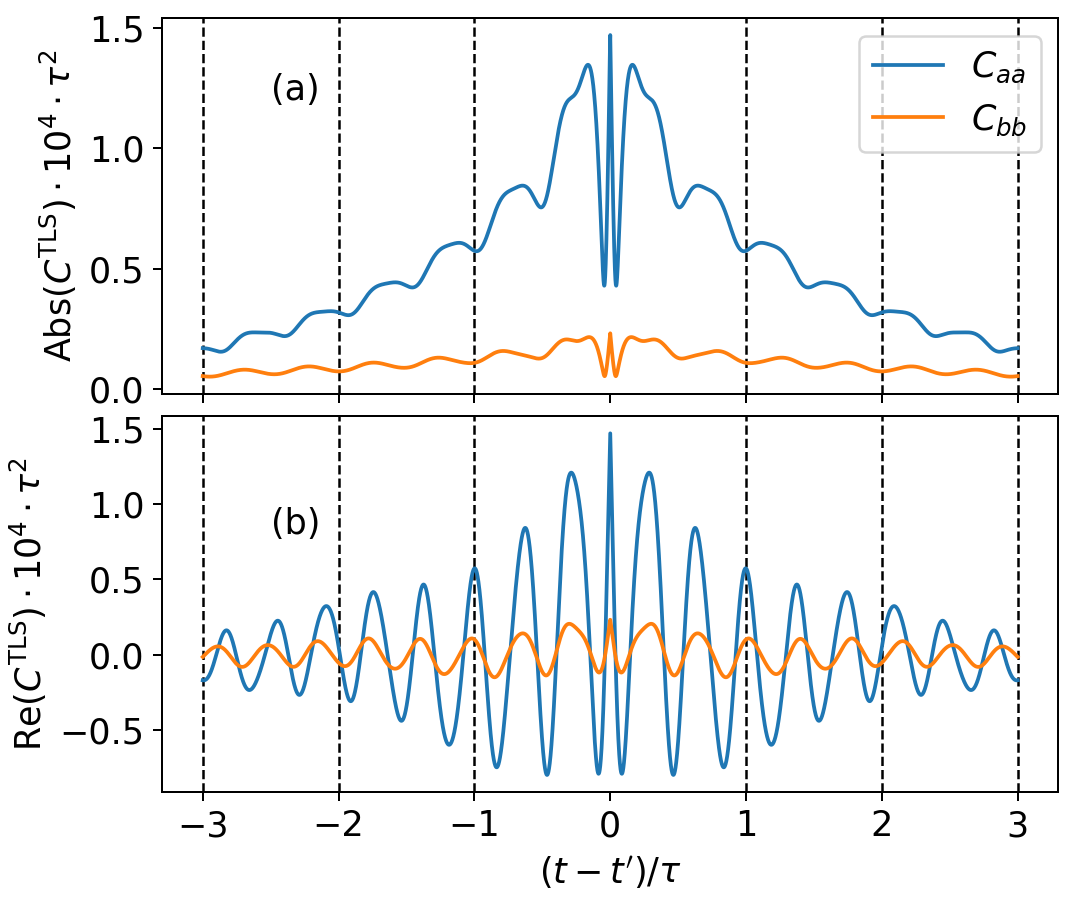}
    \caption{Diagonal terms of TLS correlation function from Eq.~\eqref{eq:tlscorrfunc} for \(N\leq 1\) for two TLSs in the dimers gaps (\(\mathbf{r}_a\in V_1\), \(\mathbf{r}_b\in V_2\)) of metal dimers serving as QNM cavities (cf.~Fig.~\ref{fig:dimer_sketch}). }
    \label{fig:tlscorr_diag}
\end{figure}

\begin{figure}
    \centering
    \includegraphics[width=1\linewidth]{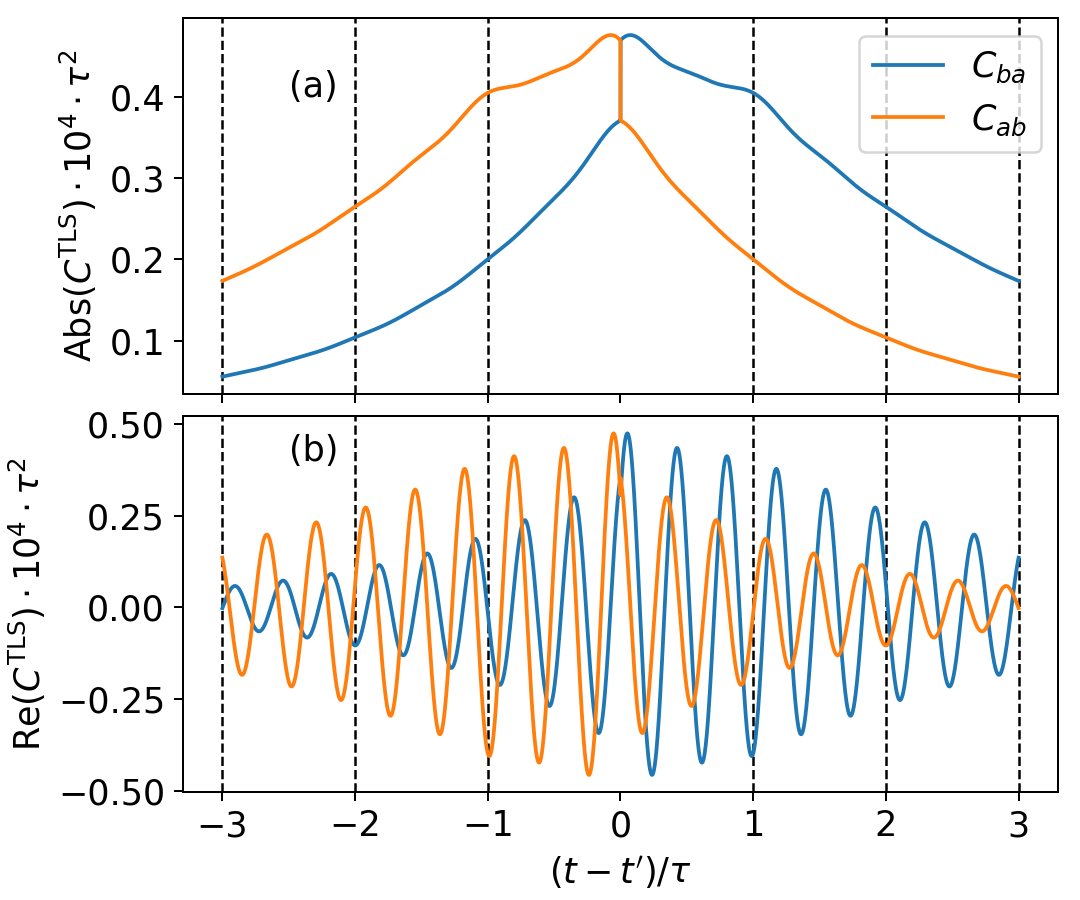}
    \caption{Off-diagonal terms of TLS correlation function from Eq.~\eqref{eq:tlscorrfunc} for \(N\leq 1\) for two TLSs in the dimers gaps (\(\mathbf{r}_a\in V_1\), \(\mathbf{r}_b\in V_2\)) of metal dimers serving as QNM cavities (cf.~Fig.~\ref{fig:dimer_sketch}). }
    \label{fig:tlscorr_ndiag}
\end{figure}

In Fig.~\ref{fig:tlscorr_diag}, we show the diagonal terms of the correlation function, which are related to the temporal decay of a TLS excitation into the bath. Accordingly, we show in Fig.~\ref{fig:tlscorr_ndiag} the off-diagonal terms of the TLS correlation [Eq.~\eqref{eq:tlscorrfunc}], which are related to bath-mediated energy exchange between the TLSs in the different cavities. Since the TLSs mostly radiate into the quasibound QNMs, the bath-mediated decay and coupling are very weak (note the scaling in Figs.~\ref{fig:tlscorr_diag} and \ref{fig:tlscorr_ndiag} with a factor of \(10^4\)). Notably, the bath-mediated coupling between the separated TLSs in Fig.~\ref{fig:tlscorr_ndiag} contains instantaneous contributions at \(t=t'\). These arise from the QNM corrections terms in Eq.~\eqref{eq:tlscorrfunc}, since the QNMs extend into the outside medium, leading to a finite coupling strength even without a delay. However, this coupling is on the order of the instantaneous intercavity coupling that was neglected in Sec.~\ref{sec:multicavityquanti}. 

We again stress that the correlation functions are an ingredient in methods for quantum bath dynamics, where products of correlation functions appear along quantum paths, whose sum yields the full dynamics \cite{caldeira1983path, chernyak1996collective, strathearn2017efficient, fuchs2023hierarchical}. Therefore, the correlation functions presented here act together to determine the state of the coupled quantum system. 

\section{Conclusions}
\red{In conclusion, we addressed the three necessary steps towards a time-depedent theory of quantized QNMs in a system involving multiple cavities, quantum emitters, and a bath of propagating photons. Firstly, we} discussed the coupling between the \red{quasi-bound} QNMs and quantum emitters placed either inside the cavities or in the surrounding homogeneous medium. We defined an area of direct influence, \(P_i(\mathbf{r}_a)\), around the cavity where direct, non-time delayed coupling between the cavity modes and emitters near the cavity is significant. \red{This direct coupling between the cavity field and quantum emitter was expressed in terms of few dominant QNMs.}

\red{As the second step, we} derived a formal solution for the dynamics of the non-bosonic bath operators \red{which arise naturally from the QNM quantization, thus} allowing \red{for a rigorous} expansion in terms of time-delayed intercavity scattering processes. 

\red{Lastly, we} introduced system-bath correlation functions that can be used in typical quantum dynamics schemes. We derived general formulas for the correlation functions for the different types of bath-mediated interactions (QNM-QNM, QNM-TLS, and TLS-TLS, cf.~Tab.~\ref{tab:summary}) and showed how practical forms of the correlations can be obtained by truncating the number of intercavity scattering processes for finite time differences. \red{We also formulated effective coupling elements defined in terms of QNM parameters that can be straight-forwardly calculated using common numerical solvers, thus connecting the quantum theory to established methods in computational photonics.}
Finally, we derived explicit forms of the correlation functions for numerical calculations, using the example of two spatially separated metal dimers with dipole emitters in each dimer gap.

These correlation functions are a central input to many standard quantum dynamics schemes; hence, the methods presented here can be implemented within existing schemes for open quantum systems. Thus, the dynamics of coupled-cavity systems can be obtained using rigorously defined and numerically calculable coupling parameters. In the future, \red{the results from this paper, in combination with an extension to structured background media (e.g., waveguide-coupled cavities)} could enable rigorous simulations of quantum devices, including on-chip quantum information technologies.

\begin{acknowledgments}
R.M.F. and M.R. acknowledge support from the Deutsche Forschungsgemeinschaft (Project number 525575745).
J.R. and S.H. acknowledge funding from Queen's University, Canada, 
the Canadian Foundation for Innovation (CFI), 
the Natural Sciences and Engineering Research Council of Canada (NSERC), and CMC Microsystems for the provision of COMSOL Multiphysics.  
S.H.  thanks the 
Alexander von Humboldt Foundation 
for support through a Humboldt Award.
\end{acknowledgments}

\appendix

\section{Direct QNM-TLS coupling}\label{appsec:qnmtlscoup}
The direct (non-time delayed) coupling between a TLS and the the QNM-generated electric field reads [cf.~Eq.~\eqref{eq:fullqnmtlscoup}],
\begin{align}\label{appeq:genqnmtlscoup}
    g_{a,i_{\mu}} = -\frac{i}{\hbar}\int_0^{\infty}\mathrm{d}\omega \mathbf{d}_a\cdot\mathbf{E}_{i_{\mu}}(\mathbf{r}_a,\omega).
\end{align}

While the field \(\mathbf{E}_{i_{\mu}}(\mathbf{r}_a,\omega)\) [Eq.~\eqref{eq:qnmgenefeld}] can be obtained numerically from the Helmholtz equation in Eq.~\eqref{eq:helmqnmgenfield}, solving the equation is generally cumbersome and unfeasible for complicated structures.
This is especially true if one wishes to know the coupling for different dipole (TLS) locations. Hence, it is useful (and often necessary) to consider approximate methods for the field calculation. Upon inserting the definition of the QNM projectors \(\mathbf{L}_{i_{\mu}}\) from Eq.~\eqref{eq:Ldef} into Eq.~\eqref{eq:qnmgenefeld}, we find
\begin{align}\label{appeq:qnmfield_imu}
    &\mathbf{E}_{i_{\mu}}(\mathbf{r}_a,\omega) = i\sqrt{\frac{\hbar}{\pi\epsilon_0}}\sum_{\eta}\sqrt{\frac{2}{\pi\omega_{i_{\eta}}}}A_{i_{\eta}}^*(\omega)\nonumber\\
    &\qquad\times\Bigg[\int_{V_i}\mathrm{d}^3r\epsilon_I(\mathbf{r},\omega)\mathbf{G}(\mathbf{r}_a,\mathbf{r},\omega)\qnm{f}{i_{\eta}}^*(\mathbf{r})\nonumber\\
    &\qquad+\int_{\overline{V}_i}\mathrm{d}^3r\epsilon_{B,I}\mathbf{G}(\mathbf{r}_a,\mathbf{r},\omega)\qnm{F}{i_{\eta}}^*(\mathbf{r},\omega)\Bigg]\left(S^{-1/2}\right)_{i_{\eta}i_{\mu}}, 
\end{align}
here, the QNMs inside the \(i\)-th cavity act as the source to the field [note that \(\qnm{F}{i_{\eta}}^*(\mathbf{r},\omega)\) originates from cavity \(i\), cf.~Eq.~\eqref{eq:regQNM}]. 

Then the QNM-generated electric field (and by extension the QNM-TLS coupling) is obtained for a specific multi-cavity system by expanding the Green's function in terms of the dominant QNMs of the different cavities \cite{fuchs2026greens}. In the following, we discuss some general properties of the QNM-TLS coupling \(g_{a,i_{\mu}}\) for a TLS inside the QNM cavity (\(\mathbf{r}_a\in V_i\)), inside another cavity (\(\mathbf{r}_a\in V_k,\, k\neq i\)), and for a TLS outside the cavities (\(\mathbf{R}_a\in V_{\rm out}\)).

\textit{TLS inside the QNM cavity}. 
Consider a TLS inside a cavity \(\mathbf{r}_a\in V_i\), coupling to the QNMs \(i_{\mu}\) of that cavity. We make the assumption that
\begin{align}\label{appeq:Greensinside}
    \mathbf{G}(\mathbf{r}_a,\mathbf{r},\omega)\big|_{\mathbf{r}_a,\mathbf{r}\in V_i} = \sum_{\mu}A_{i_{\mu}}(\omega)\qnm{f}{i_{\mu}}(\mathbf{r}_a)\qnm{f}{i_{\mu}}(\mathbf{r}),
\end{align}
i.e., that the Green's function inside the cavity is dominated by the QNMs of that cavity. This assumption holds well for cavities in a 3D homogeneous background medium. Furthermore, we perform the separation \cite{fuchs2026greens}
\begin{align}\label{appeq:cavgreensandothers}
    \mathbf{G}(\mathbf{r}_a,\mathbf{r},\omega)\big|_{\mathbf{r}_a\in V_i, \mathbf{r}\notin V_i} &= \sum_{\mu}A_{i_{\mu}}(\omega)\qnm{f}{i_{\mu}}(\mathbf{r}_a)\qnm{F}{i_{\mu}}(\mathbf{r},\omega) \nonumber\\
    &\;+ \mathbf{G}^{\rm others}(\mathbf{r}_a,\mathbf{r},\omega).
\end{align}

Here, \(\mathbf{G}^{\rm others}\) contains the contributions from the background medium and all other cavities via scattering between the separated cavities. 

We insert the expansions from Eq.~\eqref{appeq:Greensinside} and Eq.~\eqref{appeq:cavgreensandothers} into Eq.~\eqref{appeq:qnmfield_imu} and obtain
\begin{align}\label{appeq:Ei_rainVi}
    &\mathbf{E}_{i_{\mu}}(\mathbf{r}_a,\omega)\big|_{\mathbf{r}_a\in V_i} \nonumber\\
    &\qquad= i\sqrt{\frac{\hbar}{2\epsilon_0}}\sum_{\lambda\eta}\sqrt{\omega_{i_{\lambda}}}\qnm{f}{i_{\lambda}}(\mathbf{r}_a)S_{i_{\lambda}i_{\eta}}(\omega)\left(S^{-1/2}\right)_{i_{\eta}i_{\mu}}\nonumber\\
    &\qquad\quad +\mathbf{E}^{\rm others}_{i_{\mu}}(\mathbf{r}_a,\omega),
\end{align}
where \(S_{i_{\lambda}i_{\eta}}(\omega)\) is defined via the overlap matrix \(S_{i_{\lambda}i_{\eta}} =\int_0^{\infty}\mathrm{d}\omega S_{i_{\lambda}i_{\eta}}(\omega)\) from Eq.~\eqref{eq:Sintradef}, and
\begin{align}
    &\mathbf{E}^{\rm others}_{i_{\mu}}(\mathbf{r}_a,\omega)\big|_{\mathbf{r}_a\in V_i} = \sqrt{\frac{\hbar}{\pi\epsilon_0}}\sum_{\eta}\sqrt{\frac{2}{\pi\omega_{i_{\eta}}}}A_{i_{\eta}}^*(\omega)\nonumber\\
    &\;\times\int_{\overline{V}_i}\mathrm{d}^3r\epsilon_{B,I}\mathbf{G}^{\rm others}(\mathbf{r}_a,\mathbf{r},\omega)\qnm{F}{i_{\eta}}^*(\mathbf{r},\omega)\left(S^{-1/2}\right)_{i_{\eta}i_{\mu}}.
\end{align}

While the form of \(\mathbf{G}^{\rm others}\) depends on the specific multi-cavity structure and the number of other cavities, the field \(\mathbf{E}^{\rm others}\) contains scattering terms that involve at least two well-separated cavities \cite{fuchs2026greens}. Such intercavity scattering is negligible if there is no time delay to match the retardation (cf.~Sec.~\ref{sec:multicavityquanti}), and hence, \(\mathbf{E}^{\rm others}\) is negligible in the instantaneous coupling \(g_{a,i_{\mu}}\) from Eq.~\eqref{appeq:genqnmtlscoup}. Thus, from Eq.~\eqref{appeq:Ei_rainVi} and Eq.~\eqref{appeq:genqnmtlscoup}, it follows that the coupling between a TLS inside a cavity and the QNMs of that cavity takes the form from Eq.~\eqref{eq:qnmtlscoup}.

For cases with a finite time delay (cf.~Sec.~\ref{sec:qnmtlscorrfunc}), the scattering at other cavities can play a significant role, and an expansion of the Green's function in terms of the dominant QNMs of the cavities yields a time-delayed coupling with effective coupling elements (cf.~Appendix.~\ref{appsec:greensexp}).

\textit{TLS inside another cavity}. Next, we consider the case of a TLS inside another cavity, \(\mathbf{r}_a\in V_k,\, k\neq i\). Since the sources of \(\mathbf{E}_{i_{\mu}}\) lie within cavity \(i\), all terms in Eq.~\eqref{appeq:qnmfield_imu} contain at least one intercavity scattering process, and therefore \(g_{a,i_{\mu}}\big|_{\mathbf{r}_a\in V_k, k\neq i}\approx 0\). 

\textit{TLS outside the cavities}.
The field outside the cavities generally consists of a rich overlap of emissions from all cavities and scattered fields. However, the contribution \(\mathbf{E}_{i_{\mu}}(\mathbf{R}_a,\omega)\) has its sources exclusively in the \(i\)-th cavity. Hence, the instantaneous coupling to a TLS in the outside volume is weak if the TLS is far away from the cavity. We wish to derive a measure for the separation between the QNMs and TLS in the outside medium similar to the cavity separation parameter \cite{fuchs2024quantization}. For this purpose, we expand the Green's function in Eq.~\eqref{appeq:qnmfield_imu} in terms of the QNMs of cavity \(i\) [cf.~Eq.~\eqref{appeq:cavgreensandothers}:
\begin{align}\label{appeq:cavgreennoothers}
    \mathbf{G}(\mathbf{R}_a,\mathbf{r},\omega)\big|_{\mathbf{R}_a\notin V_i, \mathbf{r}\in V_i}^{\rm no inter} = \sum_{\mu}A_{i_{\mu}}(\omega)\qnm{f}{i_{\mu}}(\mathbf{R}_a)\qnm{F}{i_{\mu}}(\mathbf{r},\omega), 
\end{align}
we neglected the contribution \(\mathbf{G}^{\rm others}\) from all other cavities, since these yield intercavity scattering terms which are negligible for non-time delayed coupling [remember that the sources of \(\mathbf{E}_{i_{\mu}}(\mathbf{R}_a,\omega)\) lie exclusively in cavity \(i\)].

Similarly, for the Green's function outside the cavities, we perform the expansion \cite{fuchs2026greens}
\begin{align}\label{appeq:greenoutsidenoothers}
    \mathbf{G}(\mathbf{R}_a,\mathbf{r},\omega)\big|_{\mathbf{R}_a, \mathbf{r}\notin V_i}^{\rm no inter}  &\approx \mathbf{G}_B(\mathbf{R}_a,\mathbf{r},\omega) \nonumber\\
    &+ \sum_{\mu}A_{i_{\mu}}(\omega)\qnm{F}{i_{\mu}}(\mathbf{R}_a,\omega)\qnm{F}{i_{\mu}}(\mathbf{r},\omega),
\end{align}
again, neglecting all contributions involving intercavity scattering. Inserting the expansion from Eq.~\eqref{appeq:cavgreennoothers} and Eq.~\eqref{appeq:greenoutsidenoothers} into Eq.~\eqref{appeq:qnmfield_imu} for \(\mathbf{R}_a\in V_{\rm out}\), we find
\begin{align}\label{appeq:qnmgenfieldoutside}
    &\mathbf{E}_{i_{\mu}}(\mathbf{R}_a,\omega)\Big|_{R_a\in V_{out}} \approx \mathbf{E}^B_{i_{\mu}}(\mathbf{R}_a,\omega)\nonumber\\
    &\quad+ i\sqrt{\frac{\hbar}{2\epsilon_0}}\sum_{\lambda\eta}\sqrt{\omega_{i_{\lambda}}}\qnm{F}{i_{\lambda}}(\mathbf{R}_a,\omega)S_{i_{\lambda}i_{\eta}}(\omega)\left(S^{-1/2}\right)_{i_{\eta}i_{\mu}},
\end{align}
with \(\mathbf{E}^B_{i_{\mu}}(\mathbf{R}_a,\omega)\) defined below in Eq.~\eqref{appeq:EB_imu}. 

In the second term on the right hand side, the matrix \(S_{i_{\lambda}i_{\eta}}(\omega)\) contains poles at the complex QNM eigenfrequencies \(\tilde{\omega}_{i_{\lambda}}\) (in the lower complex half plane) and \(\tilde{\omega}^*_{i_{\eta}}\) (in the upper complex half plane). At the same time, \(\qnm{F}{i_{\lambda}}(\mathbf{R}_a,\omega)\) contains an exponential of the form \(\mathrm{e}^{i\omega \tau_{ai}}\), with \(\tau_{ai} = n_{\rm B}|\mathbf{R}_i-\mathbf{R}_a|/c\) being the retardation time between the cavity and the TLS, and \(\mathbf{R}_i\) denotes the location of cavity \(i\) (a similar decomposition of \(\qnm{F}{i_{\lambda}}\) was used in Sec.~\ref{sec:qnmcorrfunc}). Integrating this term over all frequencies yields
\begin{align}\label{appeq:fieldoutside_second}
    &i\sqrt{\frac{\hbar}{2\epsilon_0}}\sum_{\lambda\eta}\sqrt{\omega_{i_{\lambda}}}\int_0^{\infty}\mathrm{d}\omega \qnm{F}{i_{\lambda}}(\mathbf{R}_a,\omega)S_{i_{\lambda}i_{\eta}}(\omega)\left(S^{-1/2}\right)_{i_{\eta}i_{\mu}}\nonumber\\
    &\qquad\approx i\sqrt{\frac{\hbar}{2\epsilon_0}}\sum_{\lambda\eta}\sqrt{\omega_{i_{\lambda}}}\qnm{F}{i_{\lambda}}(\mathbf{R}_a,\tilde{\omega}^*_{i_{\eta}})S_{i_{\lambda}i_{\eta}}\left(S^{-1/2}\right)_{i_{\eta}i_{\mu}},
\end{align}
where the exponential term \(\mathrm{e}^{i\tilde{\omega}^*_{i_{\eta}}n_{\rm B}R_a/c}\) contained in \(\qnm{F}{i_{\lambda}}(\mathbf{R}_a,\tilde{\omega}^*_{i_{\eta}})\) causes an exponential decay that scales with with the distance between the TLS and the cavity (Note that, for simplicity, we chose the origin at the center of the cavity \(i\) so that \(\tau_{ai} = n_{\rm B} R_a/c\)).  

Furthermore, the angular emission profile of the regularized mode \(\qnm{F}{i_{\lambda}}\) is highly important, characterized by the fraction of total radiated power that is emitted towards the TLS, i.e., the directivity \(D_{i_{\lambda}}(\hat{\mathbf{R}}_a)\) \cite{balanis2016antenna}, with the unit vector \(\hat{\mathbf{R}}_a\) in the direction of the TLS.
Combined, this yields a scaling for the field strength of 
\begin{align} \label{appeq:fieldscaling}
    \qnm{F}{i_{\lambda}}(\mathbf{R}_a,\tilde{\omega}^*_{i_{\eta}})\sim \mathcal{O}\big[\mathrm{e}^{-\gamma_{i_{\eta}}n_{\rm B} R_a/c}D_{i_{\lambda}}(\hat{\mathbf{R}}_a)\big].
\end{align}

Next, we consider the first term on the right-hand side of Eq.~\eqref{appeq:qnmgenfieldoutside}, which reads,
\begin{align}\label{appeq:EB_imu}
    &\mathbf{E}^{B}_{i_{\mu}}(\mathbf{R}_a,\omega)\big|_{\mathbf{R}_a\in V_{\rm out}} = \sqrt{\frac{\hbar}{\pi\epsilon_0}}\sum_{\eta}\sqrt{\frac{2}{\pi\omega_{i_{\eta}}}}A_{i_{\eta}}^*(\omega)\nonumber\\
    &\;\times\int_{\overline{V}_i}\mathrm{d}^3r\epsilon_{B,I}\mathbf{G}_B(\mathbf{R}_a,\mathbf{r},\omega)\qnm{F}{i_{\eta}}^*(\mathbf{r},\omega)\left(S^{-1/2}\right)_{i_{\eta}i_{\mu}},
\end{align}
and contains the propagation via the background Green's function. Using the definition of the regularized QNM field \(\qnm{F}{i_{\eta}}^*\) [cf.~Eq.~\eqref{eq:regQNM}], the Helmholtz equation for \(\mathbf{G}_B\), and Green's second identity, we turn the integral over \(\overline{V}_i\) in Eq.~\eqref{appeq:EB_imu} into a surface integral over the cavity surface \(\mathcal{S}_i\) (as shown in Ref.~\onlinecite{fuchs2024quantization}):
\begin{align}\label{appeq:EB_imu_surface}
    &\mathbf{E}^{B}_{i_{\mu}}(\mathbf{R}_a,\omega)\big|_{\mathbf{R}_a\in V_{\rm out}} \nonumber\\
    &\;= \frac{1}{2i}\sqrt{\frac{\hbar}{\pi\epsilon_0}}\sum_{\eta}\sqrt{\frac{2}{\pi\omega_{i_{\eta}}}}A_{i_{\eta}}^*(\omega)\qnm{F}{i_{\eta}}^*(\mathbf{R}_a,\omega)\left(S^{-1/2}\right)_{i_{\eta}i_{\mu}}\nonumber\\
    &\;+\sqrt{\frac{\hbar}{\pi\epsilon_0}}\sum_{\eta}\sqrt{\frac{2}{\pi\omega_{i_{\eta}}}}A_{i_{\eta}}^*(\omega)\frac{c^2}{2i\omega^2}\left(S^{-1/2}\right)_{i_{\eta}i_{\mu}}\nonumber\\
    &\qquad\times\oint_{\mathcal{S}_i}\mathrm{d}A_s \Big\{\big[\opvec{n}_s\times\mathbf{G}_B(\mathbf{s},\mathbf{R}_a,\omega)\big]^T\cdot\big[\nabla_s\times\qnm{F}{i_{\eta}}^*(\mathbf{s},\omega)\big]\nonumber\\
    &\qquad\qquad-\big[\nabla_s\times\mathbf{G}_B(\mathbf{s},\mathbf{R}_a,\omega)\big]^T\cdot\big[\opvec{n}_s\times\qnm{F}{i_{\eta}}^*(\mathbf{s},\omega)\big]\Big\},
\end{align}
where \(\opvec{n}_s\) is the outward pointing surface vector on \(\mathcal{S}_i\). In the first term on the right hand side of Eq.~\eqref{appeq:EB_imu_surface}, \(A_{i_{\eta}}^*(\omega)\) contains a pole at \(\tilde{\omega}^*_{i_{\eta}}\) in the upper half of the complex place, while \(\qnm{F}{i_{\eta}}^*(\mathbf{R}_a,\omega)\) contains an exponential of the form \(\mathrm{e}^{-i\omega \tau_{ai}}\). Hence, this term does not contribute to the instantaneous coupling \(g_{a,i_{\mu}}\) from Eq.~\eqref{appeq:genqnmtlscoup} due to causality.

In the second term on the right hand side of Eq.~\eqref{appeq:EB_imu_surface}, meanwhile, \(\mathbf{G}_B(\mathbf{s},\mathbf{R}_a,\omega)\) contains an exponential of the form \(\mathrm{e}^{i\omega \tau_{ai}}\), so that we obtain, via the residue theorem,
\begin{align}\label{appeq:EB_int}
    &\int_0^{\infty}\mathrm{d}\omega \mathbf{E}^{B}_{i_{\mu}}(\mathbf{R}_a,\omega)\big|_{\mathbf{R}_a\in V_{\rm out}}\nonumber\\
    &\;\approx \sqrt{\frac{\hbar}{2\epsilon_0}}\sum_{\eta} \sqrt{\omega_{i_{\eta}}}\frac{c^2}{(\tilde{\omega}^*_{i_{\eta}})^2}\left(S^{-1/2}\right)_{i_{\eta}i_{\mu}}\nonumber\\
    &\qquad\times\oint_{\mathcal{S}_i}\mathrm{d}A_s \Big\{\big[\opvec{n}_s\times\mathbf{G}_B(\mathbf{s},\mathbf{R}_a,\tilde{\omega}^*_{i_{\eta}})\big]^T\cdot\big[\nabla_s\times\qnm{f}{i_{\eta}}^*(\mathbf{s})\big]\nonumber\\
    &\qquad\qquad-\big[\nabla_s\times\mathbf{G}_B(\mathbf{s},\mathbf{R}_a,\tilde{\omega}^*_{i_{\eta}})\big]^T\cdot\big[\opvec{n}_s\times\qnm{f}{i_{\eta}}^*(\mathbf{s})\big]\Big\},
\end{align}
where we used \(\qnm{F}{i_{\eta}}^*(\mathbf{s},\tilde{\omega}^*_{i_{\eta}}) = [\qnm{F}{i_{\eta}}(\mathbf{s},\tilde{\omega}_{i_{\eta}})]^* = \qnm{f}{i_{\eta}}^*(\mathbf{s})\).

The surface integral in Eq.~\eqref{appeq:EB_int} has the same shape as the regularized QNM field \(\qnm{F}{i_{\eta}}(\mathbf{R}_a,\tilde{\omega}^*_{i_{\eta}})\) from Eq.~\eqref{eq:regQNM} (except that the complex conjugated QNM \(\qnm{f}{i_{\eta}}^*\) appears in the surface integral instead of \(\qnm{f}{i_{\eta}}\)), and in fact shows the same scaling as \(\qnm{F}{i_{\lambda}}(\mathbf{R}_a,\tilde{\omega}^*_{i_{\eta}})\) from Eq.~\eqref{appeq:fieldscaling}. 

Thus, in combination, we find the following scaling for the coupling between the QNM \(i_{\mu}\) and a TLS in the outside medium:
\begin{align}\label{appeq:gcoupscale}
    g_{a,i_{\mu}}\big|_{\mathbf{R}_a\in V_{\rm out}}\sim \mathcal{O}[\mathrm{e}^{-\gamma^{\min}_in_{\rm B}R_a/c}D_i^{\max}(\opvec{R}_a)],
\end{align}
where \(\gamma^{\min}_i = \min_{\mu}(\gamma_{i_{\mu}})\) and \(D_i^{\max}(\opvec{R}_a) = \max_{\mu}[D_{i_{\mu}}(\opvec{R}_a)]\). We note that, as discussed in Ref.~\onlinecite{fuchs2024quantization}, the amplitude of the regularized QNM fields \(\qnm{F}{i_{\mu}}\) (and therefore of the coupling to a TLS outside the cavities) decreases in the high-\(Q\) limit like \(1/\sqrt{Q_{i_{\mu}}}\), where \(Q_{i_{\mu}} =\omega_{i_{\mu}}/(2\gamma_{i_{\mu}})\) is the quality factor of the mode. We omit this scaling here for generality, since the scaling from Eq.~\eqref{appeq:gcoupscale} holds also in the low-\(Q\) case.

Thus, we obtain the \textit{area of direct influence} \(P_i(\mathbf{R}_a)\) from Eq.~\eqref{eq:tlsqnmseparation}.

\section{Calculation of the QNM scattering}\label{appsec:QNMscatter}
The intercavity scattering from Eq.~\eqref{eq:singleK} reads,
\begin{align*}
    &\mathbb{K}^{1}_{i_{\mu}j_{\nu}}(t-t') \nonumber\\
    &\qquad= i\int_0^{\infty}\mathrm{d}\omega \int\mathrm{d}^3r \mathbf{g}_{i_{\mu}}(\mathbf{r},\omega)\cdot\mathbf{L}^*_{j_{\nu}}(\mathbf{r},\omega)\mathrm{e}^{-i\omega(t-t')}.
\end{align*}

To derive QNM-QNM coupling elements, we consider the spatial integral
\begin{align}\label{appeq:spatint}
    &\int\mathrm{d}^3r \mathbf{g}_{i_{\mu}}(\mathbf{r},\omega)\cdot\mathbf{L}^*_{j_{\nu}}(\mathbf{r},\omega)=\sum_{\mu'\nu'} \left(S^{-1/2}\right)_{i_{\mu}i_{\mu'}} \nonumber\\
    &\qquad\times\frac{i(\tilde{\omega}_{i_{\mu'}}-\tilde{\omega}^*_{j_{\nu'}})}{2\pi(\omega-\tilde{\omega}^*_{j_{\nu'}})} S^{\rm inter}_{i_{\mu'}j_{\nu'}}(\omega)\left(S^{-1/2}\right)_{j_{\nu'}j_{\nu}},
\end{align}
with
\begin{align}
   &S^{\rm inter}_{i_{\mu'}j_{\nu'}}(\omega)=\int_{\overline{V}_{ij}}\mathrm{d}^3r\frac{\omega^2\epsilon_{B,I}\qnm{F}{i_{\mu'}}(\mathbf{r},\omega)\cdot\qnm{F}{j_{\nu'}}^*(\mathbf{r},\omega)}{i(\tilde{\omega}_{i_{\mu'}}-\tilde{\omega}^*_{j_{\nu'}})\sqrt{\omega_{i_{\mu'}}\omega_{j_{\nu'}}}},
\end{align}
where we inserted the definitions of \(\mathbf{L}^*_{j_{\nu}}\) from Eq.~\eqref{eq:Ldef} and of \(\mathbf{g}_{i_{\mu}}\) from Eq.~\eqref{eq:gcoup}, and multiplied by \(i(\tilde{\omega}_{i_{\mu'}}-\tilde{\omega}^*_{j_{\nu'}})\) in the numerator and denominator. Furthermore, \(\overline{V}_{ij}\) denotes the complement of the cavity volumes \(V_i\cup V_j\). 

As shown in Ref.~\cite{fuchs2024quantization}, we turn the spatial integral into surface integrals over the two cavity surfaces so that
\begin{align}\label{appeq:spatint_inter}
    &S^{\rm inter}_{i_{\mu}j_{\nu}}(\omega)=\oint_{\mathcal{S}_i\cup\mathcal{S}_j}\mathrm{d}A_s \Bigg\{\frac{\omega\Big[\qnm{F}{i_{\mu'}}(\mathbf{s},\omega)\times\qnm{H}{j_{\nu'}}^*(\mathbf{s},\omega)\Big]\cdot\opvec{n}_s}{i(\tilde{\omega}_{i_{\mu'}}-\tilde{\omega}^*_{j_{\nu'}})\sqrt{\omega_{i_{\mu'}}\omega_{j_{\nu'}}}}\nonumber\\
    &\qquad\qquad\qquad\qquad\qquad\qquad+\mathrm{c.c.}(i_{\mu'}\leftrightarrow j_{\nu'})\Bigg\},
\end{align}
where \(\qnm{H}{i_{\mu}}(\mathbf{s},\omega) = \nabla\times\qnm{F}{i_{\mu}}(\mathbf{s},\omega)/(i\omega\mu_0)\) are the regularized magnetic QNMs.

Now, we separate \(\qnm{F}{i_{\mu'}}(\mathbf{s},\omega)\big|_{\mathbf{s}\in\mathcal{S}_j} = \qnm{F}{i_{\mu'}}'(\mathbf{s},\omega)\mathrm{e}^{i\omega\tau_{ij}}\) into a slow-rotating envelope function \(\qnm{F}{i_{\mu'}}'(\mathbf{s},\omega)\) and a fast-rotating exponential \(\mathrm{e}^{i\omega\tau_{ij}}\) in the integral over the surface \(\mathcal{S}_j\). Note that the slow-rotating part is defined such that \(\qnm{F}{i_{\mu'}}(\mathbf{s},\omega)\big|_{\mathbf{s}\in\mathcal{S}_i} = \qnm{F}{i_{\mu'}}'(\mathbf{s},\omega)\). Similarly, we separate fast-rotating and slow-rotating parts of \(\qnm{H}{j_{\nu'}}^*(\mathbf{s},\omega)\) in the integral over \(\mathcal{S}_i\). Then, we pull the fast-rotating parts out of the integrals and evaluate the rest of the integrals via the pole approximation \cite{ren2020near}:
\begin{align}
    &\omega \to \sqrt{\omega_{i_{\mu'}}\omega_{j_{\nu'}}},\nonumber\\
    &\qnm{F}{i_{\mu'}}'(\mathbf{s},\omega)\to \qnm{F}{i_{\mu'}}'(\mathbf{s},\omega_{i_{\mu'}}).
\end{align}

Thus, we obtain
\begin{align}
    S^{\rm inter}_{i_{\mu'}j_{\nu'}}(\omega) = (1-\delta_{ij})\big[S_{i_{\mu'}\rightarrow j_{\nu'}}\mathrm{e}^{-i\omega\tau_{ij}}+S_{i_{\mu'}\leftarrow j_{\nu'}}\mathrm{e}^{i\omega\tau_{ij}}\big],
\end{align}
with the retarded overlap matrices from Eq.~\eqref{eq:retardedSmatrix}. We then reinsert \( S^{\rm inter}_{i_{\mu}j_{\nu}}(\omega)\) into Eq.~\eqref{appeq:spatint} and obtain the coupling matrices \(\chi^{(-)}_{i_{\mu}\rightarrow j_{\nu}}\) from Eq.~\eqref{eq:QNMcoupmatrix} by employing the relation \(\sum_{\eta} \left(S^{1/2}\right)_{i_{\mu}i_{\eta}}\left(S^{-1/2}\right)_{i_{\eta}i_{\nu}} = \delta_{\mu\nu}\). Thus, we arrive at the relation from Eq.~\eqref{eq:singleKapprox}.

For the relation for \(C^{\rm bos}_{i_{\mu}j_{\nu}}\) from Eq.~\eqref{eq:Cbosapprox}, we use the relation between \(\mathbf{g}^*_{j_{\nu}}\) and \(\mathbf{L}^*_{j_{\nu}}\) [cf.~Eq.~\eqref{eq:gcoup}], and then follow the same steps for the spatial integral as for \(\mathbb{K}^1_{i_{\mu}j_{\nu}}\). 

In contrast to \(\mathbb{K}^1_{i_{\mu}j_{\nu}}\), where always \(i\neq j\), \(C^{\rm bos}_{i_{\mu}j_{\nu}}\) also contains intracavity terms \(i=j\). The derivation for the intracavity case follows similar steps as for the intercavity case, by treating the spatial integrals within a pole approximation to obtain effective coupling elements, except that there are no fast-rotating exponentials \(\mathrm{e}^{\pm i\omega \tau}\) in the intracavity case since all modes originate from the same cavity. A detailed derivation is given in Ref.~\onlinecite{franke2020quantized}. We defined the retarded overlap matrix from Eq.~\eqref{eq:retardedSmatrix} in such a way that \(S_{i_{\mu}\rightarrow i_{\nu}}=S_{i_{\mu}\leftarrow i_{\nu}} = S_{i_{\mu}i_{\nu}}/2\) holds, and as a result \(\chi^{(-)}_{i_{\mu}\rightarrow i_{\nu}} = \chi^{(-)}_{i_{\mu}\leftarrow i_{\nu}} = \chi^{(-)}_{i_{\mu}i_{\nu}}\), which is the single-cavity dissipator matrix from Ref.~\onlinecite{franke2020quantized}. Hence, the relation from Eq.~\eqref{eq:Cbosapprox} holds also in the intracavity case \(i=j\), and the known single-cavity result is recovered. 

\section{Derivation of the correlation functions}\label{appsec:corrfuncs}
\subsection{Derivation of the QNM correlation function}\label{appsec:QNMcorr}
With the QNM-bath coupling Hamiltonian from Eq.~\eqref{eq:HQNMbath} together with the initial scattering \(\hat{C}_{i_{\mu}}(t)\) from Eq.~\eqref{eq:initscat}, the correlation function from Eq.~\eqref{eq:gencorrfunc} becomes
\begin{align}\label{appeq:QNmcorrcommutator}
    &C^{\rm QNM}_{i_{\mu}j_{\nu}}(t-t') \nonumber\\
    &= \sum_{N=0}^{\infty}\sum_{M=0}^{\infty}\sum_{k,l}\sum_{\eta,\kappa}\sum_{pq}\int_{t_0}^t\mathrm{d}t_1\int_{t_0}^{t'}\mathrm{d}t_2 \mathbb{K}^N_{i_{\mu}k_{\eta}}(t-t_1)\nonumber\\
    &\times \int_0^{\infty}\int\mathrm{d}^3r g_{k_{\eta},p}(\mathbf{r},\omega)\mathrm{e}^{-i\omega (t_1-t_0)}\nonumber\\
    &\times \mathrm{tr}_B\left[\hat{c}_p(\mathbf{r},\omega,t_0)\hat{c}^{\dagger}_q(\mathbf{r}',\omega',t_0)\rho_B\right]\nonumber\\
    &\times \int_0^{\infty}\mathrm{d}\omega'\int\mathrm{d}^3r'g^*_{l_{\kappa},q}(\mathbf{r}',\omega')\mathrm{e}^{i\omega' (t_2-t_0)}\big[\mathbb{K}^M_{j_{\nu}l_{\kappa}}(t'-t_2)\big]^*. 
\end{align}

For an initial vacuum bath state \(\rho_B = |{\rm vac}\rangle\langle {\rm vac}|\), the trace over the bath degrees of freedom can be replaced with the commutator from Eq.~\eqref{eq:ccomm}, so that
\begin{align}\label{appeq:CQNMcommutator}
    &C^{\rm QNM}_{i_{\mu}j_{\nu}}(t-t') \nonumber\\
    &\; = \sum_{N=0}^{\infty}\sum_{M=0}^{\infty}\sum_{k,l}\sum_{\eta,\kappa}\int_{t_0}^t\mathrm{d}t_1\int_{t_0}^{t'}\mathrm{d}t_2 \mathbb{K}^N_{i_{\mu}k_{\eta}}(t-t_1)\nonumber\\
    &\qquad\qquad\qquad\times C^{\rm bos}_{k_{\eta}l_{\kappa}}(t_1-t_2)\big[\mathbb{K}^M_{j_{\nu}l_{\kappa}}(t'-t_2)\big]^*\nonumber\\
    &\; - \sum_{N=1}^{\infty}\sum_{M=1}^{\infty}\sum_{k,\eta}\mathbb{K}^N_{i_{\mu}k_{\eta}}(t-t_0)\big[\mathbb{K}^M_{j_{\nu}k_{\eta}}(t'-t_0)\big]^*,
\end{align}
with \(C^{\rm bos}_{i_{\mu}j_{\nu}}(t-t')\) as defined in Eq.~\eqref{eq:Cbos} and \(\mathbb{K}^N_{i_{\mu}j_{\nu}}(t-t')\) from Eq.~\eqref{eq:doubleK}. Note that, while \(t>t'\) is implicit for \(\mathbb{K}^N_{i_{\mu}j_{\nu}}(t-t')\), \(C^{\rm bos}_{i_{\mu}j_{\nu}}(t-t')\) has no such time order. Furthermore, \(C^{\rm bos}_{i_{\mu}j_{\nu}}\) may contain intercavity (\(i\neq j\)) or intracavity (\(i=j\)) scattering. We separate these different contributions and reorder them in the correlation function accordingly, to obtain
\begin{align}\label{appeq:QNMcorrreordered}
    &C^{\rm QNM}_{i_{\mu}j_{\nu}}(t-t') \nonumber\\
    &\; = \sum_{N=0}^{\infty}\sum_{M=0}^{\infty}\sum_{l,\eta\kappa}\int_{t_0}^t\mathrm{d}t_1\int_{t_0}^{t'}\mathrm{d}t_2 \mathbb{K}^N_{i_{\mu}l_{\eta}}(t-t_1)\nonumber\\
    &\qquad\qquad\qquad\times C^{\rm bos}_{l_{\eta}l_{\kappa}}(t_1-t_2)\big[\mathbb{K}^M_{j_{\nu}l_{\kappa}}(t'-t_2)\big]^*\nonumber\\
    &\; +\sum_{N=0}^{\infty}\sum_{k\neq j}\sum_{\eta}\int_{t'}^t\mathrm{d}t_1 \mathbb{K}^N_{i_{\mu}k_{\eta}}(t-t_1) C^{\rm bos}_{k_{\eta}j_{\nu}}(t_1-t')\nonumber\\
    &\; +\sum_{M=0}^{\infty}\sum_{l\neq i}\sum_{\kappa}\int_t^{t'}\mathrm{d}t_2\big[ \mathbb{K}^M_{j_{\nu}l_{\kappa}}(t'-t_1) C^{\rm bos}_{l_{\kappa}i_{\mu}}(t_1-t)\big]^*\nonumber\\
    &\; + \sum_{N=0}^{\infty}\sum_{M=1}^{\infty}\sum_{k,l\neq k}\sum_{\eta,\kappa}\int_{t_0}^t\mathrm{d}t_1\int_{t_0}^{t'}\mathrm{d}t_2 \mathbb{K}^N_{i_{\mu}k_{\eta}}(t-t_1)\nonumber\\
    &\qquad\qquad\qquad\times\Theta(t_1-t_2) C^{\rm bos}_{k_{\eta}l_{\kappa}}(t_1-t_2)\big[\mathbb{K}^M_{j_{\nu}l_{\kappa}}(t'-t_2)\big]^*\nonumber\\
    &\; + \sum_{N=1}^{\infty}\sum_{M=0}^{\infty}\sum_{k,l\neq k}\sum_{\eta,\kappa}\int_{t_0}^t\mathrm{d}t_1\int_{t_0}^{t'}\mathrm{d}t_2 \mathbb{K}^N_{i_{\mu}k_{\eta}}(t-t_1)\nonumber\\
    &\qquad\qquad\qquad\times\Theta(t_2-t_1) \big[\mathbb{K}^M_{j_{\nu}l_{\kappa}}(t'-t_2)C^{\rm bos}_{l_{\kappa}k_{\eta}}(t_2-t_1)\big]^*\nonumber\\
    &\; - \sum_{N=1}^{\infty}\sum_{M=1}^{\infty}\sum_{k,\eta}\mathbb{K}^N_{i_{\mu}k_{\eta}}(t-t_0)\big[\mathbb{K}^M_{j_{\nu}k_{\eta}}(t'-t_0)\big]^*,
\end{align}
we also used \(C^{\rm bos}_{i_{\mu}j_{\nu}}(t-t') = \big[C^{\rm bos}_{j_{\nu}i_{\mu}}(t'-t)\big]^*\) [cf.~Eq.~\eqref{eq:Cbos}], for \(t'>t\). 

Now, we employ the relation
\begin{align}\label{appeq:CbosK1}
    &(1-\delta_{ij})\Theta(t-t')C^{\rm bos}_{i_{\mu}j_{\nu}}(t-t') \nonumber\\
    & = (1-\delta_{ij})\Theta(t-t')\int_0^{\infty}\mathrm{d}\omega\int\mathrm{d}^3r \mathbf{g}_{i_{\mu}}(\mathbf{r},\omega)\nonumber\\
    &\qquad\qquad\qquad\qquad\qquad\qquad\qquad\times\mathbf{g}^*_{j_{\nu}}(\mathbf{r},\omega)\mathrm{e}^{-i\omega(t-t')} \nonumber\\
    &= \sum_{\nu'}(1-\delta_{ij})\Theta(t-t')\int_0^{\infty}\mathrm{d}\omega\int\mathrm{d}^3r \mathbf{g}_{i_{\mu}}(\mathbf{r},\omega)\nonumber\\
    &\qquad\qquad\qquad\times(\delta_{\nu\nu'}\omega - \chi^*_{j_{\nu}j_{\nu'}})\mathbf{L}^*_{j_{\nu'}}(\mathbf{r},\omega)\mathrm{e}^{-i\omega(t-t')}\nonumber\\
    &= \sum_{\nu'}(-\delta_{\nu\nu'}\partial_{t'}+i\chi^*_{j_{\nu}j_{\nu'}})\mathbb{K}^1_{i_{\mu}j_{\nu'}}(t-t'),
\end{align}
where we used the definitions of \(C^{\rm bos}\) from Eq.~\eqref{eq:Cbos} and \(\mathbb{K}^1\) from Eq.~\eqref{eq:singleK} together with the definition of the noise-coupling elements \(\mathbf{g}^*_{j_{\nu}}\) [analogous to Eq.~\eqref{eq:gcoup}]. Thus, using the definition of \(\mathbb{K}^N\) from Eq.~\eqref{eq:doubleK} and renaming the indices accordingly, we obtain
\begin{align}\label{appeq:QNMcorralmost}
    &C^{\rm QNM}_{i_{\mu}j_{\nu}}(t-t') \nonumber\\
    &\; = \sum_{N=0}^{\infty}\sum_{M=0}^{\infty}\sum_{l,\eta\kappa}\int_{t_0}^t\mathrm{d}t_1\int_{t_0}^{t'}\mathrm{d}t_2 \mathbb{K}^N_{i_{\mu}l_{\eta}}(t-t_1)\nonumber\\
    &\qquad\qquad\qquad\times C^{\rm bos}_{l_{\eta}l_{\kappa}}(t_1-t_2)\big[\mathbb{K}^M_{j_{\nu}l_{\kappa}}(t'-t_2)\big]^*\nonumber\\
    &\; +\Theta(t-t')\sum_{N=1}^{\infty}\sum_{\eta}(-\delta_{\nu\eta}\partial_{t'}{+}i\chi^*_{j_{\nu}j_{\eta}})\mathbb{K}^N_{i_{\mu}j_{\eta}}(t-t') \nonumber\\
    &\; +\Theta(t'-t)\sum_{M=1}^{\infty}\sum_{\eta}\big[(-\delta_{\mu\eta}\partial_t{+}i\chi^*_{i_{\mu}i_{\eta}}) \mathbb{K}^M_{j_{\nu}i_{\eta}}(t'-t) \big]^*\nonumber\\
    &\; - \sum_{N=1}^{\infty}\sum_{M=1}^{\infty}\sum_{k,\eta}\int_{t_0}^t\mathrm{d}t_1\Theta(t'-t_1)\partial_{t_1}\Big\{ \mathbb{K}^N_{i_{\mu}k_{\eta}}(t-t_1)\nonumber\\
    &\qquad\qquad\qquad\qquad\qquad\qquad\times\big[\mathbb{K}^M_{j_{\nu}k_{\eta}}(t'-t_1)\big]^*\Big\}\nonumber\\
    &\; - \sum_{N=1}^{\infty}\sum_{M=1}^{\infty}\sum_{k,\eta}\int_{t_0}^t\mathrm{d}t_1\Theta(t'-t_1)2\chi^{(-)}_{k_{\eta}k_{\eta'}}\mathbb{K}^N_{i_{\mu}k_{\eta}}(t-t_1)\nonumber\\
    &\qquad\qquad\qquad\qquad\qquad\qquad\times\big[\mathbb{K}^M_{j_{\nu}k_{\eta'}}(t'-t_1)\big]^*\nonumber\\
    &\; - \sum_{N=1}^{\infty}\sum_{M=1}^{\infty}\sum_{k,\eta}\mathbb{K}^N_{i_{\mu}k_{\eta}}(t-t_0)\big[\mathbb{K}^M_{j_{\nu}k_{\eta}}(t'-t_0)\big]^*.
\end{align}

Here, \(\chi^{(-)}_{k_{\eta}k_{\eta'}} = i(\chi_{k_{\eta}k_{\eta'}}-\chi^*_{k_{\eta'}k_{\eta}})/2\) is the imaginary part of the symmetrized QNM frequency \(\chi_{k_{\eta}k_{\eta'}}\). Note that here and in the following, we neglect instantaneous intercavity scattering \(\mathbb{K}_{i_{\mu}j_{\nu}}^N(0)\approx 0\) assuming well-separated cavities (cf.~Sec.~\ref{sec:multicavityquanti}). Using the same assumption, we note that, in Eq.~\eqref{appeq:QNMcorralmost}:
\begin{align*}
    & \sum_{N=1}^{\infty}\sum_{M=1}^{\infty}\sum_{k,\eta}\int_{t_0}^t\mathrm{d}t_1\Theta(t'-t_1)\partial_{t_1}\Big\{ \mathbb{K}^N_{i_{\mu}k_{\eta}}(t-t_1)\nonumber\\
    &\qquad\qquad\qquad\qquad\qquad\qquad\times\big[\mathbb{K}^M_{j_{\nu}k_{\eta}}(t'-t_1)\big]^*\Big\}\nonumber\\
    &\; + \sum_{N=1}^{\infty}\sum_{M=1}^{\infty}\sum_{k,\eta}\mathbb{K}^N_{i_{\mu}k_{\eta}}(t-t_0)\big[\mathbb{K}^M_{j_{\nu}k_{\eta}}(t'-t_0)\big]^* \approx 0,
\end{align*}
which we obtain by evaluating the integral at the lower limit and upper limit (distinguishing the cases \(t>t'\) and \(t<t'\)) together with \(\mathbb{K}_{i_{\mu}j_{\nu}}^N(0)\approx 0\).

Lastly, we employ the approximation
\begin{align}\label{appeq:Cbosapprox}
    C^{\rm bos}_{l_{\eta}l_{\kappa}}(t_1-t_2) \approx 2\chi^{(-)}_{l_{\eta}l_{\kappa}}\delta(t_1-t_2),
\end{align}
which holds well in the intracavity case (this was demonstrated in Ref.~\onlinecite{franke2020quantized} for a single cavity, and can be seen from Fig.~\ref{fig:QNMcorr}), and use it to rewrite in Eq.~\eqref{appeq:QNMcorralmost},
\begin{align*}
    &\sum_{N=0}^{\infty}\sum_{M=0}^{\infty}\sum_{l,\eta\kappa}\int_{t_0}^t\mathrm{d}t_1\int_{t_0}^{t'}\mathrm{d}t_2 \mathbb{K}^N_{i_{\mu}l_{\eta}}(t-t_1)\nonumber\\
    &\qquad\qquad\qquad\times C^{\rm bos}_{l_{\eta}l_{\kappa}}(t_1-t_2)\big[\mathbb{K}^M_{j_{\nu}l_{\kappa}}(t'-t_2)\big]^*\nonumber\\
    &\; - \sum_{N=1}^{\infty}\sum_{M=1}^{\infty}\sum_{k,\eta}\int_{t_0}^t\mathrm{d}t_1\Theta(t'-t_1)2\chi^{(-)}_{k_{\eta}k_{\eta'}}\mathbb{K}^N_{i_{\mu}k_{\eta}}(t-t_1)\nonumber\\
    &\approx \delta_{ij}C^{\rm bos}_{i_{\mu}j_{\nu}}(t-t')+\sum_{N=1}^{\infty}\sum_{\eta}\mathbb{K}^N_{i_{\mu}j_{\eta}}(t-t')2\chi^{(-)}_{j_{\eta}j_{\nu}}\nonumber\\
    &\qquad\qquad\qquad\qquad+\sum_{M=1}^{\infty}\sum_{\eta}\Big[\mathbb{K}^M_{j_{\nu}i_{\eta}}(t-t')2\chi^{(-)}_{i_{\eta}i_{\mu}}\Big]^*,
\end{align*}
using  the definition of \(\mathbb{K}^0\) from Eq.~\eqref{eq:Kzero}.

In conclusion, we rewrite Eq.~\eqref{appeq:QNMcorralmost} into
\begin{align}
    &C^{\rm QNM}_{i_{\mu}j_{\nu}}(t-t') = \delta_{ij}C^{\rm bos}_{i_{\mu}j_{\nu}}(t-t')\nonumber\\
    &\qquad - \sum_{N=1}^{\infty}\sum_{\eta}(\delta_{\eta\nu}\partial_{t'}{-}i\chi_{j_{\eta}j_{\nu}})\mathbb{K}^N_{i_{\mu}j_{\eta}}(t-t') \nonumber\\
    &\qquad - \sum_{N=1}^{\infty}\sum_{\eta}\big[(\delta_{\eta\mu}\partial_{t}{-}i\chi_{i_{\eta}i_{\mu}})\mathbb{K}^N_{j_{\nu}i_{\eta}}(t'-t)\big]^*,
\end{align}
which is precisely the correlation function from Eq.~\eqref{eq:QNMcorrfunc}.

\subsection{Derivation of the QNM-TLS correlation function}\label{appsec:qnmtlscorr}
Using the Hamiltonian of QNM-bath coupling from Eq.~\eqref{eq:HQNMbath} and the Hamiltonian for TLS-bath coupling from Eq.~\eqref{eq:HTLSbath}, the correlation function for the bath-mediated coupling between a QNM and TLS reads [cf.~Eq.~\eqref{eq:gencorrfunc}]
\begin{align}\label{appeq:CQTcommutator}
    &C^{\rm Q-T}_{i_{\mu} a}(t-t') = \sum_{j,\nu}\sum_{N=0}^{\infty}\int_{t_0}^t\mathrm{d}t_1 \mathbb{K}^N_{i_{\mu}j_{\nu}}(t-t_1)\nonumber\\
    &\qquad\times\sum_p \int\mathrm{d}^3r\int_0^{\infty}\mathrm{d}\omega \mathbf{g}^*_{a,p}(\mathbf{r},\omega) \mathrm{e}^{i\omega(t'-t_0)}\nonumber\\
    &\qquad\qquad\times\mathrm{tr}_B\big[\hat{C}_{j_{\nu}}(t_1)\hat{c}_p^{\dagger}(\mathbf{r},\omega,t_0)\rho_B\big]\nonumber\\
    &+\sum_{j,\nu}\int_{t_0}^{t'}\mathrm{d}t_1 C^{\rm QNM}_{i_{\mu}j_{\nu}}(t-t_1) \left[\frac{\mathbf{d}_a\cdot\mathbf{E}_{j_{\nu}}(\mathbf{r}_a,t'-t_1)}{i\hbar}\right]^*.
\end{align}
The QNM-bath noise operators \(\hat{C}_{i_{\mu}}(t)\) are defined in Eq.~\eqref{eq:initscat}, the retarded QNM-QNM scattering elements \(\mathbb{K}^N_{i_{\mu}j_{\nu}}(t-t')\) are defined in Eq.~\eqref{eq:doubleK}, and the time-dependent QNM-generated electric field is defined in Eq.~\eqref{eq:timedepQNMgenefield}. Furthermore, we used
\begin{align}
    &C^{\rm QNM}_{i_{\mu}j_{\nu}}(t-t') \nonumber\\
    &\qquad= \sum_{kl,\eta\kappa}\sum_{N=0}^{\infty}\sum_{M=0}^{\infty}\int_{t_0}^t\mathrm{d}t_1\int_{t_0}^{t'}\mathrm{d}t_2 \mathbb{K}^N_{i_{\mu}k_{\eta}}(t-t_1)\nonumber\\
    &\qquad\qquad\times\Big[\mathbb{K}^M_{j_{\nu}l_{\kappa}}(t'-t_2)\Big]^*\mathrm{tr}_B\big[\hat{C}_{k_{\eta}}(t_1)\hat{C}^{\dagger}_{l_{\kappa}}(t_2)\rho_B\big],
\end{align}
which follows from Eq.~\eqref{appeq:QNmcorrcommutator} together with the definition of \(\hat{C}_{i_{\mu}}(t)\) from Eq.~\eqref{eq:initscat}. 

Next, we again assume an initial vacuum bath state \(\rho_B = |{\rm vac}\rangle\langle {\rm vac}|\), so that 
\begin{align}
    &\mathrm{tr}_B\big[\hat{C}_{j_{\nu}}(t_1)\hat{c}_p^{\dagger}(\mathbf{r},\omega,t_0)\rho_B\big] = \big[\hat{C}_{j_{\nu}}(t_1),\hat{c}_p^{\dagger}(\mathbf{r},\omega,t_0)\big]_-\nonumber\\
    &= ig_{j_{\nu},p}(\mathbf{r},\omega)\mathrm{e}^{-i\omega (t_1-t_0)}-\sum_{k,\eta}\mathbb{K}^1_{j_{\nu}k_{\eta}}(t_1-t_0)L_{k_{\eta},p}(\mathbf{r},\omega), 
\end{align}
using the commutator from Eq.~\eqref{eq:ccomm}. Here, \(\mathbf{L}_{k_{\eta}}(\mathbf{r},\omega)\) is the QNM projector kernel from Eq.~\eqref{eq:Ldef}, \(\mathbf{g}_{j_{\nu}}(\mathbf{r},\omega)\) is the QNM-bath noise coupling element from Eq.~\eqref{eq:gcoup}, and \(\mathbb{K}^1_{j_{\nu}k_{\eta}}(t_1-t_0)\) is the first-order retarded QNM-QNM scattering element from Eq.~\eqref{eq:singleK}. 

Combining this with the QNM correlation function \(C^{\rm QNM}_{i_{\mu}j_{\nu}}(t-t_1)\) from Eq.~\eqref{eq:QNMcorrfunc}, we thus obtain from Eq.~\eqref{appeq:CQTcommutator}:
\begin{align}
    &C^{\rm Q-T}_{i_{\mu} a}(t-t') = \sum_{j,\nu\eta}\sum_{N=0}^{\infty}\int_{t_0}^t\mathrm{d}t_1 \mathbb{K}^N_{i_{\mu}j_{\nu}}(t-t_1)\nonumber\\
    &\qquad \times(\delta_{\nu\eta}\partial_{t'}{-}i\chi_{j_{\nu}j_{\eta}})\left[\frac{\mathbf{d}_a\cdot\mathbf{E}_{j_{\eta}}(\mathbf{r}_a,t'-t_1)}{i\hbar}\right]^*\nonumber\\
    &-\sum_{j,\nu}\sum_{N=1}^{\infty}\mathbb{K}^N_{i_{\mu}j_{\nu}}(t-t_0)\left[\frac{\mathbf{d}_a\cdot\mathbf{E}_{j_{\eta}}(\mathbf{r}_a,t'-t_0)}{i\hbar}\right]^*\nonumber\\
    & +\sum_{j,\nu\eta}\int_{t_0}^{t'}\mathrm{d}t_1 \left[\frac{\mathbf{d}_a\cdot\mathbf{E}_{j_{\eta}}(\mathbf{r}_a,t'-t_1)}{i\hbar}\right]^*\nonumber\\
    &\quad\times\Bigg\{\delta_{ij}\delta_{\nu\eta}C^{\rm bos}_{i_{\mu}j_{\eta}}(t-t_1)\nonumber\\
    &\qquad-\Theta(t-t_1)\sum_{N=1}^{\infty}(\delta_{\nu\eta}\partial_{t_1}{-}i\chi_{j_{\eta}j_{\nu}})\mathbb{K}^N_{i_{\mu}j_{\eta}}(t-t_1)\nonumber\\
    &\qquad -\Theta(t_1-t)\sum_{N=1}^{\infty}\Big[(-\delta_{\mu\eta}\partial_{t_1}{-}i\chi_{i_{\eta}i_{\mu}})\mathbb{K}^N_{j_{\nu}i_{\eta}}(t_1-t)\Big]^*\Bigg\}.
\end{align}

Next, we separate in the first term on the right-hand side the cases \(t'>t_1\) and \(t'<t_1\), and rearrange the terms to obtain, 
\begin{align}\label{appeq:CQTreordered}
    &C^{\rm Q-T}_{i_{\mu} a}(t-t') \nonumber\\
    &= \sum_{\nu}(\delta_{\nu\mu}\partial_{t'}{-}i\chi_{i_{\mu}i_{\nu}})\left[\frac{\mathbf{d}_a\cdot\mathbf{E}_{i_{\nu}}(\mathbf{r}_a,t'-t)}{i\hbar}\right]^*\nonumber\\
    &+\sum_{\nu}\int_{t_0}^{t'}\mathrm{d}t_1 C^{\rm bos}_{i_{\mu}i_{\nu}}(t-t_1)\left[\frac{\mathbf{d}_a\cdot\mathbf{E}_{i_{\nu}}(\mathbf{r}_a,t'-t_1)}{i\hbar}\right]^*\nonumber\\
    &-\sum_{j,\nu\eta}\sum_{N=1}^{\infty}\Theta(t'-t)\int_{t}^{t'}\mathrm{d}t_1 \Bigg[\frac{\mathbf{d}_a\cdot\mathbf{E}_{j_{\nu}}(\mathbf{r}_a,t'-t_1)}{i\hbar}\nonumber\\
    &\qquad\qquad\qquad\qquad\times(\delta_{\mu\eta}\partial_{t}{-}i\chi_{i_{\eta}i_{\mu}})\mathbb{K}^N_{j_{\nu}i_{\eta}}(t_1-t)\Bigg]^*\nonumber\\
    &+\sum_{j,\nu\eta}\sum_{N=1}^{\infty}\int_{t_0}^{t}\mathrm{d}t_1 \mathbb{K}^N_{i_{\mu}j_{\eta}}(t-t_1)(\delta_{\nu\eta}\partial_{t'}{-}i\chi_{j_{\eta}j_{\nu}})\nonumber\\
    &\qquad\qquad\qquad\qquad\qquad\times\left[\frac{\mathbf{d}_a\cdot\mathbf{E}_{j_{\nu}}(\mathbf{r}_a,t'-t_1)}{i\hbar}\right]^*\nonumber\\
    &-\sum_{j,\nu\eta}\sum_{N=1}^{\infty}\int_{t_0}^{t}\mathrm{d}t_1 \Big[(\delta_{\nu\eta}\partial_{t_1}{-}i\chi_{j_{\eta}j_{\nu}})\mathbb{K}^N_{i_{\mu}j_{\eta}}(t-t_1)\Big]\nonumber\\
    &\qquad\qquad\qquad\qquad\times\Theta(t'-t_1)\left[\frac{\mathbf{d}_a\cdot\mathbf{E}_{j_{\nu}}(\mathbf{r}_a,t'-t_1)}{i\hbar}\right]^*\nonumber\\
    &-\sum_{j,\nu}\sum_{N=1}^{\infty}\mathbb{K}^N_{i_{\mu}j_{\nu}}(t-t_0)\left[\frac{\mathbf{d}_a\cdot\mathbf{E}_{j_{\eta}}(\mathbf{r}_a,t'-t_0)}{i\hbar}\right]^*.
\end{align}

Via partial integration and neglecting instantaneous intercavity scattering \(\mathbb{K}^N_{i_{\mu}j_{\nu}}(0)\approx 0\) (cf.~Sec.~\ref{sec:multicavityquanti}), we obtain, in Eq.~\eqref{appeq:CQTreordered},
\begin{align*}
    &\sum_{j,\nu\eta}\sum_{N=1}^{\infty}\int_{t_0}^{t}\mathrm{d}t_1 \mathbb{K}^N_{i_{\mu}j_{\eta}}(t-t_1)(\delta_{\nu\eta}\partial_{t'}{-}i\chi_{j_{\eta}j_{\nu}})\nonumber\\
    &\qquad\qquad\qquad\qquad\times\left[\frac{\mathbf{d}_a\cdot\mathbf{E}_{j_{\nu}}(\mathbf{r}_a,t'-t_1)}{i\hbar}\right]^*\nonumber\\
    &-\sum_{j,\nu\eta}\sum_{N=1}^{\infty}\int_{t_0}^{t}\mathrm{d}t_1 \Big[(\delta_{\nu\eta}\partial_{t_1}{-}i\chi_{j_{\eta}j_{\nu}})\mathbb{K}^N_{i_{\mu}j_{\eta}}(t-t_1)\Big]\nonumber\\
    &\qquad\qquad\qquad\qquad\times\Theta(t'-t_1)\left[\frac{\mathbf{d}_a\cdot\mathbf{E}_{j_{\nu}}(\mathbf{r}_a,t'-t_1)}{i\hbar}\right]^*\nonumber\\
    &-\sum_{j,\nu}\sum_{N=1}^{\infty}\mathbb{K}^N_{i_{\mu}j_{\nu}}(t-t_0)\left[\frac{\mathbf{d}_a\cdot\mathbf{E}_{j_{\eta}}(\mathbf{r}_a,t'-t_0)}{i\hbar}\right]^*\nonumber\\
    &\approx \sum_{j,\nu\eta}\sum_{N=1}^{\infty}\Theta(t-t')\int_{t'}^t\mathrm{d}t_1 \left[\frac{\mathbf{d}_a\cdot\mathbf{E}_{j_{\nu}}(\mathbf{r}_a,t'-t_1)}{i\hbar}\right]^*\nonumber\\
    &\qquad\qquad\qquad\qquad\qquad\times\Big[(\delta_{\nu\eta}\partial_{t_1}{-}i\chi_{j_{\eta}j_{\nu}})\mathbb{K}^N_{i_{\mu}j_{\eta}}(t-t_1)\Big].
\end{align*}

With this and Eq.~\eqref{appeq:Cbosapprox}, we rewrite Eq.~\eqref{appeq:CQTreordered} to read:
\begin{align}
    &C^{\rm Q-T}_{i_{\mu} a}(t-t') \nonumber\\
    &= \Theta(t-t')\sum_{\nu}(\delta_{\nu\mu}\partial_{t'}{-}i\chi_{i_{\mu}i_{\nu}})\left[\frac{\mathbf{d}_a\cdot\mathbf{E}_{i_{\nu}}(\mathbf{r}_a,t'-t)}{i\hbar}\right]^*\nonumber\\
    &+\Theta(t'-t)\sum_{\nu}(\delta_{\nu\mu}\partial_{t'}{-}i\chi^*_{i_{\nu}i_{\mu}})\left[\frac{\mathbf{d}_a\cdot\mathbf{E}_{i_{\nu}}(\mathbf{r}_a,t'-t)}{i\hbar}\right]^*\nonumber\\
    &+\sum_{j,\nu\eta}\sum_{N=1}^{\infty}\Theta(t-t')\int_{t'}^t\mathrm{d}t_1 \left[\frac{\mathbf{d}_a\cdot\mathbf{E}_{j_{\nu}}(\mathbf{r}_a,t'-t_1)}{i\hbar}\right]^*\nonumber\\
    &\qquad\qquad\qquad\qquad\qquad\times \Big[(\delta_{\nu\eta}\partial_{t_1}{-}i\chi_{j_{\eta}j_{\nu}})\mathbb{K}^N_{i_{\mu}j_{\eta}}(t-t_1)\Big]\nonumber\\
    &-\sum_{j,\nu\eta}\sum_{N=1}^{\infty}\Theta(t'-t)\int_{t}^{t'}\mathrm{d}t_1 \Bigg[\frac{\mathbf{d}_a\cdot\mathbf{E}_{j_{\nu}}(\mathbf{r}_a,t'-t_1)}{i\hbar}\nonumber\\
    &\qquad\qquad\qquad\times(\delta_{\mu\eta}\partial_{t}{-}i\chi_{i_{\eta}i_{\mu}})\mathbb{K}^N_{j_{\nu}i_{\eta}}(t_1-t)\Bigg]^*.
\end{align}

Combining the terms with \(t>t'\) and \(t'>t\), respectively, yields the final form from Eq.~\eqref{eq:QTcorr}.

\subsection{Derivation of the TLS correlation function} \label{appsec:tlscorrfunc}
Using the Hamiltonian of the TLS-bath interaction from Eq.~\eqref{eq:HTLSbath}, the correlation function for the bath-mediated interaction between two TLSs \(a\) and \(b\) reads, 
\begin{align}\label{appeq:CTLScommutator}
    &C^{\rm TLS}_{ab}(t-t') = \sum_{pq}\int\mathrm{d}^3r\int_0^{\infty}\mathrm{d}\omega g_{a,p}(\mathbf{r},\omega)\mathrm{e}^{-i\omega(t-t_0)}\nonumber\\
    &\qquad\qquad\qquad\times\mathrm{tr}_B\big[\hat{c}_p(\mathbf{r},\omega,t_0)\hat{c}^{\dagger}_q(\mathbf{r}',\omega',t_0)\rho_B\big]\nonumber\\
    &\qquad\qquad\qquad\times\int\mathrm{d}^3r'\int_0^{\infty}\mathrm{d}\omega' g^*_{b,q}(\mathbf{r}',\omega')\mathrm{e}^{i\omega'(t'-t_0)}\nonumber\\
    &+\sum_{i,\mu}\int_{t_0}^t\mathrm{d}t_1\frac{\mathbf{d}_a\cdot\mathrm{E}_{i_{\mu}}(\mathbf{r}_a,t-t_1)}{i\hbar}C^{\rm Q-T}_{i_{\mu} b}(t_1-t')\nonumber\\
    &+\sum_{j,\nu}\int_{t_0}^{t'}\mathrm{d}t_1\Bigg[\frac{\mathbf{d}_b\cdot\mathbf{E}_{j_{\nu}}(\mathbf{r}_b,t'-t_1)}{i\hbar}C^{\rm Q-T}_{j_{\nu} a}(t_1-t)\Bigg]^*\nonumber\\
    &-\sum_{ij,\mu\nu}\int_{t_0}^t\int_{t_0}^{t'}\mathrm{d}t_2 \frac{\mathbf{d}_a\cdot\mathrm{E}_{i_{\mu}}(\mathbf{r}_a,t-t_1)}{i\hbar}C^{\rm QNM}_{i_{\mu}j_{\nu}}(t_1-t_2)\nonumber\\
    &\qquad\qquad\qquad\qquad\times\Bigg[\frac{\mathbf{d}_b\cdot\mathbf{E}_{j_{\nu}}(\mathbf{r}_b,t'-t_1)}{i\hbar}\Bigg]^*,
\end{align}
also employing the definitions of the QNM correlation function \(C^{\rm QNM}_{i_{\mu}j_{\nu}}\) from Eq.~\eqref{appeq:CQNMcommutator} and the QNM-TLS correlation function \(C^{\rm Q-T}_{i_{\mu}a}\) from Eq.~\eqref{appeq:CQTcommutator}. Under the assumption of an initial vacuum bath state \(\rho_B = |{\rm vac}\rangle\langle{\rm vac}|\), we replace the trace in the first line on the right-hand side of Eq.~\eqref{appeq:CTLScommutator} with the commutator of the bath operators \(\opvec{c}(\mathbf{r},\omega)\) from Eq.~\eqref{eq:ccomm}. We furthermore insert the results of the QNM correlation function from Eq.~\eqref{eq:QNMcorrfunc} and of the QNM-TLS correlation function from Eq.~\eqref{eq:QTcorr}. Thus, we rewrite Eq.~\eqref{appeq:CTLScommutator} into (canceling some terms):
\begin{align}
    &C^{\rm TLS}_{ab}(t-t') \nonumber\\
    &= \frac{1}{\pi\hbar\epsilon_0}\int_0^{\infty}\mathrm{d}\omega \mathbf{d}_a\cdot\mathrm{Im}\big[\mathbf{G}(\mathbf{r}_a,\mathbf{r}_b,\omega)\big]\cdot\mathbf{d}_b^* \mathrm{e}^{-i\omega(t-t')}\nonumber\\
    &-\sum_{i,\mu}\frac{\mathbf{d}_a\cdot\mathbf{E}_{i_{\mu}}(\mathbf{r}_a,t-t_0)}{i\hbar}\Bigg[\frac{\mathbf{d}_b\cdot\mathbf{E}_{i_{\mu}}(\mathbf{r}_b,t'-t_0)}{i\hbar}\Bigg]^*\nonumber\\
    &-\Theta(t-t')\sum_{ij,\mu\nu\eta}\sum_{N=0}^{\infty}\int_{t'}^t\mathrm{d}t_1\int_{t'}^{t_1}\mathrm{d}t_2 \frac{\mathbf{d}_a\cdot\mathbf{E}_{i_{\mu}}(\mathbf{r}_a,t-t_1)}{i\hbar}\nonumber\\
    &\;\times (\delta_{\nu\eta}\partial_{t_1}{+}i\chi_{j_{\eta}j_{\nu}})\mathbb{K}^N_{i_{\mu}j_{\nu}}(t_1-t_2)\Bigg[\frac{\mathbf{d}_b\cdot\mathbf{E}_{j_{\nu}}(\mathbf{r}_b,t'-t_2)}{i\hbar}\Bigg]^*\nonumber\\
    &-\Theta(t'-t)\sum_{ij,\mu\nu\eta}\sum_{N=0}^{\infty}\int_{t}^{t'}\mathrm{d}t_1\int_{t}^{t_1}\mathrm{d}t_2 \frac{\mathbf{d}_a\cdot\mathbf{E}_{i_{\mu}}(\mathbf{r}_a,t-t_2)}{i\hbar}\nonumber\\
    &\;\times\Bigg[\frac{\mathbf{d}_b\cdot\mathbf{E}_{j_{\nu}}(\mathbf{r}_b,t'-t_1)}{i\hbar}(\delta_{\mu\eta}\partial_{t_1}{+}i\chi_{i_{\eta}i_{\mu}})\mathbb{K}^N_{j_{\nu}i_{\eta}}(t_1-t_2)\Bigg]^*\nonumber\\
    &-\sum_{i,\mu\nu}\int_{t_0}^t\mathrm{d}t_1\int_{t_0}^{t'}\mathrm{d}t_2\frac{\mathbf{d}_a\cdot\mathbf{E}_{i_{\mu}}(\mathbf{r}_a,t-t_1)}{i\hbar}\nonumber\\
    &\qquad\times C^{\rm bos}_{i_{\mu}i_{\nu}}(t_1-t_2)\Bigg[\frac{\mathbf{d}_b\cdot\mathbf{E}_{i_{\nu}}(\mathbf{r}_b,t'-t_2)}{i\hbar}\Bigg]^*\nonumber\\
    &-\sum_{i,\mu\nu}\int_{t_0}^{t}\Theta(t'-t_1)\frac{\mathbf{d}_a\cdot\mathbf{E}_{i_{\mu}}(\mathbf{r}_a,t-t_1)}{i\hbar}\nonumber\\
    &\qquad\times\Bigg[(\delta_{\mu\nu}\partial_{t_1}{-}i\chi_{i_{\nu}i_{\mu}})\frac{\mathbf{d}_b\cdot\mathbf{E}_{i_{\nu}}(\mathbf{r}_b,t'-t_1)}{i\hbar}\Bigg]^*\nonumber\\
    &-\sum_{i,\mu\nu}\int_{t_0}^{t}\Theta(t'-t_1)\Bigg[\frac{\mathbf{d}_b\cdot\mathbf{E}_{i_{\nu}}(\mathbf{r}_b,t'-t_1)}{i\hbar}\Bigg]^*\nonumber\\
    &\qquad\times (\delta_{\mu\nu}\partial_{t_1}{-}i\chi_{i_{\mu}i_{\nu}})\frac{\mathbf{d}_a\cdot\mathbf{E}_{i_{\mu}}(\mathbf{r}_a,t-t_1)}{i\hbar}.
\end{align}

Next, we employ partial integration on the last two terms, as well as the approximation of the bosonic QNM correlation function \( C^{\rm bos}_{i_{\mu}i_{\nu}}\) from Eq.~\eqref{appeq:Cbosapprox} to find (neglecting instantaneous intercavity scattering terms \(\mathbb{K}_{i_{\mu}j_{\nu}}(0)\) in the process)
\begin{align}
    &C^{\rm TLS}_{ab}(t-t') \nonumber\\
    &= \frac{1}{\pi\hbar\epsilon_0}\int_0^{\infty}\mathrm{d}\omega \mathbf{d}_a\cdot\mathrm{Im}\big[\mathbf{G}(\mathbf{r}_a,\mathbf{r}_b,\omega)\big]\cdot\mathbf{d}_b^* \mathrm{e}^{-i\omega(t-t')}\nonumber\\
    &-\sum_{ij,\mu\nu\eta}\sum_{N=0}^{\infty}\int_{t'}^t\mathrm{d}t_1\int_{t'}^{t_1}\mathrm{d}t_2 \frac{\mathbf{d}_a\cdot\mathbf{E}_{i_{\mu}}(\mathbf{r}_a,t-t_1)}{i\hbar}\nonumber\\
    &\;\times (\delta_{\nu\eta}\partial_{t_1}{+}i\chi_{j_{\eta}j_{\nu}})\mathbb{K}^N_{i_{\mu}j_{\nu}}(t_1-t_2)\Bigg[\frac{\mathbf{d}_b\cdot\mathbf{E}_{j_{\nu}}(\mathbf{r}_b,t'-t_2)}{i\hbar}\Bigg]^*\nonumber\\
    &-\sum_{ij,\mu\nu\eta}\sum_{N=0}^{\infty}\int_{t}^{t'}\mathrm{d}t_1\int_{t}^{t_1}\mathrm{d}t_2 \frac{\mathbf{d}_a\cdot\mathbf{E}_{i_{\mu}}(\mathbf{r}_a,t-t_2)}{i\hbar}\nonumber\\
    &\;\times\Bigg[\frac{\mathbf{d}_b\cdot\mathbf{E}_{j_{\nu}}(\mathbf{r}_b,t'-t_1)}{i\hbar}(\delta_{\mu\eta}\partial_{t_1}{+}i\chi_{i_{\eta}i_{\mu}})\mathbb{K}^N_{j_{\nu}i_{\eta}}(t_1-t_2)\Bigg]^*\nonumber\\
    &-\Theta(t-t')\sum_{i,\mu} \frac{\mathbf{d}_a\cdot\mathbf{E}_{i_{\mu}}(\mathbf{r}_a,t-t')}{i\hbar}\Bigg[\frac{\mathbf{d}_b\cdot\mathbf{E}_{j_{\nu}}(\mathbf{r}_b,0)}{i\hbar}\Bigg]^*\nonumber\\
    &-\Theta(t'-t)\sum_{i,\mu} \frac{\mathbf{d}_a\cdot\mathbf{E}_{i_{\mu}}(\mathbf{r}_a,0)}{i\hbar}\Bigg[\frac{\mathbf{d}_b\cdot\mathbf{E}_{j_{\nu}}(\mathbf{r}_b,t'-t)}{i\hbar}\Bigg]^*.
\end{align}

Finally, by separating the terms \(N=0\) and \(N>0\), we arrive at the form from Eq.~\eqref{eq:tlscorrfunc}.

\section{Multi-cavity QNM Green's function expansion}\label{appsec:greensexp}
For the calculation of the QNM-TLS coupling [Eq.~\eqref{appeq:genqnmtlscoup}], the QNM-TLS correlation function [Eq.~\eqref{eq:QTcorr}], and the TLS correlation function [Eq.~\eqref{eq:tlscorrfunc}], we require the full, multi-cavity Green's function \(\mathbf{G}(\mathbf{r},\mathbf{r}',\omega)\). We employ the multi-cavity QNM expansion from Ref.~\onlinecite{fuchs2026greens}, where a set of scattering equations is used to obtain the QNM expansion of the Green's function by iteratively adding more cavities until all QNM cavities are included. Since the derivation of the QNM expansion was shown in Ref.~\onlinecite{fuchs2026greens} for the same model system of two coupled metal dimers with dipole emitters in the dimer gaps, we show here only the relevant expansions without discussing their derivation. 

We consider two cavities with one dominant QNM each (\(i_{\mu} \to 1,2\)). In all cases considered in the main text, one of the positions \(\mathbf{r},\mathbf{r}'\) lies within one of the cavities. Without loss of generality, we assume \(\mathbf{r}\in V_1\), with the volume of cavity 1 \(V_1\). All other cases are obtained by switching the indices \(1\leftrightarrow 2\) or employing the relation \(\mathbf{G}(\mathbf{r}',\mathbf{r},\omega)=[\mathbf{G}(\mathbf{r},\mathbf{r}',\omega)]^T\) and \(T\) denoting the transpose. 

For \(\mathbf{r}'\in V_1\), we assume that an expansion in terms of the QNM of cavity 1,
\begin{align}
    \mathbf{G}(\mathbf{r},\mathbf{r}',\omega)\big|_{\mathbf{r},\mathbf{r}'\in V_1} = A_1(\omega)\qnm{f}{1}(\mathbf{r})\qnm{f}{1}(\mathbf{r}'),
\end{align}
yields a good approximation of the full Green's function.

For \(\mathbf{r}'\in V_2\), we obtain,
\begin{align}
    \mathbf{G}(\mathbf{r},\mathbf{r}',\omega)\big|_{\mathbf{r},V_1, \mathbf{r}'\in V_2} = B_{12}(\omega)\qnm{f}{1}(\mathbf{r})\qnm{f}{2}(\mathbf{r}'),
\end{align}
where 
\begin{align}\label{eq:Bsurface}
    &B_{12}(\omega) = \frac{c^2}{\omega^2}A_1(\omega)A_2(\omega)\nonumber\\
    &\quad\qquad\quad\times\oint_{\mathcal{S}_2}\mathrm{d}A_s \Big\{\Big[\nabla_s\times\qnm{F}{1}(\mathbf{s},\omega)\Big]\cdot\Big[\opvec{n}_s\times\qnm{f}{2}(\mathbf{s})\Big]\nonumber\\
    &\qquad\qquad\qquad-\Big[\opvec{n}_s\times\qnm{F}{1}(\mathbf{s},\omega)\Big]\cdot\Big[\nabla_s\times\qnm{f}{2}(\mathbf{s})\Big]\Big\}
\end{align}
describes intercavity scattering, and \(B_{12}(\omega)=B_{21}(\omega)\) \cite{fuchs2026greens}. 

Finally, for \(\mathbf{r}'\in V_{\rm out}\), we find
\begin{align}
    \mathbf{G}(\mathbf{r},\mathbf{r}',\omega)\big|_{\mathbf{r},V_1, \mathbf{r}'\in V_{\rm out}} &= A_1(\omega)\qnm{f}{1}(\mathbf{r})\qnm{F}{1}(\mathbf{r}',\omega)\nonumber\\
    &+B_{12}(\omega)\qnm{f}{1}(\mathbf{r})\qnm{F}{2}(\mathbf{r}',\omega).
\end{align}

For an efficient numerical calculation of \(B_{12}(\omega)\), we separate \(\qnm{F}{1}(\mathbf{s},\omega)\big|_{\mathbf{s}\in V_2} = \qnm{F}{1}'(\mathbf{s},\omega)\mathrm{e}^{i\omega\tau_{12}}\) into slow-rotating envelope and fast-rotating exponential (we performed the same separation in Appendix~\ref{appsec:QNMscatter} in the derivation of the retarded overlap matrix \(S_{i_{\mu}\leftarrow j_{\nu}}\)). Then, we perform a pole approximation \(\omega\to\omega_1\) on the slow-rotating envelope, to obtain \(B_{12}(\omega) \approx N_{12} \frac{c^2}{\omega^2}A_1(\omega)A_2(\omega)\mathrm{e}^{i\omega \tau_{12}}\), 
with
\begin{align}\label{appeq:N21}
    N_{12} &= \oint_{\mathcal{S}_2}\mathrm{d}A_s \Big\{\Big[\nabla_s\times\qnm{F}{1}'(\mathbf{s},\omega_1)\Big]\cdot\Big[\opvec{n}_s\times\qnm{f}{2}(\mathbf{s})\Big]\nonumber\\
    &\qquad\qquad-\Big[\opvec{n}_s\times\qnm{F}{1}'(\mathbf{s},\omega_1)\Big]\cdot\Big[\nabla_s\times\qnm{f}{2}(\mathbf{s})\Big]\Big\}.
\end{align}

\section{Numerical details on the metal dimer example calculations}\label{appsec:numerics}

\begin{figure*}
    \centering
    \includegraphics[width = 1.99\columnwidth]{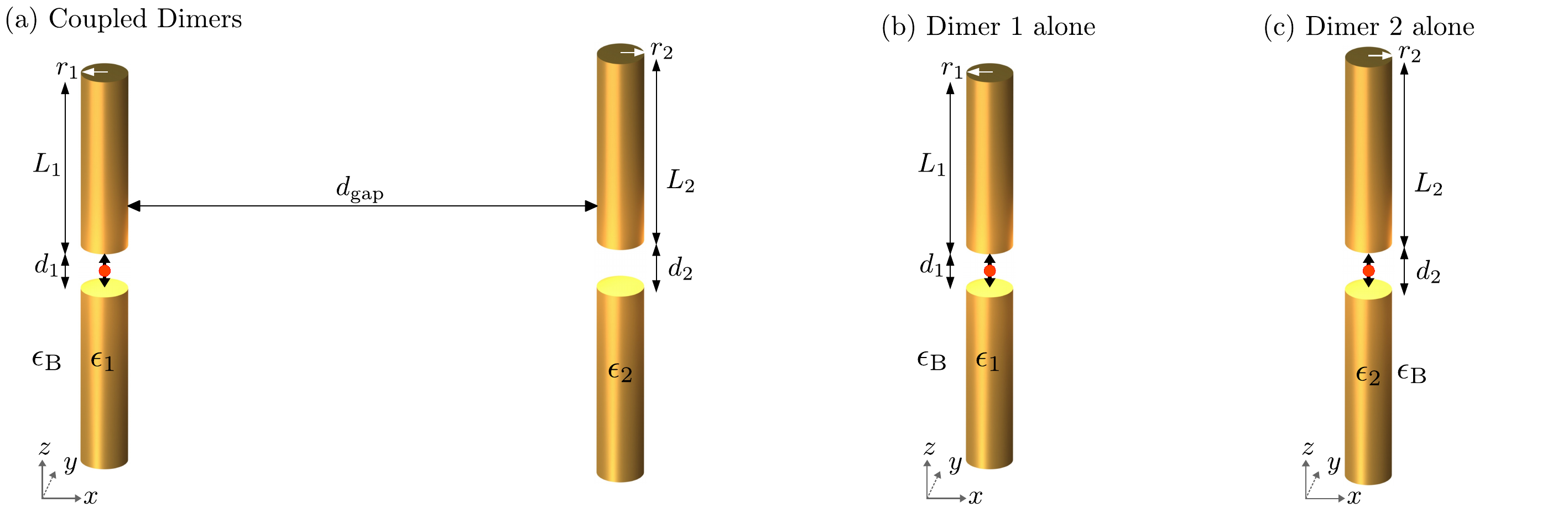}
    \caption{(a) Schematic for coupled dimers. $r_1=10~$nm, $L_1=80~$nm, $d_1=10~$nm. $r_2=10~$nm, $L_2=90~$nm, $d_2=20~$nm. The surface-to-surface gap distance between two dimers is $d_{\rm gap}$. Schematic for (b) dimer 1 alone and (c) dimer 2 alone.
    }
    \label{fig: Tdimer_sche_ver}
\end{figure*}

\subsection{Model set up}
We consider two coupled metallic dimers placed in free space ($\epsilon_{\rm B}=1$), as shown in Fig.\ref{fig: Tdimer_sche_ver} (a). Each dimer consists of two identical cylindrical nanorods. For the first dimer, the radius (length) of the nanorods is $r_{1}=10~$nm ($L_{1}=80~$nm), and the surface-to-surface distance between two nanorods (along the $z$ axis direction)  is $d_{1}=10~$nm. For the second dimer, we have $r_{2}=10~$nm, $L_{2}=90~$nm, and $d_{2}=20~$nm. The surface-to-surface gap distance between two dimers is $d_{\rm gap}$ (the line connecting the centers of two dimers is along the $x$-axis direction). The center-to-center gap distance between two dimers is $(d_{\rm gap}+r_1+r_2)$.

Drude models are used for the dielectric functions of two metallic (gold-like) dimers
\begin{equation}\label{eq:Drude}
    \epsilon_{\rm 1}=1-\frac{\omega_{\rm p}^{2}}{\omega^{2}+i\omega\gamma_{\rm p1}},~~~\epsilon_{\rm 2}=1-\frac{\omega_{\rm p}^{2}}{\omega^{2}+i\omega\gamma_{\rm p2}},
\end{equation}
using $\hbar\omega_{\rm p}=8.2934$ eV, with  $\hbar\gamma_{\rm p1}=0.0928$ eV for dimer 1 and
$\hbar\gamma_{\rm p2}=0.3\hbar\gamma_{\rm p1}=0.0278$ eV
for dimer 2.

An inverse Green's function approach in complex frequency space is used to obtain the QNMs~\cite{bai_efficient_2013-1}, with the aid of COMSOL Multiphysics (a finite element solver). The computational model is a cylinder with a diameter of $3$~$\mu$m and a length of $5.2$~$\mu$m, resulting in a total volume of around $37$ $\mu$m$^{3}$, which contains perfectly matched layers (PMLs) with a thickness of $600~$nm to minimize boundary reflections efficiently. 
High-resolution mesh settings are used, with the maximum mesh element sizes around the dipole point, inside and outside the metallic nanorods, set to $0.1$ nm, $2$ nm, and $ 80$ nm, respectively.

\subsection{Bare QNMs}

\begin{figure*}
    \centering
    \includegraphics[width = 1.79\columnwidth]{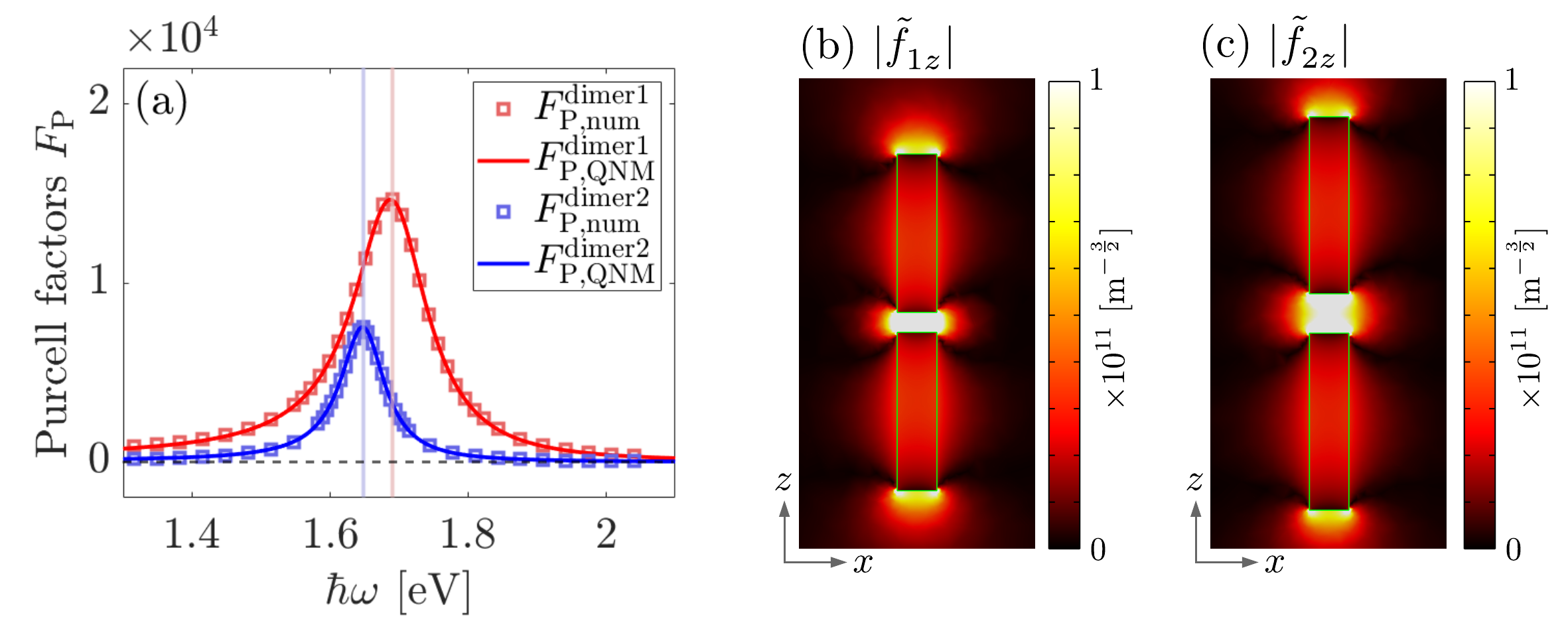}
    \caption{(a) Red (blue) solid curve and red (blue) squares show the Purcell factors at the gap center of the dimer 1 (dimer 2). Red (blue) vertical line shows the resonance position $\hbar\omega_1$ ($\hbar\omega_2$) of the single QNM for dimer 1 alone (dimer 2 alone). (b) Mode distribution $|\tilde{f}_{1z}|$ for dimer 1 only. (c) Mode distribution $|\tilde{f}_{2z}|$ for dimer 2 only.
    }
    \label{fig: Tdimer_sche_ver_modePF}
\end{figure*}

\begin{figure*}
    \centering
    \includegraphics[width = 1.79\columnwidth]{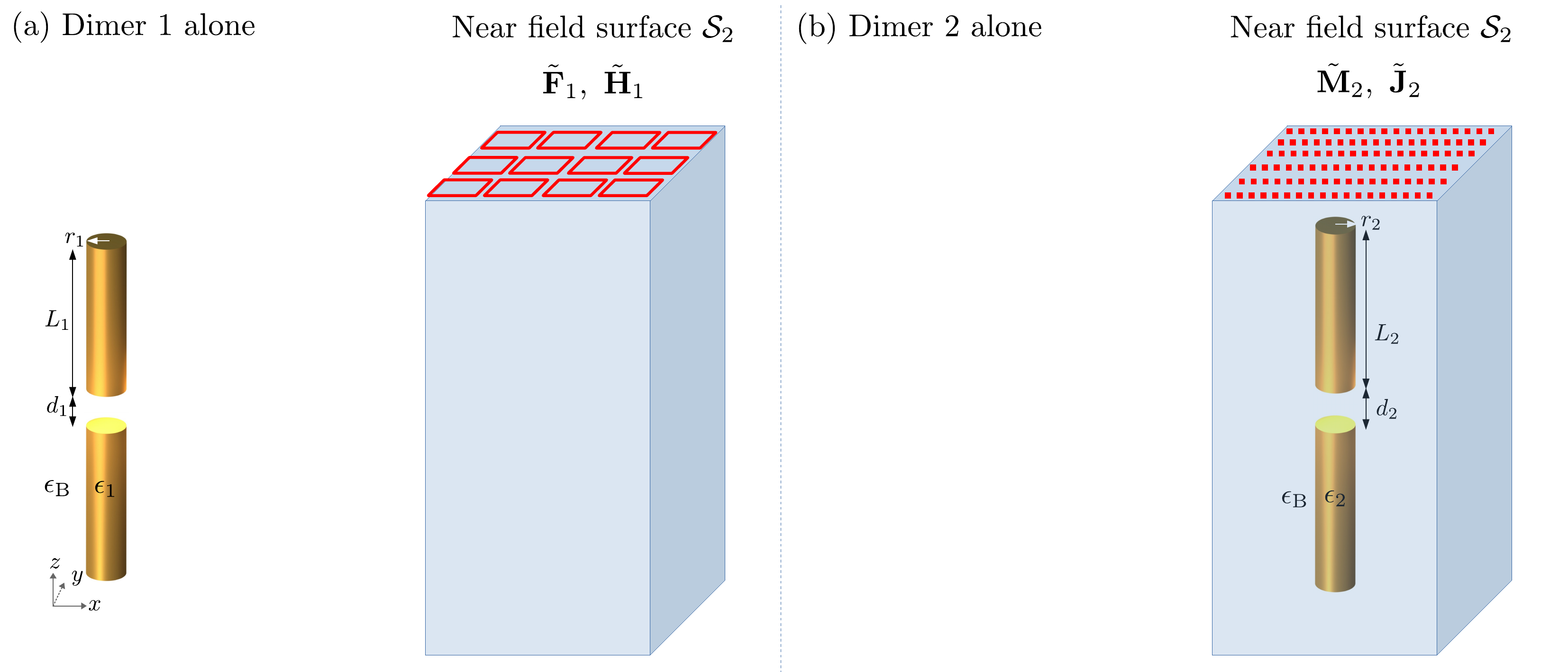}
    \caption{Schematic diagram of integral over near field surface $\mathcal{S}_2$ of dimer 2, where $\mathcal{S}_2$ is a closed cuboid surface and $50~$nm away from the surface of the dimer 2. The grid element of surface integral for regularized QNMs $\mathbf{\tilde{F}}_1$ and $\mathbf{\tilde{H}}_1$ is $(5~{\rm nm})^2$ (see the red box in (a), not to scale). While for surface currents $\mathbf{\tilde{M}}_2$ and $\mathbf{\tilde{J}}_2$, the grid element of the surface integral is $(0.5~{\rm nm})^2$ (see the red dot in (b), not to scale).
    }
    \label{fig: integral}
\end{figure*}

First, we consider dimer 1 alone [see Fig.~\ref{fig: Tdimer_sche_ver} (b)], a single QNM $\mathbf{\tilde{f}}_{1}^{\rm }$ dominates in the frequency range of interest when the dipole is placed at the center of the gap, with an eigenfrequency of $\hbar\tilde{\omega}_1=\hbar\omega_1-i\hbar\gamma_1=(1.6904-0.0652i)$ eV, and a quality factor of $Q_1=\omega_1/(2\gamma_1)\approx13$. The distribution of the dominated $z-$component $|\tilde{f}_{1z}|$ of the mode is shown in Fig.~\ref{fig: Tdimer_sche_ver_modePF} (b), where a hot spot occurs within the gap. With a QNMs expansion, the photon Green's function close to dimer 1 can be approximated as follows,
~\cite{leung1994completeness,ge2014quasinormal},
\begin{align}\label{eq:GF}
\begin{split}
\mathbf{G}_{\rm dimer1}\left(\mathbf{r},\mathbf{r}_{0},\omega\right)\approx A_{\rm 1}(\omega)\tilde{\mathbf{f}}_{\rm 1}\left({\bf r}\right)\tilde{\mathbf{f}}_{\rm 1}\left({\bf r}_{0}\right),
\end{split}
\end{align}
using the single mode approximation, and the expansion factor $A_{\rm 1}(\omega)=\omega/[2(\tilde{\omega}_{\rm 1}-\omega)]$.

We consider a $z$ polarized dipole with real dipole moment $\mathbf{d}=\mathbf{n}_{z}|\mathbf{d}|$ at the center of the gap of dimer 1 ($\mathbf{r}_{\rm a1}$), which is $5~$ nm from the surface of the top/bottom nanorods [see Fig.~\ref{fig: Tdimer_sche_ver} (b)], and the Purcell factors (enhanced spontaneous emission decay rates due to the strong local fields within the gap) for this dipole are shown in
Fig.~\ref{fig: Tdimer_sche_ver_modePF}(a). The QNM result $F_{\rm P,QNM}^{\rm dimer1}$ (red solid curve) is obtained through the reconstructed Green's function~\cite{kristensen_modes_2014},
 \begin{align}
 \begin{split}\label{eq:Gamma_QNM}
 \Gamma^{\rm dimer1}(\mathbf{r}_{\rm a1},\omega)&=\frac{2}{\hbar\epsilon_{0}}\mathbf{d}\cdot{\rm Im}\{\mathbf{G}_{\rm dimer1}(\mathbf{r}_{\rm a1},\mathbf{r}_{\rm a1},\omega)\}\cdot\mathbf{d},\\
 F_{\rm P,QNM}^{\rm dimer1}(\mathbf{r}_{\rm a1},\omega)&=\frac{\Gamma^{\rm dimer1}(\mathbf{r}_{\rm a1},\omega)}{\Gamma_0(\omega)},\\
 \end{split}  
 \end{align}
with the decay rate 
\begin{equation}
\Gamma_0(\omega)=\frac{2}{\hbar\epsilon_{0}}\mathbf{d}\cdot{\rm Im}\{\mathbf{G}_{\rm B}(\mathbf{r}_{\rm a1},\mathbf{r}_{\rm a1},\omega)\}\cdot\mathbf{d},
\end{equation}
for emission into the homogeneous background medium, and simplifies to $\mathrm{Im}\{\mathbf{G}_{\rm B}({\bf r}_{\rm a1},{\bf r}_{\rm a1},\omega)\}=(\omega^3n_{\rm B}/6\pi c^3)\mathbb{1}$.

The full dipole result (i.e., with no mode approximations) $F_{\rm P, num}^{\rm dimer1}$ (red squares) is given by the enhanced power flow from the dipole,
\begin{equation}\label{eq:Purcellfulldipole}
    F_{\rm P,num}^{\rm dimer1/dimer2}(\mathbf{r}_{\rm a1}/\mathbf{r}_{\rm a2},\omega)=\frac{\int_{ \Sigma_{\rm d}}\hat{\mathbf{n}}\cdot {\bf S}_{\rm }(\mathbf{r},\omega)d\mathbf{r} }{\int_{ \Sigma_{\rm d}}\hat{\mathbf{n}}\cdot {\bf S}_{\rm 0}(\mathbf{r},\omega)d\mathbf{r} },
\end{equation}
where ${\bf S}$ and $\mathbf{S}_0$ are the Poynting vectors with and without the resonator.
Here, $\Sigma_{\rm d}$ represents a closed spherical surface enclosing the dipole at $\mathbf{r}_{\rm a1}$ or $\mathbf{r}_{\rm a2}$ (the center of the gap of dimer 1 or dimer 2).
As shown in Fig.~\ref{fig: Tdimer_sche_ver_modePF} (a), the QNM results (red solid curve, $F_{\rm P,QNM}^{\rm dimer1}$) agree very well with full-dipole numerical results (red squares, $F_{\rm P,num}^{\rm dimer1}$), which verifies the validity of the single QNM approximation for metallic dimer 1 in the frequency regime of interest.

Similarly, we find for dimer 2 alone a single QNM at $\hbar\tilde{\omega}_2=\hbar\omega_2-i\hbar\gamma_2=(1.6482-0.0388i)$ eV with a quality factor of $Q_2=\omega_2/(2\gamma_2)\approx21$, when the dipole is placed at the gap center of the dimer 2 (see Fig.~\ref{fig: Tdimer_sche_ver} (c)). The mode distribution (dominant $z$ component $|\tilde{f}_{2z}|$) is shown in the Fig.~\ref{fig: Tdimer_sche_ver_modePF} (c). The Purcell factors for a $z-$polarized dipole at the gap center of the dimer 2 ($\mathbf{r}_{\rm a2}$) are shown in the Fig.~\ref{fig: Tdimer_sche_ver_modePF} (a), where solid blue curve shows QNM result $F_{\rm P, QNM}^{\rm dimer2}$ ($A_{\rm 2}(\omega)=\omega/[2(\tilde{\omega}_{\rm 2}-\omega)]$),
 \begin{align}
 \begin{split}\label{eq:Gamma_QNM2}
\mathbf{G}_{\rm dimer2}\left(\mathbf{r},\mathbf{r}_{0},\omega\right)&\approx A_{\rm 2}(\omega)\tilde{\mathbf{f}}_{\rm 2}\left({\bf r}\right)\tilde{\mathbf{f}}_{\rm 2}\left({\bf r}_{0}\right),\\
 \Gamma^{\rm dimer2}(\mathbf{r}_{\rm a2},\omega)&=\frac{2}{\hbar\epsilon_{0}}\mathbf{d}\cdot{\rm Im}\{\mathbf{G}_{\rm dimer2}(\mathbf{r}_{\rm a2},\mathbf{r}_{\rm a2},\omega)\}\cdot\mathbf{d},\\
 F_{\rm P,QNM}^{\rm dimer2}(\mathbf{r}_{\rm a2},\omega)&=\frac{\Gamma^{\rm dimer2}(\mathbf{r}_{\rm a2},\omega)}{\Gamma_0(\omega)},\\
 \end{split}  
 \end{align}
which also agrees very well with full dipole numerical results, $F_{\rm P,num}^{\rm dimer2}$ (blue squares, Eq.~\eqref{eq:Purcellfulldipole}), thus 
verifying the single-mode approximation of dimer 2 alone in the frequency regime of interest.

\subsection{Overlap matrix}
As shown in the main text, the overlap matrix 
\begin{align}
S_{i_{\mu}j_{\eta}}= \delta_{ij}S^{\rm intra}_{i_{\mu}j_{\eta}}+(1-\delta_{ij})S^{\rm inter}_{i_{\mu}j_{\eta}},
\end{align}
contains intra-cavity contributions $S^{\rm intra}_{i_{\mu}j_{\eta}}$ and inter-cavity contributions $S^{\rm inter}_{i_{\mu}j_{\eta}}$. For the system, we consider [cf. Fig.~\ref{fig: Tdimer_sche_ver} (a)] only one QNM is dominating each cavity in the frequency regime of interest, so we simplify the notation: $i_\mu\rightarrow i$, $j_\eta\rightarrow j$ and yield the overlap matrix
\begin{align}
\mathbf{S}=
\begin{bmatrix}
S_{11}^{\rm intra} & S_{12}^{\rm inter} \\
S_{21}^{\rm inter} & S_{22}^{\rm intra} 
\end{bmatrix}.
\end{align}
The full expression of the diagonal element ($S_{11/22}^{\rm intra}$) is shown in Eq.~\eqref{eq:Sintradef}, where the nonradiative contribution (the first term) and the radiative contribution (the second term) are included. We use the numerically efficient pole approximation for the non-radiative (radiative) contribution as given in Eq. (67) (Eq.~(69), with near field to far field transformation) of Ref.~\cite{ren2020near} in the single-mode limit.

As shown in Table~\ref{tab: s_intra_inter}, for dimer 1 alone, we got $S_{\rm 11,nrad}^{\rm intra}=0.6022$, and $S_{\rm 11,rad}^{\rm intra}=0.3973$. When calculating $S_{\rm 11,rad}^{\rm intra}$, the chosen near field surface a closed cuboid surface $50~$nm away from the surface of dimer 1. The far-field surface is a closed spherical surface in the very far-field region, and the angled grid is chosen to be very small, around $\pi/200$, to achieve convergence of $S_{\rm 11,rad}^{\rm intra}$. As a result, $S_{11}^{\rm intra}=S_{\rm 11,nrad}^{\rm intra}+S_{\rm 11,rad}^{\rm intra}=0.9995$ is close to unity. Similarly, for dimer 2 alone, we got $S_{22}^{\rm intra}=S_{\rm 22,nrad}^{\rm intra}+S_{\rm 22,rad}^{\rm intra}=0.3042+0.6966=1.0008$.

As for the off-diagonal element, the full expression is shown in Eq.~\eqref{eq:Sinterdef} (also see Eq.~(15) of Ref.~\cite{fuchs2024quantization}). Analog to our previous work, we also employed the pole approximation (see Appendix F of Ref.~\cite{fuchs2024quantization} for more details),

\begin{align} \label{eq:Sinterdef_pole}
    &S^{\rm inter}_{\rm 12,pole} \nonumber\\
    &= -\frac{1}{2\epsilon_0}\frac{\tilde{\omega}_1}{\sqrt{\omega_{1}\omega_{2}}}\frac{1}{i(\tilde{\omega}_{1}-\tilde{\omega}_{2}^{*})}\oint_{\mathcal{S}_{1}}\mathrm{d}\mathbf{s}\Big[\qnm{
    J}{1}(\mathbf{s})\cdot\qnm{F}{2}^*(\mathbf{s},\tilde{\omega}_1)\nonumber\\
    &\qquad\qquad\qquad\qquad\qquad\qquad\qquad+\qnm{M}{1}(\mathbf{s})\cdot\qnm{H}{2}^*(\mathbf{s},\tilde{\omega}_1)\Big]\nonumber\\
    &-\frac{1}{2\epsilon_0}\frac{\tilde{\omega}_2^{*}}{\sqrt{\omega_{1}\omega_{2}}}\frac{1}{i(\tilde{\omega}_{1}-\tilde{\omega}_{2}^{*})}\oint_{\mathcal{S}_{2}}\mathrm{d}\mathbf{s}\Big[\qnm{M}{2}^*(\mathbf{s})\cdot\qnm{H}{1}(\mathbf{s},\tilde{\omega}_2^{*})\nonumber\\
    &\qquad\qquad\qquad\qquad\qquad\qquad\qquad+\qnm{J}{2}^*(\mathbf{s})\cdot\qnm{F}{1}(\mathbf{s},\tilde{\omega}_2^{*})\Big],
\end{align}
where $\mathcal{S}_{1/2}$ is the near field surface of dimer 1/2. $\tilde{\mathbf{J}}_{\rm 1/2}$ ($\tilde{\mathbf{M}}_{\rm 1/2}$) are electric (magnetic) surface currents on the near field surface $\mathcal{S}_{1/2}$~\cite{Neartofar_1992},
\begin{align}
\tilde{\mathbf{J}}_{\rm 1/2}(\mathbf{s})&=\mathbf{\hat{n}}_{1/2}\times\tilde{\mathbf{h}}_{1/2}(\mathbf{s}),\\
\tilde{\mathbf{M}}_{\rm 1/2}(\mathbf{s})&=-\mathbf{\hat{n}}_{1/2}\times\tilde{\mathbf{f}}_{1/2}(\mathbf{s}),\\
\mathbf{\tilde{h}}_{1/2}(\mathbf{s})&=\frac{1}{i\tilde{\omega}_{1/2}\mu_{0}}\nabla\times\mathbf{\tilde{f}}_{1/2}(\mathbf{s}).
\end{align}
Here $\mathbf{\tilde{h}}_{1/2}$ are magnetic QNMs and $\mathbf{\hat{n}}_{1/2}$ are unit vectors normal to the surface $\mathcal{S}_{1/2}$, pointing outward.

In our previous work (coupled horizontal nanorods) (Ref.~\cite{fuchs2024quantization}), the axial-symmetry of the QNMs to the nanorod’s longitudinal
axis is used to simplify the integral. Here, with the coupled vertical dimers, one can still use the symmetry of $\left(\mathbf{\tilde{M}}_2^{*}\cdot\mathbf{\tilde{H}}_1\right)$ and $\left(\mathbf{\tilde{J}}_2^{*}\cdot\mathbf{\tilde{F}}_1\right)$ along the y and z directions to partly simplify the calculation.
The same applies to the integral over $\mathcal{S}_1$. 
But, we chose to sum the value of every point on the closed near field surface $\mathcal{S}_{1/2}$  to obtain the integral $S_{\rm 12,pole}^{\rm inter}$ (Eq.~\eqref{eq:Sinterdef_pole}).

However, we can apply the following optimization: For example, for the integral over $\mathcal{S}_2$, as shown in Fig.~\ref{fig: integral},  we use relatively coarse grid size for $\tilde{\mathbf{F}}_{1}$ and $\tilde{\mathbf{H}}_{1}$ over $\mathcal{S}_2$, which are far away from dimer 1 and thus $\tilde{\mathbf{F}}_{1}$ and $\tilde{\mathbf{H}}_{1}$ varies slowly spatially. In contrast, the surface currents $\mathbf{\tilde{M}}_2$ and $\mathbf{\tilde{J}}_2$ over $\mathcal{S}_2$, change fast in the near field region and require a relatively fine grid size. As a result, we use an interpolation of the regularized QNMs $\tilde{\mathbf{F}}_{1}$,  $\tilde{\mathbf{H}}_{1}$ from the coarse grid to the fine grid, to calculate the dot products of the regularized QNMs and surface currents:  $\left(\mathbf{\tilde{M}}_2^{*}\cdot\mathbf{\tilde{H}}_1\right)$ and $\left(\mathbf{\tilde{J}}_2^{*}\cdot\mathbf{\tilde{F}}_1\right)$.

In the calculation, $\mathcal{S}_2$ is chosen as a cuboid surface enclosing dimer 2 is $50~$nm away from the surface of dimer 2 (see Fig.~\ref{fig: integral}). The grid element of surface integral for $\tilde{\mathbf{F}}_{1}$ and  $\tilde{\mathbf{H}}_{1}$ is $(5~{\rm nm})^2$. While for $\mathbf{\tilde{M}}_2$ and $\mathbf{\tilde{J}}_2$, it's $(0.5~{\rm nm})^2$. Similarly, $\mathcal{S}_1$ is also chosen as a cuboid surface but enclosing dimer 1, which is $50~$nm away from the surface of dimer 1. The grid element of surface integral for regularized QNMs $\tilde{\mathbf{F}}_{2}$ and  $\tilde{\mathbf{H}}_{2}$ is $(5~{\rm nm})^2$. While for surface currents $\mathbf{\tilde{M}}_1$ and $\mathbf{\tilde{J}}_1$, it's $(0.5~{\rm nm})^2$. As a result, for $d_{\rm gap}=2000~$nm, we got \blue{$S_{\rm 12,pole}^{\rm inter}=(-0.0205 + 0.0149i)$} (see Table~\ref{tab: s_intra_inter}).

\begin{table}
    \centering
    \caption{Intra-cavity overlap matrix element $S_{11/22}^{\rm intra}$ from Eq.~\eqref{eq:Sintradef} for metallic dimer 1/2 (cf.~Fig.~\ref{fig: Tdimer_sche_ver}) and inter-cavity overlap $S_{\rm 12,pole}^{\rm inter}$ from Eq.~\eqref{eq:Sinterdef_pole} for $d_{\rm gap}=2000~$nm. The matrix elements are calculated within a pole approximation for the frequency integral \cite{ren2020near, fuchs2024quantization}.}
    \begin{tabular}{c c}
    \hline \hline
    Matrix element & Pole \\ \hline
    ${S_{11}^{\rm intra}}={S_{\rm 11, nrad}^{\rm intra}}+{S_{\rm 11, rad}^{\rm intra}}$ & ${0.9995}={(0.6022)}+{(0.3973)}$\\
    ${S_{22}^{\rm intra}}={S_{\rm 22, nrad}^{\rm intra}}+{S_{\rm 22, rad}^{\rm intra}}$ & ${1.0008}={(0.3042)}+{(0.6966)}$ \\
    $S_{\rm 12,pole}^{\rm inter}$ for $d_{\rm gap}=2000~$nm & \blue{${(-0.0205 + 0.0149i)}$} \\\hline
    \end{tabular}
    \label{tab: s_intra_inter}
\end{table}

\subsection{Retarded overlap matrix}

Eq.~\eqref{eq:retardedSmatrix} shows the retarded overlap matrix. We focus on  the case $i\neq j$, which reads
\begin{align}
    &{S}_{i_{\mu}\leftarrow j_{\eta}} = \frac{1}{2\epsilon_0}\frac{1}{i(\Tilde{\omega}_{i_{\mu}}-\Tilde{\omega}^*_{j_{\eta}})}\nonumber\\
    &\times\oint_{\mathcal{S}_j}\mathrm{d}A_s\Big[{(\qnm{H}{i_{\mu}}'(\mathbf{s},\omega_{i_{\mu}})\times\opvec{n}_s)\cdot\qnm{F}{j_{\eta}}^*(\mathbf{s},\omega_{j_{\eta}})}\nonumber\\
    &\qquad\qquad\red{+}{(\qnm{H}{j_{\eta}}^*(\mathbf{s},\omega_{j_{\eta}})\times\opvec{n}_s)\cdot\qnm{F}{i_{\mu}}'(\mathbf{s},\omega_{i_{\mu}})} \Big].
\end{align}
Since $\mathcal{S}_j$ is the near field surface of resonator j, so the regularized QNMs $\tilde{\mathbf{F}}_{j\eta}$ and $\tilde{\mathbf{H}}_{j\eta}$ over $\mathcal{S}_j$ could be approximated as QNMs $\tilde{\mathbf{f}}_{j\eta}$ and $\tilde{\mathbf{h}}_{j\eta}$, which then could be described as the surface currents $\tilde{\mathbf{M}}_{j\eta}$ and $\tilde{\mathbf{J}}_{j\eta}$ as follows,

\begin{align}
    &{S}_{i_{\mu}\leftarrow j_{\eta}} = \frac{1}{2\epsilon_0}\frac{1}{i(\Tilde{\omega}_{i_{\mu}}-\Tilde{\omega}^*_{j_{\eta}})}\nonumber\\
    &\times\oint_{\mathcal{S}_j}\mathrm{d}A_s\Big[{(\opvec{n}_s \times\qnm{F}{j_{\eta}}^*(\mathbf{s},\omega_{j_{\eta}}))\cdot \qnm{H}{i_{\mu}}'(\mathbf{s},\omega_{i_{\mu}})}\nonumber\\
    &\qquad\qquad{+}{(\qnm{H}{j_{\eta}}^*(\mathbf{s},\omega_{j_{\eta}})\times\opvec{n}_s)\cdot\qnm{F}{i_{\mu}}'(\mathbf{s},\omega_{i_{\mu}})} \Big],\\
    \approx&\frac{1}{2\epsilon_0}\frac{1}{i(\Tilde{\omega}_{i_{\mu}}-\Tilde{\omega}^*_{j_{\eta}})}\nonumber\\
    &\times\oint_{\mathcal{S}_j}\mathrm{d}A_s\Big[{(\opvec{n}_s \times\qnm{f}{j_{\eta}}^*(\mathbf{s}))\cdot \qnm{H}{i_{\mu}}'(\mathbf{s},\omega_{i_{\mu}})}\nonumber\\
    &\qquad\qquad{+}{(\qnm{h}{j_{\eta}}^*(\mathbf{s})\times\opvec{n}_s)\cdot\qnm{F}{i_{\mu}}'(\mathbf{s},\omega_{i_{\mu}})} \Big],\\
    =&\frac{1}{2\epsilon_0}\frac{1}{i(\Tilde{\omega}_{i_{\mu}}-\Tilde{\omega}^*_{j_{\eta}})}\nonumber\\
    &\times\oint_{\mathcal{S}_j}\mathrm{d}A_s\Big[{(-\qnm{M}{j_{\eta}}^*(\mathbf{s}))\cdot \qnm{H}{i_{\mu}}'(\mathbf{s},\omega_{i_{\mu}})}\nonumber\\
    &\qquad\qquad{+}{(-\qnm{J}{j_{\eta}}^*(\mathbf{s}))\cdot\qnm{F}{i_{\mu}}'(\mathbf{s},\omega_{i_{\mu}})} \Big].
\end{align}

Applied to our system with one mode in each cavity, the expression becomes

\begin{align} \label{appeq:S1left2}
    &{S}_{1\leftarrow 2} \approx -\frac{1}{2\epsilon_0}\frac{1}{i(\Tilde{\omega}_{1}-\Tilde{\omega}^*_{2})}\nonumber\\
    &\oint_{\mathcal{S}_2}\mathrm{d}A_s\Bigg[\qnm{F}{1}'(\mathbf{s},\omega_{1})\cdot\qnm{J}{2}^*(\mathbf{s})\red{+}{\qnm{H}{1}'(\mathbf{s},\omega_{1})\cdot\qnm{M}{2}^*(\mathbf{s})}\Bigg].
\end{align}
Here, slow-
varying envelope functions are defined as \(\qnm{F}{1}'(\mathbf{s},\omega)|_{\mathbf{s}\in \mathcal{S}_2} = \mathrm{e}^{-i\omega\tau_{12}}\qnm{F}{1}(\mathbf{s},\omega)\) and \(\qnm{H}{1}'(\mathbf{s},\omega)|_{\mathbf{s}\in \mathcal{S}_2} = \mathrm{e}^{-i\omega\tau_{12}}\qnm{H}{1}(\mathbf{s},\omega)\) with \(\tau_{12}=n_{\rm B}R_{12}/c\)
and $R_{12}$ is the center-to-center gap distance.
Using \(S_{i_{\mu}\rightarrow j_{\eta}}=\left(S_{j_{\eta}\leftarrow i_{\mu} }\right)^*\), we also have
\begin{align} \label{appeq:S1right2}
    &{S}_{1\rightarrow 2}= (\Tilde{S}_{2\leftarrow 1})^* \approx -\frac{1}{2\epsilon_0}\frac{1}{i(\Tilde{\omega}_{1}-\Tilde{\omega}^*_{2})}\nonumber\\
    &\times\oint_{\mathcal{S}_1}\mathrm{d}A_s\Bigg[\qnm{F}{2}'^*(\mathbf{s},\omega_{2})\cdot\qnm{J}{1}(\mathbf{s})\red{+}{\qnm{H}{2}'^*(\mathbf{s},\omega_{2})\cdot\qnm{M}{1}(\mathbf{s})}\Bigg].
\end{align}

In the calculation of ${S}_{1\leftarrow 2}$ and ${S}_{1\rightarrow 2}$, similar to the calculation of $S_{\rm 12,pole}^{\rm inter}$ (Eq.~\eqref{eq:Sinterdef_pole}), near field surface $\mathcal{S}_2$ ($\mathcal{S}_1$) is chosen as a cuboid surface and is $50~$nm away from dimer 2 (dimer 1). In addition, we also used different surface integral discretizations for regularized fields and surface currents, which are $(5~{\rm nm})^2$ and $(0.5~{\rm nm})^2$, respectively. 

As a result, for gap distance $d_{\rm dap}=2000$nm (corresponding to $R_{12}=2020$nm), we got \blue{${S}_{1\leftarrow 2}=(-0.0068 - 0.0205i)$} 
and \blue{${S}_{1\rightarrow 2}=(0.0095 + 0.0190i)$} as shown in Table~\ref{tab: s_retarded}.

\begin{table}
    \centering
    \caption{Retarded and advanced overlap matrix elements between the modes of two separated metal dimers for $d_{\rm gap}=2000~$nm, calculated using a pole approximation [cf. Eq.~\eqref{appeq:S1left2} and Eq.~\eqref{appeq:S1right2}].}
    \begin{tabular}{c c}\hline\hline
    Matrix element & Pole\\\hline
    $\tilde{S}_{1\leftarrow 2}$ &\blue{$(-0.0068 - 0.0205i)$} \\ 
    $\tilde{S}_{1\rightarrow 2}$ & \blue{$(0.0095 + 0.0190i)$}\\ \hline \hline
    \end{tabular}
    \label{tab: s_retarded}
\end{table}

\subsection{QNMs and regularized QNMs at dipole location}

In the calculation of coupling strength, QNMs and regularized QNMs at the dipole location are required. $\mathbf{r}_{\rm a1}$ ($\mathbf{r}_{\rm a2}$) is at gap center of the dimer 1 (dimer 2). For dimer 1 alone, $z-$component $\tilde{f}_{1z}^{\rm }(\mathbf{r}_{\rm a1})$ of QNM at the gap center dimer 1 is shown in Table~\ref{table: cQNM}, which won't change with the gap distance, same for the QNM $\tilde{f}_{2z}^{\rm }(\mathbf{r}_{\rm a2})$ with dimer 2 alone. 

While the regularized QNM $\tilde{F}_{2z}^{\rm }(\mathbf{r}_{\rm a1},\tilde{\omega}_2^*)$ ($\tilde{F}_{1z}^{\rm }(\mathbf{r}_{\rm a2},\tilde{\omega}_1^*)$) for dimer 2 (dimer 1) alone,  at the $\mathbf{r}_{\rm a1}$ ($\mathbf{r}_{\rm a2}$) will change with the gap distance. With $d_{\rm gap}=2000~$nm, the regularized QNMs are obtained using NF2FF (the near field surface is a cuboid surface $50~$nm away from the surface of dimer 1 or dimer 2)~\cite{ren2020near}, as shown in Table~\ref{table: cQNM}.

\begin{table}[htb]
\caption {Dipole location $\mathbf{r}_{\rm a1}$ ($\mathbf{r}_{\rm a2}$) is fixed at the gap center of the dimer 1 (dimer 2). Bare QNMs $\tilde{f}_{1z}^{\rm }(\mathbf{r}_{\rm a1})$ and $\tilde{f}_{2z}^{\rm }(\mathbf{r}_{\rm a2})$ don't change with gap distance.  Regularized QNMs $\tilde{F}_{2z}^{\rm }(\mathbf{r}_{\rm a1},\tilde{\omega}_2^*)$ and $\tilde{F}_{1z}^{\rm }(\mathbf{r}_{\rm a2},\tilde{\omega}_1^*)$ will change with the gap distance. Here they are shown with a gap distance of $d_{\rm gap}=2000~$nm. 
} \label{table: cQNM} 
    \centering
    \begin{tabular}{|c|c|}
 \hline
~ & $\mathbf{r}_{\rm a1}$   \\
 \hline
$\tilde{f}_{1z}^{\rm }(\mathbf{r}_{\rm a1})$ & $(1.94\times10^{11} + 1.78\times10^{9}i)$ ${\rm [m]}^{-3/2}$   \\
 \hline
$\tilde{F}_{2z}^{\rm }(\mathbf{r}_{\rm a1},\tilde{\omega}_2^*)$ &  \blue{$(-2.12\times10^{7} - 5.79\times10^{7}i)$} ${\rm [m]}^{-3/2}$\\
 \hline
 ~ & $\mathbf{r}_{\rm a2}$   \\
 \hline
$\tilde{f}_{2z}^{\rm }(\mathbf{r}_{\rm a2})$ & $(1.05\times10^{11} + 6.67\times10^{8}i)$ ${\rm [m]}^{-3/2}$   \\
 \hline
$\tilde{F}_{1z}^{\rm }(\mathbf{r}_{\rm a2},\tilde{\omega}_1^*)$ &   \blue{$(3.72\times10^{6} - 4.58\times10^{7}i)$} ${\rm [m]}^{-3/2}$ \\
 \hline

 \end{tabular}
\end{table}

\bibliography{thebiblio}
\end{document}